\documentclass[journal]{IEEEtran}
\usepackage{amsmath,graphicx}
\usepackage{acronym}
\usepackage{accents}
\usepackage{ifthen}
\def\estDim{D}

\def\linewidthA{1.2pt}
\def\marksizeA{3pt}

\acrodef{AWGN}{additive white Gaussian noise}
\acrodef{DNR}{dense-multipath-to-noise ratio}
\acrodef{RIS}{reconfigurable intelligent surface}
\acrodef{OFDM}{orthogonal frequency division multiplexing}
\acrodef{MIMO}{multiple-input-multiple-output}
\acrodef{SIMO}{single-input-multiple-output}
\acrodef{LoS}{line-of-sight}
\acrodef{CDF}{cumulative distribution function}
\acrodef{ECDF}{empirical cumulative distribution function}
\acrodef{AP}{access point}
\acrodef{CP}{carrier phase}
\acrodef{NCP}{no carrier phase}
\acrodef{NLoS}{non-line-of-sight}
\acrodef{SMC}{specular multipath component}
\acrodef{DMC}{dense multipath component}
\acrodef{SNR}{signal-to-noise ratio}
\acrodef{DNR}{dense-multipath-to-noise ratio}   % [TUG]
\acrodef{SDNR}{signal-to-dense multipath-plus-noise ratio}
\acrodef{RS}{radio stripe}
\acrodef{SINR}{signal-to-interference-noise ratio}
\acrodef{BS}{base station}
\acrodef{UE}{user equipment}
\acrodef{ELAA}{extremely large-scale antenna array}
\acrodef{NR}{new radio}
\acrodef{FR2}{frequency range 2}
\acrodef{DL}{downlink}
\acrodef{CPU}{central processing unit}
\acrodef{ISAC}{integrated sensing and communications}
\acrodef{UL}{uplink}
\acrodef{TDoA}{time-difference-of-arrival}
\acrodef{KPI}{key performance indicator}
\acrodef{AoA}{angle-of-arrival}
\acrodef{AoD}{angle-of-departure}
\acrodef{ULA}{uniform linear array}
\acrodef{ML}{maximum likelihood}
\acrodef{PEB}{position error bound}
\acrodef{CEB}{clock error bound}
\acrodef{RMSE}{root-mean-squared error}
\acrodef{FIM}{Fisher information matrix}
\acrodef{EFIM}[EFIM]{equivalent \ac{FIM}}   % [TUG]
\acrodef{RCS}{radar cross section}
\acrodef{CRB}[CRLB]{Cram\'er-Rao lower bound}
\acrodef{LMR}{LoS-to-multipath ratio}
\acrodef{SRS}{sounding reference signal}
\acrodef{ILS}{iterative least squares}
\acrodef{SP}{scatter point}
\acrodef{S-parameter}{scattering parameter}
\acrodef{SDNR}{signal-to-dense multipath-plus-noise ratio}
\acrodef{RP}{reflection point}
\acrodef{PDF}{probability density function}

\usepackage{amsmath,amssymb,epsfig,psfrag,cite,subfigure}
\include{macros}
\usepackage{graphicx}
\usepackage{epstopdf}
\usepackage{balance}        % [TUG] for \balance at the end of the document

\usepackage{bm}
\usepackage{upgreek}        % [TUG] for some alternative greek like \upkappa (?)
\usepackage{tabularx, boldline, multirow} % [TUG] for some fancy tables
\usepackage{array, booktabs} % [TUG] for some fancy tables

\usepackage{geometry}
\usepackage{tcolorbox}

\usepackage{accents}
\newcommand*{\dt}[1]{%
	\accentset{\mbox{\large .}}{#1}}

\allowdisplaybreaks
\usepackage{soul}

\usepackage{amsthm}

\usepackage{nicefrac}

\usepackage{lipsum,multicol}

\usepackage{algorithm}
\usepackage{algpseudocode}
\algdef{SE}[SUBALG]{Indent}{EndIndent}{}{\algorithmicend\ }%
\algtext*{Indent}
\algtext*{EndIndent}

\usepackage{enumitem}
\usepackage{mwe}    
\usepackage{epsfig,psfrag}
\usepackage{subfigure}
\usepackage{color}
\usepackage{url}
\usepackage{mathtools,xparse}

\usepackage{gensymb}

\usepackage[multiple]{footmisc}

\usepackage{extarrows}      % [TUG]     REMOVE THIS!
\usepackage{cuted}          % [TUG]     REMOVE THIS!

\theoremstyle{remark}

\newtheoremstyle{mytheoremstyle} % name
    {\topsep}                    % Space above
    {\topsep}                    % Space below
    {\upshape}                   % Body font
    {.5em}                           % Indent amount
    {\itshape}                   % Theorem head font
    {.}                          % Punctuation after theorem head
    {.5em}                       % Space after theorem head
    {}  % Theorem head spec (can be left empty, meaning ‘normal’)

\theoremstyle{plain}

\newtheoremstyle{iremark}
  {\topsep}   % ABOVESPACE
  {\topsep}   % BELOWSPACE
  {\upshape}  % BODYFONT
  {0.2in}       % INDENT (empty value is the same as 0pt)
  {\itshape}  % HEADFONT
  {.}         % HEADPUNCT
  {5pt plus 1pt minus 1pt} % HEADSPACE
  {\thmname{#1}\thmnumber{ \itshape#2}\thmnote{ (#3)}} % CUSTOM-HEAD-SPEC

\theoremstyle{definition}
\newtheorem*{proof}{Proof}

\DeclareFontFamily{U}{mathx}{\hyphenchar\font45}
\DeclareFontShape{U}{mathx}{m}{n}{
	<5> <6> <7> <8> <9> <10>
	<10.95> <12> <14.4> <17.28> <20.74> <24.88>
	mathx10
}{}
\DeclareSymbolFont{mathx}{U}{mathx}{m}{n}
\DeclareFontSubstitution{U}{mathx}{m}{n}
\newcommand{\rmv}{\hspace*{-.3mm}}

\DeclareMathOperator*{\E}{\mathbb{E}}

\makeatletter \renewcommand\d[1]{\ensuremath{%
		\;\mathrm{d}#1\@ifnextchar\d{\!}{}}}
\makeatother

\makeatletter
\newcommand*\rel@kern[1]{\kern#1\dimexpr\macc@kerna}
\newcommand*\widebar[1]{%
  \begingroup
  \def\mathaccent##1##2{%
    \rel@kern{0.8}%
    \overline{\rel@kern{-0.8}\macc@nucleus\rel@kern{0.2}}%
    \rel@kern{-0.2}%
  }%
  \macc@depth\@ne
  \let\math@bgroup\@empty \let\math@egroup\macc@set@skewchar
  \mathsurround\z@ \frozen@everymath{\mathgroup\macc@group\relax}%
  \macc@set@skewchar\relax
  \let\mathaccentV\macc@nested@a
  \macc@nested@a\relax111{#1}%
  \endgroup
}
\makeatother

\usepackage{scalerel,stackengine}
\stackMath
\newcommand\widecheck[1]{%
\savestack{\tmpbox}{\stretchto{%
  \scaleto{%
    \scalerel*[\widthof{\ensuremath{#1}}]{\kern-.6pt\bigwedge\kern-.6pt}%
    {\rule[-\textheight/2]{1ex}{\textheight}}%WIDTH-LIMITED BIG WEDGE
  }{\textheight}% 
}{0.5ex}}%
\stackon[1pt]{#1}{\scalebox{-1}{\tmpbox}}%
}

\newcommand{\LOS}{\text{\tiny LoS}}

\newcommand{\DMC}{\text{\tiny DMC}}
\newcommand{\RS}{\text{\tiny RS}}
\newcommand{\RP}{\text{\tiny RP}}
\newcommand{\SP}{\text{\tiny SP}}
\newcommand{\UE}{\text{\tiny UE}}

\newcommand{\normbig}[1]{\Big\lVert#1\Big\rVert}

\newcommand{\pp}{\bm{p}}
\newcommand{\ppSP}{\pp_{\scriptscriptstyle \nSP}^{\SP}}     % Position of \nSP-th Scatter Point (SP)
\newcommand{\ppRP}{\pp_{\scriptscriptstyle n,\nRP}^{\RP}}   % Position of \nRP-th Reflection Point (RP) from the perspective of the n-th RS
\newcommand{\ppSPx}[1]{\pp_{#1}^{\SP}}      % Position of #1-th Scatter Point (SP)
\newcommand{\ppRPx}[1]{\pp_{\scriptscriptstyle n,#1}^{\RP}}   % Position of #1-th Reflection Point (RP) from the perspective of the n-th RS
\newcommand{\ppncx}[1]{\pp_{\scriptscriptstyle n,#1}^{\text{\tiny{c}}}}   % Position of #1-th Reflection Point (RP) from the perspective of the n-th RS

\newcommand{\hh}{\bm{h}}
\newcommand{\ww}{\bm{w}}
\newcommand{\aaa}{\bm{a}}
\newcommand{\bb}{\bm{b}}
\newcommand{\cc}{\bm{c}}

\newcommand{\yy}{\bm{y}}
\newcommand{\sss}{\bm{s}}            % transmit signal
\newcommand{\zz}{\bm{z}}
\newcommand{\xx}{\bm{x}}
\newcommand{\xxhat}{\widehat{\xx}}

\newcommand{\BB}{\bm{B}}

\newcommand{\ppPrimecn}{\pp_{\scriptscriptstyle n,\ncomponent}^{\prime}}   % Position of \nRP-th Reflection Point (RP) in local RS coordinates
\newcommand{\yyp}{\yy^{\prime}}

\newcommand{\yypb}{\accentset{\circ}{\yy}^{\prime}}
\newcommand{\ccp}{\cc^{\prime}}

\newcommand{\llr}{\mathcal{L}}

\newcommand{\yyb}{\bm{Y}}

\newcommand{\wwb}{\bm{W}}
\newcommand{\zzb}{\bm{Z}}
\newcommand{\rrb}{\bm{R}}

\newcommand{\mmb}{\bm{M}}                   % azimuth rotation Matrix
\newcommand{\mmbp}{\bm{M}'_{\scriptscriptstyle 0}} % azimuth rotation Matrix canceling the z-component

\newcommand{\ccbp}{\bm{C}^{\prime}}

\newcommand{\deltaf}{\Delta_f}
\newcommand{\deltaphi}{\delta_{\phi}}                   % CP: scalar phase offset
\newcommand{\deltaphib}{\bm{\delta}_{\phi}}             % NCP: N-vector phase offset
\newcommand{\deltaphin}{\delta_{\phi,n}}                % NCP: scalar phase offset of the n-th RS
\newcommand{\deltaphinx}[1]{\delta_{\phi,#1}}           % NCP: scalar phase offset of the #1 RS
\newcommand{\deltatau}{\delta_{\tau}}                   % scalar clock offset

\newcommand{\tauu}{\widetilde{\tau}}

\newcommand{\nw}{\bm{n}_{\scriptscriptstyle\text{\tiny{w}},\ell}}
\newcommand{\ppwl}{\bm{p}^{\text{\tiny w}}_{\ell}}

\newcommand{\er}{ \bm{e}_{\scriptscriptstyle n,\ell} }

\newcommand{\deltac}{\bm{\delta}}    % clock parameter vector including 1 clock offset and 1 phase offset (CP) / N phase offsets (NCP)

\newcommand{\pprs}{\pp^{\RS}}                               % general notation of an RS position
\newcommand{\pprsx}[1]{\pp^{\RS}_{\scriptscriptstyle #1}}   % position of the #1-th RS
\newcommand{\xrs}{p_{\scriptscriptstyle n,x}^{\RS}}
\newcommand{\yrs}{p_{\scriptscriptstyle n,y}^{\RS}}
\newcommand{\zrs}{p_{\scriptscriptstyle n,z}^{\RS}}

\newcommand{\wwdmc}{\ww^{\DMC}}
\newcommand{\wwbdmc}{\wwb^{\DMC}}
\newcommand{\rrbdmc}{\rrb^{\DMC}}
\newcommand{\rrbn}{\rrb_{\scriptstyle n}} % Covariance matrix at the n-th RS

\newcommand{\toep}{{\rm{Toep}}}

\newcommand{\trp}{\mathsf{T}}
\newcommand{\hermit}{\mathsf{H}}
\newcommand{\mtCN}[1]{{\mathcal{CN}_{\!\scriptscriptstyle#1}}}      % following the new "Notation": addint the dimension in the subscript

\newcommand{\etadmc}{\etab_{\DMC}}
\newcommand{\etach}{\etab^{\scriptscriptstyle \rm{ch}}}                                        % per-RS channel parameter vector (dimension 4 * \Nc)
\newcommand{\etachtrp}{\etab^{\scriptscriptstyle \rm{ch}\trp}}
\newcommand{\complexset}[2]{ \mathbb{C}^{#1 \times #2}  }

\newcommand{\complexsett}{ \mathbb{C}  }
\newcommand{\realset}[2]{ \mathbb{R}^{#1 \times #2}  }
\newcommand{\realsetone}[1]{ \mathbb{R}^{#1\times1}  }
\newcommand{\realsett}{ \mathbb{R}  }
\newcommand{\binset}[2]{ \binsett^{#1 \times #2}  }                         % binary set - martix
\newcommand{\binsetone}[1]{ \binsett^{#1}  }                                % binary set - vector
\newcommand{\binsett}{ \mathbb{B}  }                                        % binary set - general

\newcommand{\Imatrix}{{ \bm{\mathrm{I}} }}                                  % Identity matrix
\newcommand{\boldzero}{{ {\bm{0}} }}
\newcommand{\boldone}{{ {\bm{1}} }}
\newcommand{\zeromatrix}[2]{\boldzero_{\scriptscriptstyle #1 \times #2}}    % (#1 x #2) matrix of all zeros
\newcommand{\onematrix}[2]{\boldone_{\scriptscriptstyle #1 \times #2}}      % (#1 x #2) matrix of all ones

\newcommand{\realp}[1]{ \Re\rmv\left\{#1\right\}  }
\newcommand{\imp}[1]{ \Im\rmv\left\{#1\right\}  }
\newcommand{\vecc}[1]{ {\rm{vec}}\left(#1\right)  }

\newcommand{\etab}{ {\boldsymbol{\eta}} }
\newcommand{\etabhat}{ \widehat{\etab} }
\newcommand{\kappab}{ {\boldsymbol{\kappa}} }
\newcommand{\alphab}{ {\boldsymbol{\alpha}} }
\newcommand{\gammab}{ {\boldsymbol{\gamma}} }

\newcommand{\gammabhat}{ {\widehat{\gammab}} }

\newcommand{\phib}{ {\boldsymbol{\phi}} }

\newcommand{\alphabar}{ {\alphab} }

\newcommand{\veccinv}[1]{ {\rm{reshape}}_{M,K}\left(#1\right)  }

\newcommand{\traceAuto}[1]{ {{{\rm{tr}}\left( #1 \right)}}  }       % automatic round brackets

\newcommand{\lightspeed}{\mathsf{c}}                                % vacuum speed of light
\newcommand{\kB}{\mathsf{k}_\mathsf{B}}                                      % Boltzmann constant
\newcommand{\wavenumber}{\mathsf{k}_{\scriptscriptstyle \mathrm{0}}}                                      % wave vector number

\newcommand{\fc}{f_\text{c}}                                        % carrier frequency
\newcommand{\Pt}{P_{\text{t}}}                                      % transmit power

\newcommand{\alphad}{\alpha_{\text{\tiny d}}}                       % DMC peak power
\newcommand{\betad}{\beta_{\text{\tiny d}}}                         % DMC normalized coherence bandwidth
\newcommand{\taud}{\tau_{\text{\tiny d}}}                           % DMC onset time

\newcommand{\hhsum}{{\hh}_{\scriptscriptstyle n,k}}                                                % sum channel vector of Los + RPs + SPs
\newcommand{\hhrp}{{\hh}^{\RP}_{\scriptscriptstyle n, \nRP}}            % sum channel vector of RPs including the LoS
\newcommand{\hhsp}{{\hh}^{\SP}_{\scriptscriptstyle n, \nSP}}            % sum channel vector of SPs
\newcommand{\pathamplitudecn}{{\rho}_{\scriptscriptstyle n, \ncomponent}}% real-valued square-root of path gain of the \ncomponent-th component of RS n
\newcommand{\pathamplitudecnx}[1]{{\rho}_{\scriptscriptstyle n, #1}}% real-valued square-root of path gain of the #1-th component of RS n
\newcommand{\refcoeff}{{\upgamma}_{\scriptscriptstyle n, \nRP}}         % reflection coefficient of the \nRP-th RP of RS n
\newcommand{\rcs}{{\varsigma}_{\scriptscriptstyle n, \nSP}}             % radar cross-section (RCS) of the \nSP-th SP
\newcommand{\radiusSPx}[1]{r^{\SP}_{#1} }                               % radius of the #1-th SP

\newcommand{\SDNRbar}{\mathrm{SDNR}}
\newcommand{\DNR}{\mathrm{DNR}}

\newcommand{\Ei}{\bm{E}^\mathsf{i}_{\scriptscriptstyle n,\nRP}}         % Incident electric field vector
\newcommand{\Er}{\bm{E}^\mathsf{r}_{\scriptscriptstyle n,\nRP}}         % Reflected electric field vector
\newcommand{\Eo}{{E}_0}                                                 % Electric field strength
\newcommand{\poli}{\polRS}                                % Polarization vector of incident wave
\newcommand{\polp}{\bm{\rho}^{\scriptscriptstyle\parallel}_{\scriptscriptstyle n,\nRP}}             % Polarization vector parallel component
\newcommand{\pols}{\bm{\rho}^{\scriptscriptstyle\perp}_{\scriptscriptstyle n,\nRP}}                 % Polarization vector orthogonal component
\newcommand{\polRS}{\bm{\rho}^\RS_{n}}                                  % Polarization vector at the n-th RS
\newcommand{\polUE}{\bm{\rho}^\UE}                                      % Polarization vector at UE

\newcommand{\Rp}{R^{\scriptscriptstyle\parallel}_{\scriptscriptstyle n,\nRP}}                       % Reflection coefficient parallel component
\newcommand{\Rs}{R^{\scriptscriptstyle\perp}_{\scriptscriptstyle n,\nRP}}                           % Reflection coefficient orthogonal component

\newcommand{\anglei}{\vartheta_\mathsf{i}}                              % Incidence angle
\newcommand{\anglet}{\vartheta_\mathsf{t}}                              % Transmission angle

\newcommand{\dimGlobal}{ \mathcal{D}_\mathrm{g} }       % dimension of global parameter vector
\newcommand{\dimLocal}{ \mathcal{D}_\mathrm{ch} }       % dimension of local parameter vector
\newcommand{\dimClock}{ \mathcal{D}_\mathrm{c} }       % dimension of clock parameter vector; NCP: N+1, CP: 2
\newcommand{\NspGlobal}{ J }                                                     % global number of ALL Scatter Points (SPs), some of which may be visible at RS n
\newcommand{\NrpGlobal}{ L }                                                     % global number of ALL Reflection Points (RPs) + the LoS!, some of which may be visible at RS n
\newcommand{\Nrp}{ \NrpGlobal}                                                    % number of Reflection Points (RPs) + the LoS! // discussed with Alessio to replace L_n
\newcommand{\Nsp}{ \NspGlobal}                            % number of Scatter Points (SPs)
\newcommand{\Nc}{ N_{\text{\tiny{c}}} }                   % number of ALL components at EVERY single RS including the Los + RPs + SPs
\newcommand{\ncomponent}{\kappa}                        % an arbitrary component (LoS, RP, SP)
\newcommand{\nRP}{\ell}                                 % an arbitrary Reflection Point (RP)
\newcommand{\nSP}{\iota }                               % an arbitrary Scatter Point (SP)

\newcommand{\fim}{ {\boldsymbol{J}} }                           % general Fisher Information Matrix symbol
\newcommand{\fimetach}{\fim_{\etach}}                           % local channel FIM (general)
\newcommand{\fimetachn}{\fimetach^{\scriptscriptstyle (n)}}     % local channel FIM contributed by the n-th RS
\newcommand{\efim}{\bm{J}_\text{\tiny e}}                       % equivalent FIM (EFIM)

\newcommand{\gammanc}{\gamma_{\scriptscriptstyle n,\ncomponent}}                % compl. amplitude for a single component \ell at RS n
\newcommand{\gammancp}{\gamma_{\scriptscriptstyle n,\ncomponent'}}              % compl. amplitude for a single component \ell-prime at RS n
\newcommand{\aaanc}{ \boldsymbol{\aaa}_{\scriptscriptstyle n,\ncomponent} }     % derivative of spatial (angular) array response
\newcommand{\bbnc}{ \boldsymbol{\bb}_{\scriptscriptstyle n,\ncomponent} }       % derivative of temporal (delay) array response
\newcommand{\aaap}{ \dt{\boldsymbol{\aaa}}_{\scriptscriptstyle n,\ncomponent} } % derivative of spatial (angular) array response
\newcommand{\aaadplain}{ \dt{\boldsymbol{\aaa}}} % derivative of spatial (angular) array response GENERAL
\newcommand{\bbd}{ \dot{\boldsymbol{\bb}}_{\scriptscriptstyle n,\ncomponent} }  % derivative of temporal (delay) response for RS n and component k
\newcommand{\bbdplain}{ \dot{\boldsymbol{\bb}} }  % derivative of temporal (delay) response GENERAL

\newcommand{\muyyni}{ \boldsymbol{\mu}_{n,\ncomponent} }    % Use this for all components (LoS, RPs, SPs)

\newcommand{\ppm}{\pp^{\text{\tiny{m}}}_\nRP}                       % position of the \nRP-th mirror UE
\newcommand{\rangem}{ {\boldsymbol{r}}_{\scriptscriptstyle n,\nRP}^{\text{\tiny{m}}} }  % (3x1) Cart. vector from the Radio Stripe to the \nRP-th mirror UE
\newcommand{\rangeUS}{ {\boldsymbol{r}}_{\scriptscriptstyle \nSP}^{\text{\tiny{US}}} } % (3x1) Cart. vector from the UE to an SP  
\newcommand{\rangeS}{ {\boldsymbol{r}}_{\scriptscriptstyle n,\nSP}^{\text{\tiny{S}}} } % (3x1) Cart. vector from the n-th RS to the \nSP-th SP
\newcommand{\rangeR}{ {\boldsymbol{r}}_{\scriptscriptstyle n,\nRP}^{\text{\tiny{R}}} }              % (3x1) vect. distance from the n-th RS to its \nRP-th RP
\newcommand{\distU}{ {d}_{\scriptscriptstyle n}^{\text{\tiny{U}}} }                     % Scalar LoS distance from the UE to the n-th RS
\newcommand{\distR}{ {d}_{\scriptscriptstyle n,\nRP}^{\text{\tiny{R}}} }                % Scalar distance from the n-th RS to its \nRP-th RP
\newcommand{\distUR}{ {d}_{\scriptscriptstyle n,\nRP}^{\text{\tiny{UR}}} }              % Scalar distance from the UE to the \nRP-th RP of the n-th RS
\newcommand{\distS}{ {d}_{\scriptscriptstyle n, \nSP}^{\text{\tiny{S}}} }               % Scalar distance from the n-th RS to the \nSP-th SP
\newcommand{\distUS}{ {d}_{\scriptscriptstyle \nSP}^{\text{\tiny{US}}} }                % Scalar distance from the UE to the \nSP-th SP

\newcommand{\thetabn}{\bm{\theta}_{\scriptscriptstyle n}}              % AoA of ALL components impinging at RS n 
\newcommand{\thetacn}{\theta_{\scriptscriptstyle n,\ncomponent}}       % AoA of a SINGLE component \ncomponent impinging at RS n 
\newcommand{\thetacnx}[1]{\theta_{\scriptscriptstyle n,#1}}            % AoA of a SINGLE component #1 impinging at RS n 
\newcommand{\thetabsp}{\bm{\theta}^{\SP}_{\scriptscriptstyle n}}       % AoAs of ALL scatter points impinging at RS n 
\newcommand{\thetabrp}{\bm{\theta}^{\RP}_{\scriptscriptstyle n}}       % AoAs of ALL reflection points impinging at RS n 
\newcommand{\thetasp}{{\theta}^{\SP}_{\scriptscriptstyle n, \nSP}}     % AoA of a SINGLE scatter point impinging at RS n 
\newcommand{\thetalos}{{\theta}^{\LOS}_{\scriptscriptstyle n}}         % LoS AoA impinging at RS n 
\newcommand{\thetarp}{{\theta}^{\RP}_{\scriptscriptstyle n, \nRP}}     % AoA of a SINGLE reflection point impinging at RS n 
\newcommand{\thetarpx}[1]{{\theta}^{\RP}_{\scriptscriptstyle n, #1}}   % AoA of a SINGLE RP #1 impinging at RS n 

\newcommand{\thetaspx}[1]{{\theta}^{\SP}_{\scriptscriptstyle n, #1}}     % AoA of a SINGLE RP #1 impinging at RS n 

\newcommand{\tauubn}{\bm{\tauu}_{\scriptscriptstyle n}}                             % Pseudodelays of ALL components impinging at RS n 
\newcommand{\tauucn}{\tauu_{\scriptscriptstyle n,\ncomponent}}                      % Pseudodelay of a SINGLE component \ncomponent impinging at RS n 
\newcommand{\taucn}{\tau_{\scriptscriptstyle n,\ncomponent}}                      % true delay of a SINGLE component \ncomponent impinging at RS n 
\newcommand{\tauubsp}{\bm{\tauu}^{\SP}_{\scriptscriptstyle n}}         % Pseudodelays of ALL scatter points impinging at RS n 
\newcommand{\tauubrp}{\bm{\tauu}^{\RP}_{\scriptscriptstyle n}}         % Pseudodelays of ALL reflection points impinging at RS n 
\newcommand{\tauusp}{\tauu^{\SP}_{\scriptscriptstyle n, \nSP}}         % Pseudodelay of a SINGLE scatter point impinging at RS n 
\newcommand{\tauurp}{\tauu^{\RP}_{\scriptscriptstyle n, \nRP}}         % Pseudodelay of a SINGLE reflection point impinging at RS n 
\newcommand{\tauurpx}[1]{\tauu^{\RP}_{\scriptscriptstyle n, #1}}       % Pseudodelay of a SINGLE RP #1 impinging at RS n 

\newcommand{\tauuspx}[1]{\tauu^{\SP}_{\scriptscriptstyle n, #1}}       % Pseudodelay of a SINGLE SP #1 impinging at RS n 

\newcommand{\tauulos}{\tauu^{\LOS}_{\scriptscriptstyle n}}       % Pseudodelay of the LoS impinging at RS n 
\newcommand{\taulos}{\tau^{\LOS}_{\scriptscriptstyle n}}               % TRUE LoS delay (NOT pseudodelay!) of the LoS component impinging at RS n 
\newcommand{\taurp}{\tau^{\RP}_{\scriptscriptstyle n, \nRP}}           % TRUE delay (NOT pseudodelay!) of a SINGLE reflection point impinging at RS n 
\newcommand{\taurpbar}{\overline{\tau}_{\RP}}                          % average RP delay (NOT pseudodelay!) over all RPs of all RSs
\newcommand{\taulosbar}{\overline{\tau}_{\LOS}}                         % average LoS delay (NOT pseudodelay!) over all RSs
\newcommand{\phibn}{\overline{\bm{\phi}}_{\scriptscriptstyle n}}                  % Phases of ALL components impinging at RS n 
\newcommand{\phicn}{\phi_{\scriptscriptstyle n,\ncomponent}}           % Phase of a SINGLE component \ncomponent impinging at RS n 
\newcommand{\phicnx}[1]{\phi_{\scriptscriptstyle n,#1}}                 % Phase of a SINGLE component #1 impinging at RS n 
\newcommand{\phibsp}{\bm{\phi}^{\SP}_{\scriptscriptstyle n}}           % Phases of ALL scatter points impinging at RS n 
\newcommand{\phibrp}{\bm{\phi}^{\RP}_{\scriptscriptstyle n}}           % Phases of ALL reflection points impinging at RS n 
\newcommand{\phibrpx}[1]{\bm{\phi}^{\RP}_{\scriptscriptstyle#1}} 

\newcommand{\phisp}{\phi^{\SP}_{\scriptscriptstyle n, \nSP}}           % Phase of a SINGLE scatter point impinging at RS n 
\newcommand{\philos}{\phi^{\LOS}_{\scriptscriptstyle n}}               % Phase of the LoS component impinging at RS n 
\newcommand{\phirp}{\phi^{\RP}_{\scriptscriptstyle n, \nRP}}           % Phase of a SINGLE reflection point impinging at RS n 
\newcommand{\phirpx}[1]{\phi^{\RP}_{\scriptscriptstyle n, #1}}        % Phase of a SINGLE RP #1 impinging at RS n 
\newcommand{\phir}{\bm{\phi}}                                     % "reduced" phase vector (excluding the LoS) of ALL RSs
\newcommand{\phirx}[1]{\bm{\phi}_{\scriptscriptstyle#1}}% "reduced" phase vector (excluding the LoS) of RS #1
\newcommand{\varphicnx}[1]{\varphi_{\scriptscriptstyle n, #1}}     % Reflection-induced phase shift of a SINGLE COMPONENT #1 impinging at RS n 
\newcommand{\alphabn}{\bm{\alpha}_{\scriptscriptstyle n}}              % Amplitudes of ALL components impinging at RS n 
\newcommand{\alphabnx}[1]{\bm{\alpha}_{\scriptscriptstyle #1}}         % Amplitudes of ALL components impinging at RS #1 
\newcommand{\alphacn}{\alpha_{\scriptscriptstyle n,\ncomponent}}       % Amplitude of a SINGLE component \ncomponent impinging at RS n 
\newcommand{\alphacnx}[1]{\alpha_{\scriptscriptstyle n,#1}}            % Amplitude of a SINGLE component #1 impinging at RS n 
\newcommand{\alphabsp}{\bm{\alpha}^{\SP}_{\scriptscriptstyle n}}       % Amplitudes of of ALL scatter points impinging at RS n 
\newcommand{\alphabrp}{\bm{\alpha}^{\RP}_{\scriptscriptstyle n}}       % Amplitudes of ALL reflection points impinging at RS n 
\newcommand{\alphasp}{\alpha^{\SP}_{\scriptscriptstyle n, \nSP}}       % Amplitude of a SINGLE scatter point impinging at RS n 
\newcommand{\alphabcn}{\bm{\alpha}_{\scriptscriptstyle n}}              % Amplitudes of ALL components impinging at RS n 

\newcommand{\alphalos}{\alpha^{\LOS}_{\scriptscriptstyle n}}           % Amplitude of the LoS component impinging at RS n 

\newcommand{\gammalos}{\gamma^{\LOS}_{\scriptscriptstyle n}}           % Amplitude of the LoS component impinging at RS n 

\newcommand{\alphabrpx}[1]{\bm{\alpha}^{\RP}_{\scriptscriptstyle#1}}       % Amplitudes of ALL reflection points impinging at RS n 

\newcommand{\rholos}{\rho^{\LOS}_{\scriptscriptstyle n}}           % Rho of the LoS component impinging at RS n 
\newcommand{\alpharp}{\alpha^{\RP}_{\scriptscriptstyle n, \nRP}}       % Amplitude of a SINGLE reflection point impinging at RS n 
\newcommand{\etabWanted}{\etab_{\text{\tiny w}}}                        % parameters of interest
\newcommand{\etabUnwanted}{\etab_{\text{\tiny u}}}                      % nuisance parameters

\newcommand{\ppspbar}{ {{\bm{p}}^{\SP}} }

\newcommand{\jacobian}{ {\boldsymbol{T}} }
\newcommand{\jacobP}{ {\boldsymbol{P}} }
\newcommand{\jacobC}{ {\boldsymbol{C}} }
\newcommand{\jacobA}{ {\boldsymbol{A}} }                    % "association matrix"
\newcommand{\ebn}{ {\boldsymbol{e}_{\scriptscriptstyle n}} }                     % unit vector with the n-th element being 1

\newcommand{\peb}{\mathcal{P}}      % position error bound
\newcommand{\ceb}{\mathcal{C}}      % clock error bound
\newcommand{\coeb}{\ceb_{\tau}}     % clock time offset bound
\newcommand{\cpeb}{\ceb_{\phi}}     % clock phase offset bound

\newcommand{\house}{ {\boldsymbol{\mathcal{H}}} }           % Householder matrix
\newcommand{\BWlow}{ B_{\text{\tiny low}} }             % low bandwidth regime limit
\newcommand{\BWhigh}{ B_{\text{\tiny high}} }           % low bandwidth regime limit
\newcommand{\Sumk}{ S_{\scriptstyle k} }                % sum over k^2
\newcommand{\Summ}{ S_{\scriptstyle m} }                % sum over m^2

\usepackage{placeins}
\usepackage{tikzscale}		% scale tikz plot
\usepackage{tcolorbox}

\usepackage{tikz}

\usepackage[framemethod=TikZ]{mdframed}	% for math boxes with 

\usepackage{pgfplots}
  \tcbset{shield externalize} %there can be problems with tikz externalize and tcolorbox
  \pgfplotsset{compat=newest}
  \usetikzlibrary{plotmarks}
  \usetikzlibrary{arrows.meta}
  
  \usetikzlibrary{arrows}
  \usetikzlibrary{mindmap,trees}%,backgrounds}
  \usetikzlibrary{positioning,spy}
  \usetikzlibrary{calc,fadings,decorations.pathreplacing}
  \usetikzlibrary{decorations.pathmorphing,decorations.text}
  \usetikzlibrary{datavisualization.polar}      % for polar plots
  \usepackage{pgfplotstable}
  \usepgfplotslibrary{polar}
  \usetikzlibrary{patterns}

  \usetikzlibrary{plotmarks}
  \usepgfplotslibrary{patchplots}
  \usepackage{grffile}
  \usepackage{amsmath}
  \usepackage{tikzscale}		% scale tikz plot
  \usepackage{tikz-3dplot} 
  \usetikzlibrary{positioning}
  \usetikzlibrary{arrows}
  \usetikzlibrary{fit,backgrounds}  % for the fit of boxes

\tikzset{%
  >=latex,
  inner sep=0pt,%
  outer sep=2pt,%
  mark coordinate/.style={inner sep=0pt,outer sep=0pt,minimum size=3pt,
  fill=black,circle}%
}
\newcommand{\externalizeFigures}{false}
\ifthenelse{\equal{\externalizeFigures}{true}}
{
  \usepackage{pgfplots}
  \pgfplotsset{compat=newest}
  \usepgfplotslibrary{external}
  \tikzexternalize[prefix=Figures/externalized]
  \usepackage{shellesc}
}
{}

\newlength{\figurewidth}
\newlength{\figureheight}

\usepackage{xcolor}

\definecolor{IEEEblue}{RGB}{0 98 155}
\definecolor{IEEElightblue}{RGB}{0 181 226}
\definecolor{IEEEturquoise}{RGB}{0 156 166}
\definecolor{IEEEred}{RGB}{186 12 47}
\definecolor{IEEEgreen}{RGB}{0 132 61}
\definecolor{IEEElightgreen}{RGB}{120 190 32}
\definecolor{IEEEorange}{RGB}{225 163 0}
\definecolor{IEEEdarkorange}{RGB}{232 119 34}
\definecolor{IEEEyellow}{RGB}{255 209 0}

\definecolor{xcolorMahogany}{rgb}{0.6627 0.2039 0.1216}

\tikzstyle{NodeGlobalUnwanted}=[circle, rounded corners=0.06cm, draw=IEEEorange,fill=IEEEorange!40,minimum size=20pt,inner sep=0pt, font=\small]
\tikzstyle{NodeGlobalWanted}=[circle, rounded corners=0.06cm, draw=IEEEgreen,fill=IEEEgreen!40,minimum size=20pt,inner sep=0pt, font=\small]
\tikzstyle{NodeLocal}=[circle, rounded corners=0.06cm, draw=IEEElightgreen,fill=IEEElightgreen!40,minimum size=20pt,inner sep=0pt, font=\small]
\tikzstyle{NodeKnown}=[circle, rounded corners=0.06cm, draw=IEEEblue,fill=IEEEblue!40,minimum size=20pt,inner sep=0pt, font=\small]
\tikzstyle{NodeUnknown}=[circle, rounded corners=0.06cm, draw=gray,fill=gray!40,minimum size=20pt,inner sep=0pt, font=\small]
\tikzstyle{Function}=[rectangle, rounded corners=0.06cm, draw=gray,fill=gray!20,minimum size=20pt,inner sep=1pt, font=\small,fill opacity=0.95, draw opacity=1,text opacity=1]

\newcommand{\ticked}{$\text{\rlap{$\checkmark$}}\square$}
\newcommand{\unticked}{{$\square$}}
\newcommand{\tick}[1]{\ifthenelse{#1=1}{\ticked}{\unticked}}    % tick marks for itemize: (by TW)

\usepackage{tcolorbox}
\newcommand{\showTodoBoxes}{false}      % SET TRUE TO SHOW TODOBOXES!
\newenvironment{todobox}[1]
{
    \ifthenelse{\equal{\showTodoBoxes}{true}}
    {
        \ifthenelse{\equal{\externalizeFigures}{true}}
        {
            \tikzexternaldisable
        }{}
        {\small
        \begin{tcolorbox}[width=\columnwidth] 
            \textbf{ToDo:}
            \begin{itemize}
                #1
            \end{itemize}
        \end{tcolorbox}}
        \ifthenelse{\equal{\externalizeFigures}{true}}
        {
            \tikzexternalenable
        }{}
    }{}
}

\newlength{\plotWidth}		% length for scaling the plot width
\newlength{\plotHeight}		% length for scaling the plot height

\definecolor{IEEEblue}{RGB}{0 98 155}
\ifthenelse{\equal{\externalizeFigures}{true}}
{
        \DeclareRobustCommand{\roundlabel}[1]{\tikzexternaldisable \tikz[baseline=(char.base)]{
                \node[rectangle, rounded corners=0.65mm,inner sep=0.65mm,fill=IEEEblue, draw=IEEEblue, text=white, font=\itshape](char) {#1};}\tikzexternalenable}
    
    \DeclareRobustCommand{\roundlabeltxt}[1]{\tikzexternaldisable \tikz[baseline=(char.base)]{
                \node[rectangle, rounded corners=0.65mm,inner sep=0.65mm,fill=white, text=IEEEblue, draw=IEEEblue, font=\itshape](char) {#1};}\tikzexternalenable}
    
}
{
    \DeclareRobustCommand{\roundlabel}[1]{\tikz[baseline=(char.base)]{
            \node[rectangle, rounded corners=0.65mm,inner sep=0.65mm,fill=IEEEblue, draw=IEEEblue, text=white, font=\itshape](char) {#1};}}
    
    \DeclareRobustCommand{\roundlabeltxt}[1]{\tikz[baseline=(char.base)]{
                \node[rectangle, rounded corners=0.65mm,inner sep=0.65mm,fill=white, text=IEEEblue, draw=IEEEblue, font=\itshape](char) {#1};}}
    
}

\DeclareRobustCommand{\missing}[1]{\tikz[baseline=(char.base)]{
            \node[rectangle, rounded corners=0.65mm,inner sep=0.65mm,fill=IEEEred, draw=IEEEred, text=white, font=\itshape](char) {Missing Content:};
            } { \color{IEEEred}#1} }

\graphicspath{{./Figures/}}

\newcolumntype{C}{@{\hskip 0.075cm}c@{\hskip 0.075cm}}
\newenvironment{bmatrixs}
  {\left[\begin{array}{*{20}{C}}}
  {\end{array}\right]}

\usepackage[per-mode=symbol]{siunitx}    		% for \SI{}{} units
\DeclareSIUnit{\dBm}{dBm}	% SI unit "dBm"

\ifx\useExternalSupplementary\undefined
    \newcommand{\useExternalSupplementary}{true}   % true: appendix moves to supp.tex // false: appendix from main.tex compiles
\else
\fi

\ifthenelse{\equal{\useExternalSupplementary}{true}}
{
    \usepackage{xcite}
    \externalcitedocument{supp}

    \usepackage{xr-hyper}
    \usepackage[hidelinks]{hyperref}
    
    \externaldocument{supp}
    
}
{
}

\def\lwWall{1pt}

\begin{document}
\bstctlcite{IEEEexample:BSTcontrol}

%%%%%%%%%%%%%%%%%% title page information %%%%%%%%%%%%%%%%%%
\title{Joint Localization, Synchronization and Mapping via Phase-Coherent Distributed Arrays}

\author{Alessio Fascista,~\IEEEmembership{Member,~IEEE}, 
Benjamin J. B. Deutschmann,~\IEEEmembership{Member,~IEEE}, 
Musa Furkan Keskin,~\IEEEmembership{Member,~IEEE},   
Thomas Wilding,~\IEEEmembership{Member,~IEEE}, 
Angelo Coluccia,~\IEEEmembership{Senior Member,~IEEE},
Klaus Witrisal,~\IEEEmembership{Senior Member,~IEEE},
Erik Leitinger,~\IEEEmembership{Senior Member,~IEEE}, 
Gonzalo Seco-Granados~\IEEEmembership{Fellow,~IEEE}, 
Henk Wymeersch,~\IEEEmembership{Fellow,~IEEE}
\thanks{This work is supported, in part, by the SNS JU project 6G-DISAC under the EU's Horizon Europe research and innovation Program under Grant Agreement No 101139130. 
The project has received funding from the European Union’s Horizon 2020 research and
innovation programme under grant agreement No 101013425 (Project ``REINDEER").
The financial support by the Christian Doppler Research Association, the Austrian Federal Ministry for Digital and Economic Affairs and the National Foundation for Research, Technology and Development is gratefully acknowledged.}}

\maketitle

% ---------- Page numbers: REMOVE THEM ALLTOGETHER FOR THE FINAL SUBMISSION ----------
\thispagestyle{empty} %--to make title page numberless
\pagestyle{empty}    % -- to make other pages numberless.
% ---------- Page numbers: REMOVE THEM ALLTOGETHER FOR THE FINAL SUBMISSION ----------

% ---------- Page numbers: put them on ALL pages ONLY for the INITIAL SUBMISSION ----------
\thispagestyle{plain}
\pagestyle{plain}
% ---------- Page numbers: put them on ALL pages ONLY for the INITIAL SUBMISSION ----------

\begin{abstract}
Extremely large-scale antenna array (ELAA) systems emerge as a promising technology in beyond 5G and 6G wireless networks to support the deployment of distributed architectures. This paper explores the use of ELAAs to enable joint localization, synchronization and mapping in sub-6 GHz uplink channels, capitalizing on the near-field effects of phase-coherent distributed arrays. We focus on a scenario where a single-antenna user equipment (UE) communicates with a network of access points (APs) distributed in an indoor environment, considering both specular reflections from walls and scattering from objects. The UE is assumed to be unsynchronized to the network, while the APs can be time- and phase-synchronized to each other. We formulate the problem of joint estimation of location, clock offset and phase offset of the UE, and the locations of scattering points (SPs) (i.e., mapping). Through comprehensive Fisher information analysis, we assess the impact of bandwidth, AP array size, wall reflections, SPs and phase synchronization on localization accuracy. Furthermore, we derive the \ac{ML} estimator for the joint localization, synchronization, and mapping problem, which optimally combines the information collected by all the distributed arrays. To overcome its intractable high dimensionality, we propose a novel three-step algorithm that first estimates phase offset leveraging carrier phase information of line-of-sight (LoS) paths, then determines the UE location and clock offset via LoS paths and wall reflections, and finally locates SPs using a null-space transformation technique. Simulation results demonstrate the effectiveness of our approach in distributed architectures supported by radio stripes (RSs) --- an innovative alternative for implementing ELAAs --- while revealing the benefits of carrier phase exploitation and showcasing the interplay between delay and angular information under different bandwidth regimes.

	\textit{Index Terms--} Extremely large-scale antenna arrays, distributed architectures, radio stripes, positioning, phase synchronization, carrier phase.
	%\vspace{-0.1in}
\end{abstract}

%%%%%%%%%%%%%%%%%%%%%%%%%%  body  %%%%%%%%%%%%%%%%%%%%%%%%%%
\section{Introduction}
%\expandafter\show\the\font
Different generations of wireless communication systems have boosted communication rates by essentially harnessing two orthogonal dimensions: frequency \cite{dang2020should} and space \cite{chen2020beam,wang2024tutorial}. In terms of frequency, larger bandwidth available at increasingly higher carrier frequencies brought commensurate surges in capacity, motivating the transition from sub-6 GHz to mmWave in 5G \cite{hong2021role}, and the exploration of sub-THz in 6G \cite{rikkinen2020thz}. In terms of space, larger antenna arrays provide spatial multiplexing and diversity gains, thus increasing either rate or reliability. This spurred the introduction of MIMO in 3G-4G  \cite{lee2009mimo}, leading to massive MIMO in 5G, and now \ac{ELAA} in 6G \cite{an2024near}. The increase in both bandwidth and antenna sizes led not only to improvements in data rates, but also, first as a side-effect and now as a key service for 6G, improvements in radio localization \cite{chen20246g}. 

This interest in localization is motivated primarily by the high resolution brought in the delay/distance domain via the bandwidth and in the angle domain via the array size \cite{del2017survey,mMIMO_CPP_TWC_2019}. Combined, this resolution unleashes extreme accuracies in support of demanding 6G use cases such as extended reality, cooperating robots, and digital twins \cite{behravan2022positioning}. At the same time, due to the large bandwidth and apertures, conventional channel models were no longer appropriate, and effects such as wavefront curvature and channel non-stationary must be properly modeled to reap the full benefits \cite{elzanaty2023near}. 

\begin{figure}[t]
    \centering
    \vskip 0pt
    \setlength{\plotWidth}{0.7\linewidth}
    \input{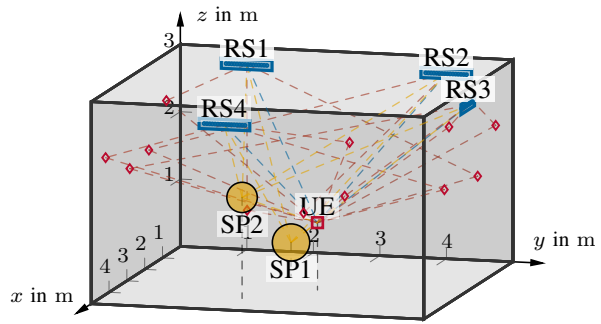}
    \vspace{-0.2cm}\caption{Representative scenario of uplink user equipment (UE) localization supported by a network of $N\!=\!4$ distributed radio stripes.
    There are $\NspGlobal\!=2\!$ dominant scatterers, $4$  walls, and a single \ac{UE}.}
     \label{fig:scenario}
\end{figure}

Despite initial excitement about sub-THz bands, the reality is that mainly lower bands %, either below 6 GHz or between 7 to  15 GHz 
will support the bulk of 6G traffic \cite{wild20236g}. This has put more focus on localization using  \ac{ELAA} systems with limited bandwidths. Such systems can be realized in various forms, such as a \emph{physically large collocated antenna array},  \emph{\ac{RIS}}, \emph{distributed arrays} or \emph{\acp{RS}} \cite{an2024near}. Among cell-free implementation alternatives, the \acp{RS} technology holds great potential as a cost-efficient distributed architecture for dense area deployments, such as stadiums and railway stations \cite{ubiquitous_cellFree_2019,seqRS_TCOM_2021}.
Serially connected \acp{RS} communicate with a \ac{CPU} via a shared bus that simultaneously provides synchronization and power supply \cite{ubiquitous_cellFree_2019}. Unlike stripe-like \acp{RIS} \cite{LOS_NLOS_NearField_2021}, \acp{RS} can actively transmit and sample/process received signals. From the viewpoint of positioning and sensing, synchronization of distributed arrays has been experimentally shown to improve \ac{AoA} estimation accuracy \cite{sync_AOA_TAP_2019}. 
In general, the different realizations of \ac{ELAA} systems vary mainly in terms of whether observations are available in element-space (i.e., at each antenna) or beam-space (i.e.,  projected into a lower-dimensional space), following the terminology from \cite{zoltowski1996closed}. Research on localization can  similarly be divided based on these different points of view \cite{elzanaty2023towards}. 

In terms of large collocated arrays, \cite{yang2023near} considers narrowband localization using a dynamic metasurface antenna array, exploiting only the wavefront curvature in beam-space. A related problem is tackled in \cite{lu2023near}, considering again narrowband beam-space observations, but also accounting for multipath propagation. A far from exhaustive list of RIS-based works includes \cite{elzanaty2023towards,ozturk2023ris,rahal2024ris,dardari2021nlos}. Finally, distributed arrays, comprising many phase-coherent \acp{AP} provide an extremely large near-field region, which was harnessed in \cite{coherentDistMIMO_loc_2019,coherentDistMIMO_loc_2018,coherentDistMIMO_loc_CRB_2021,SA_3D_Loc_TSP_2022,fascista2023uplink}. Specifically, the works \cite{coherentDistMIMO_loc_2019,coherentDistMIMO_loc_2018,coherentDistMIMO_loc_CRB_2021} consider wideband element-space observations at mmWave, combining the benefits of a wide aperture with a high delay resolution. Unfortunately, such systems are impractical since (i) phase synchronization among distributed arrays at mmWave bands is extremely challenging due to hardware imperfections (e.g., phase noise and frequency errors) \cite[Sec.~6.2.1]{hexax_d22} and (ii) location calibration at a fraction of the wavelength (which is \SI{1}{\centi\metre} at \SI{30}{\giga\hertz}) is also extremely difficult. Localization using space angle (SA) measurements at distributed linear arrays was studied in \cite{SA_3D_Loc_TSP_2022} in the absence of phase synchronization between the arrays and without considering any specific frequency band. Surprisingly, there are only limited research activities on the very practical setting of distributed cell-free MIMO localization under realistic propagation conditions at lower bands. % (below 15 GHz). 
We note that this problem bears some resemblance to the carrier phase positioning problem \cite{nikonowicz2024indoor}. 
Our previous work \cite{fascista2023uplink}, which forms the basis of the current paper, considered the impact of phase-coherent and phase-incoherent operation on uplink positioning at sub-6 GHz in realistic channels exhibiting \acp{DMC} \cite{richter2005estimation}.

Starting from our preliminary analysis in \cite{fascista2023uplink}, this study addresses the problem of uplink joint positioning, synchronization, and environment mapping using distributed radio stripes (RSs) at sub-6 GHz frequencies, and provides a comprehensive performance investigation by deriving novel estimators and fundamental bounds under various settings, considering \acp{RS} different levels of phase/time synchronization and diverse multipath characteristics in indoor sub-6 GHz environments. The main contributions are as follows:

\emph{(i)} We investigate the problem of joint uplink positioning, synchronization and mapping with distributed \acp{RS}, considering the distinctive properties of sub-6 GHz operation, including phase and time synchronization capabilities \cite{REINDEER_D2_1}, dense multipath environment \cite{GreLeiWitFle:TWC2024,VenLeiTerWit:TWC2023} and \ac{CP} exploitation. For this challenging scenario, we propose a detailed generative model that accounts for the peculiar characteristics of the channels over the distributed \ac{RS} network and allows for computation of the theoretical lower bounds under various operational assumptions.

\emph{(ii)} We derive the \ac{ML} estimator for joint localization, synchronization, and mapping, which optimally combines signal observations collected at all \acp{RS}. Through a convenient re-parameterization, we are able to obtain ML estimates of the channel amplitudes for all involved paths in closed form, thereby reducing the dimensionality of the estimation process. To make the estimation task feasible, we propose novel reduced-complexity estimation algorithms for the remaining parameters in three steps.  First, we estimate the phase offset by leveraging \ac{CP} information of \ac{LoS} paths. Subsequently, this is used to determine an initial estimate of the \ac{UE} location and clock offset through a suitable relaxed \ac{ML} estimator. Finally, such initial estimates are exploited for low-complexity mapping of dominant scatterers in the environment.

\emph{(iii)} We derive the \acp{CRB} on position, clock and phase offsets estimation, including mapping of \acp{SP}, both with and without exploiting \ac{CP} information. Our investigation highlights the impact of multipath overlap on estimation performance.
    
\emph{(iv)} We carry out extensive simulation analyses to showcase the impact of various system parameters (bandwidth, aperture size, \ac{SDNR}, presence/absence of phase synchronization, reflector walls and \acp{SP}) on position and clock offset estimation accuracy, offering insights into practical \ac{RS} deployments towards 6G networks.

\textit{Notations:} Vectors and matrices are denoted by boldface lower-case and upper-case letters, respectively.
Symbols $(\cdot)^{-1}$, 
 $(\cdot)^\trp$,  $(\cdot)^\hermit$, and $(\cdot)^\dag$  denote the inverse, transpose,
conjugate transpose (hermitian), and left pseudoinverse, respectively. $j=\sqrt{-1}$ is the imaginary unit and we denote with $\realp{\cdot}$ and $\imp{\cdot}$ the real and imaginary part operators. As to numerical sets, 
$\mathbb{R}$ and $\mathbb{C}$ represent real and
complex numbers sets, respectively. We use $\binsett\!=\!\{0,1\}$ to denote the binary set.  $\toep(\xx, \xx^\hermit)$ denotes a Hermitian Toeplitz matrix with the first column $\xx$ and the first row $\xx^\hermit$. $\veccinv{\cdot}$ reshapes a vector into an $M \times K$ matrix.
% TUG: ----------
We use $\zeromatrix{M}{N}$ and $\onematrix{M}{N}$ to denote $M \times N$ matrices of all zeros and all ones, respectively.
The identity matrix is $\Imatrix$, where the size is left implicit. We use  $\bm{x} \sim \mtCN{N}(\bm{0}, \bm{M})$ if $\bm{x} \in \mathbb{C}^N$ is a
complex normal ($N$-dimensional) vector with zero mean and 
(Hermitian)
positive definite covariance matrix $\bm{M}\!\in\!\mathbb{C}^{N \times N}$. $\mathbb{E}[\cdot]$ denotes the expected value, $\|\cdot\|$ is the Euclidean norm of a vector, whereas $\odot$ and $\otimes$ stand for Hadamard and Kronecker products.

\todobox{
    \item[\tick{0}] Decide on: ``$n$-th'' RS or ``$n$\textsuperscript{th}'' RS
    \item[\tick{0}] Possibly introduce $k \in \{0 \ \cdots \ K\}$ ?
    \item[\tick{0}] Possibly change $\rrb \rightarrow \rrb_n$ ?.
}

%%%%%%%%%%%%%%%%%%%%%%%%%%%%%%%%%%%%%%%%%%%%%%%%%%%%%%%%
%%%%%%%%%%%%%%%%%%%%%%%%%%%%%%%%%%%%%%%%%%%%%%%%%%%%%%%%
\section{System Model}
Consider a \acp{RS} network composed of $N$ stripes, each consisting of $M$ antennas, and a single \ac{UE} equipped with a single antenna and communicating with the network through the \ac{UL} channel \cite{seqRS_TCOM_2021,RS_WPT_TWC_2022}. 
We will address both cases of presence or absence of \textit{phase synchronization} between the \ac{UE} and the \acp{RS} network. In the former case, phase coherence effectively turns the \acp{RS} network into a large multiple-antenna access point \cite[Sec.~3.1]{ubiquitous_cellFree_2019}, while the \ac{UE} has an unknown \textit{phase offset} $\deltaphi$ and unknown \textit{clock offset} $\deltatau$ with respect to the \acp{RS} network. 
In addition, we assume that the wavefront of the signal transmitted by the UE is planar over each individual RS (i.e., individual RSs lie in the far-field of the UE). However, we do not assume a planar but spherical wavefront over the entire RS network 
%it might be located in the near-field of the whole RSs network 
due to a large aperture distributed over a wide area.
That is, we use a planar wavefront model per \ac{RS} but a spherical wavefront model for the whole \ac{RS} network.
Moreover, the \acp{RS} are
%equally distributed 
deployed around a rectangle 
%perimeter of length $L$, the latter 
 at a specific height, as depicted in Fig.\,\ref{fig:scenario}. Each individual \ac{RS} is placed at a known phase center position $\pprsx{n} = [\xrs \ \yrs \ \zrs]^\trp$ with known azimuth orientation $\beta_n$ around the $z$-axis, which is measured counter-clockwise from the $x$-axis\footnote{The orientation of each RS is defined by a single angle representing the rotation around the $z$-axis, meaning that the RSs are parallel to the $x$-$y$ plane.}, 
 % V1: 
 while the \ac{UE} is located at an unknown position $\pp = [p_x \ p_y \ p_z]^\trp$.
 % V2: 
 
We consider a \ac{UL} communication scenario where the \ac{UE} transmits \ac{OFDM} pilots $\sss = [ s_{\scriptscriptstyle 1} \ \cdots \ s_{\scriptscriptstyle K} ]^\trp \in \complexset{K}{1}$ over $K$ subcarriers with  spacing $\deltaf$, e.g., \ac{SRS} for 5G \ac{NR} \ac{UL} positioning \cite{3GPP_1810532},  $\lVert \sss \rVert \triangleq 1$. Assuming quasi-static block fading and transmit power $\Pt$ (in $\SI{}{\watt}$), the UL received signal  %\unit[power-half-as-sqrt]{\Hz\tothe{0.5}} 
at the $n$-th RS over subcarrier $k$ is \cite{mMIMO_CPP_TWC_2019}
\begin{align}\label{eq_ynk}
    \yy_{\scriptscriptstyle n,k} = \hhsum s_k \sqrt{\Pt} + \wwdmc_{\scriptscriptstyle n,k} s_{\scriptscriptstyle k} + \zz_{\scriptscriptstyle n,k} \quad \in \complexset{M}{1} ~,
\end{align}
where $\wwdmc_{\scriptscriptstyle n,k}\!\in\!\complexset{M}{1}$ is due to \acp{DMC} and $\zz_{\scriptscriptstyle n,k} \!\in\! \complexset{M}{1} $ denotes circularly symmetric complex Gaussian thermal noise with $\zz_{\scriptscriptstyle n,k} \sim \mtCN{M}(\boldzero, \allowbreak \frac{\sigma^2}{K} \Imatrix )$. 
%
%\subsection{Thermal Noise}
The total noise power over all $K$ subcarriers is 
%\begin{align}
    $\sigma^2 = \kB T B$
%\end{align}
where $\kB$ %in \SI{}{\joule\per\kelvin} 
is the Boltzmann constant, $T$ the absolute temperature, %in \SI{}{\kelvin} 
and $B=\deltaf  K$ the  bandwidth.
$%\begin{align}
    \hhsum = \sum_{\nRP = 0}^{L-1} \hhrp + \sum_{\nSP = 1}^{J} \hhsp  \in \complexset{M}{1}
$ %\end{align}
 denotes the multipath channel consisting of a sum of \acp{SMC} at \acp{RP} $\hhrp$ including the \ac{LoS} $(\nRP= 0)$, and reflections at \acp{SP} $\hhsp$.
To ease notation, we define the sum channel as
\begin{align}\label{eq_SMC_channel}
    \hhsum = \sum_{\ncomponent = 0}^{\Nc-1} \pathamplitudecn e^{j\phicn} \aaa(\thetacn) e^{-j 2 \pi k \deltaf \tauucn} 
\end{align} 
where $\ncomponent \in \{0 \ \ldots \ \Nc\!-\!1 \}$ indicates any component out of $\Nc = \Nrp + \Nsp$ (with $\mathcal{N}_c=\mathcal{\Nrp}\cup \mathcal{\Nsp}$)
components impinging on \ac{RS} $n$, encompassing the \roundlabeltxt{\ac{LoS}} ($\kappa=0$), all \roundlabeltxt{\acp{RP}} ($\kappa\in\mathcal{\Nrp} \setminus \{0\}$), and all \roundlabeltxt{\acp{SP}} ($\kappa\in\mathcal{\Nsp}$). $\mathcal{\Nrp}$  contains the indices of \ac{LoS} and \acp{RP}, and $\mathcal{\Nsp}=\{\Nrp+1\ \ldots\ \Nrp + \Nsp\}$ contains \acp{SP}, with $\Nrp=|\mathcal{\Nrp}|$ and $\Nsp=|\mathcal{\Nsp}|$ being the corresponding set cardinalities. 
The \textit{unknown} channel parameters for each component $\ncomponent$ are: amplitude $\pathamplitudecn$, phase $\phicn$, (pseudo-) delay $\tauucn$, and  \ac{AoA} $\thetacn$.

\section{Signal and Channel Models}\label{sec_signal_channel_model}

In this section, we first develop a suitable formalization for the channel models of the distributed \acp{RS}, emphasizing the fundamental differences between \ac{RP} and \ac{SP} channels. Then, our contribution will be to bridge such models to the per-\ac{RS} observed signal model used to infer the parameters relevant to the localization problem.
From the channel parameters in~\eqref{eq_SMC_channel},

$\bullet$ the amplitude $\pathamplitudecn \in \realsett$ represents the magnitude of the forward transmission coefficient\footnote{The complex channel amplitude $\pathamplitudecn e^{j\phicn}$ represents a \ac{S-parameter} such that its squared magnitude corresponds to the path gain between the antenna ports (i.e., the reciprocal of the path loss).} 
between the \ac{UE} antenna and a hypothetical isotropic antenna at the phase center position $\pprsx{n}$.
    Note that we lump together the unknown path amplitude $\pathamplitudecn$ and the known transmit power $\Pt$ into a single nuisance parameter $\alphacn=\sqrt{\Pt}\pathamplitudecn$, where for notational convenience the \ac{LoS} amplitude is $\alphalos \triangleq \alphacnx{0}$ (similarly, $\rholos \triangleq \pathamplitudecnx{0}$).

$\bullet$ $\phicn$ is the phase term involving the effects of one-way signal propagation, phase shift induced by reflection, and phase offset between the $n$\textsuperscript{th} \ac{RS} %\acp{RS} network 
    and the \ac{UE}, given by
    \begin{equation} \label{eq_phirp}
        \phicn = -2\pi\fc \taucn + \varphicnx{\ncomponent} + \deltaphin    \,,
    \end{equation}
    where $\philos \triangleq \phirpx{0}$ is the \ac{LoS} component phase and $\taucn$  the one-way delay between signal transmission at the \ac{UE} and  reception of component $\ncomponent$ at \ac{RS} $n$.
     $\taulos \triangleq \tau^{\RP}_{\scriptscriptstyle n, 0}$ denotes the delay of the \ac{LoS} component. 
    %
    %\footnote{We consider the most general scenario in which the reflection points are different for each \ac{RS}.}. 
    In addition, the reflection-induced phase shift $\varphicnx{\ncomponent}$ is assumed unknown for \ac{NLoS} components $\ncomponent > 0$, while $\varphicnx{0} \triangleq 0$ for the \ac{LoS}\footnote{The integer phase ambiguity in $\phicn$ in \eqref{eq_phirp} can be entirely absorbed into the unknown phase offset $\deltaphi$, thus $\varphicnx{0} = 0$.}.

 $\bullet$ $\tauucn$ is the pseudo-delay including the effect of one-way propagation and the clock offset of the \ac{UE}, namely
\begin{align} \label{eq_tau_pseudo}
    \tauucn = \taucn + \deltatau \,.
\end{align}
where the component delay is defined as
% \begin{align*}
%     \taucn = \begin{cases} \frac{1}{\lightspeed}  \lVert \ppRP - \pprsx{n} \rVert \,, & \kappa=0, \roundlabeltxt{LoS}
%     \\
%     \frac{1}{\lightspeed} \left( \|\pp - \ppRP  \| + \|\ppRP - \pprsx{n}  \| \right) \, , & \kappa\in\mathcal{\Nrp}, \roundlabeltxt{RP} \\
%     \frac{1}{\lightspeed} \left( \|\pp - \ppSP  \| + \|\ppSP - \pprsx{n}  \| \right) \, ,& \kappa\in\mathcal{\Nsp}, \roundlabeltxt{SP}
%     \end{cases} 
% \end{align*}
\begin{align}\label{eq_taurp}%\color{red}
 \taucn  %\tau(\bm{p}_{n,\kappa}) 
 = \frac{1}{\lightspeed} \left( \|\pp - \ppncx{\ncomponent} \| + \|\ppncx{\ncomponent} - \pprsx{n}  \| \right)
\end{align}
%\com{
inserting for the position $\ppncx{0}=\bm{p}$ for $\ncomponent=0$ (\ac{LoS}), and $\ppncx{\ncomponent}=\ppRPx{\nRP} $ for $\ncomponent\in \mathcal{\Nrp}$ or $\ppncx{\ncomponent}=\ppSPx{\nSP}$ for $\ncomponent\in \mathcal{\Nsp}$ for an \ac{RP} or \ac{SP}, respectively. Indices $\ncomponent$ are mapped to  \ac{RP} and \ac{SP} indices using  respective functions $\nRP(\ncomponent) = \ncomponent$ and $\nSP(\ncomponent) = \ncomponent - \Nrp$.%}
%where again we pose         
 
    $\bullet$ $\aaa(\thetacn) \in \complexset{M}{1}$ is the RS array response to a signal impinging with \ac{AoA} $\thetacn$ (azimuth relative to the boresight of the $n$-th RS array). Without loss of generality, we assume  each \ac{RS} is equipped with a \ac{ULA} with  spacing %\footnote{For \ac{RS} deployments, the element spacing can be larger than the standard half-wavelength spacing to increase spatial resolution \cite{RW_Indoor_2022}, \cite[Ch.~2.2.4]{REINDEER_D3_1}.} 
    $d$,
%antenna elements spaced a half-wavelength apart
so that the array response vector takes the form
\begin{equation}\label{eq:spatial-response}
    \aaa(\theta) \triangleq %\frac{1}{\sqrt{M}}
    \left[1 \ e^{j \frac{2\pi}{\lambda}d \sin \theta} \ \cdots \ e^{j\frac{2\pi}{\lambda}d (M-1) \sin \theta}\right]^\trp~,
\end{equation}
with $\lambda = \lightspeed/\fc$ denoting the wavelength, and $\lightspeed$ and $\fc$ denoting the speed of light and carrier frequency, respectively. 
The \ac{AoA} $\thetacn$ relates the known position and orientation of the $n$-th RS and the unknown \ac{UE}, \ac{RP}, or \ac{SP} position according to
\begin{equation} \label{eq_thetanl}
    \thetacn = \frac{\pi}{2} - \mathrm{atan2}\left( [\ppPrimecn]_2 , [\ppPrimecn]_1 \right) ~,
\end{equation}
where $\mathrm{atan2}(y,x)$ denotes the four-quadrant arc tangent function, and we implicitly make the distinction between \acp{AoA} of the \ac{LoS}, \acp{RP}, and \acp{SP} such that
% \begin{align} \label{eq_ppsnl}
%     \ppPrimecn = \begin{cases} \mmb^{-1}(\beta_n)(\pp - \pprsx{n}) \,, & %\kappa=0, 
%     \roundlabeltxt{LoS}
%     \\
%     \mmb^{-1}(\beta_n)(\ppRP - \pprsx{n}) \, , & %\kappa\in\mathcal{\Nrp}\setminus \{0\}, 
%     \roundlabeltxt{RP} \\
%     \mmb^{-1}(\beta_n)(\ppSP - \pprsx{n}) \, , & %\kappa\in\mathcal{\Nsp}, 
%     \roundlabeltxt{SP}
%     \end{cases} 
% \end{align}
% \begin{align}\color{red}
%     \ppPrimecn = \begin{cases} \mmb^{-1}(\beta_n)(\pp - \pprsx{n}) \,, & \kappa=0, \roundlabeltxt{LoS}
%     \\
%     \mmb^{-1}(\beta_n)(\bm{p}^\mathrm{RP}_{n,\ell(\kappa)} - \pprsx{n}) \, , & \kappa\in\mathcal{\Nrp} \setminus \{0\}, \roundlabeltxt{RP} \\
%     \mmb^{-1}(\beta_n)(\bm{p}^\mathrm{SP}_{\iota(\kappa)} - \pprsx{n}) \, , & \kappa\in\mathcal{\Nsp}, \roundlabeltxt{SP}
%     \end{cases} 
% \end{align}
\begin{align}\label{eq_ppsnl}%\color{red}
 \ppPrimecn = f(\ppncx{\ncomponent}) = \mmb^{-1}(\beta_n)(\ppncx{\ncomponent} - \pprsx{n})
\end{align}
is the \ac{UE} ($\ncomponent=0$), \ac{RP} ($\ncomponent\in\mathcal{\Nrp}\setminus \{0\}$), or \ac{SP} ($\ncomponent\in\mathcal{\Nsp}$) position in the local reference frame of the $n$-th \ac{RS}, and $\mmb(\beta)$ is the rotation matrix around the $z$-axis (see Supplementary Material~\ref{app_rot_mat}) and $\beta_n$ is the azimuth rotation angle of the \ac{RS}.

\subsection{Specular Multipath Components}\label{sec_SMC_model}
The peculiarity of a specular reflection at a surface is that the incident and reflected rays form equal angles with the surface normal. 
Such specular reflections typically occur when the wavelength is large compared to the surface roughness~\cite{Kulmer2018RoughSurface}. 
Specular reflections at large planar surfaces are particularly relevant indoors, which we model through a geometric channel~\cite{LeiVenTeaMey:TSP2023}.
The point at a specularly reflecting surface where a ray cast by the \ac{UE} impinges before getting reflected to the $n$-th \ac{RS} is denoted $\ppRP$ and defined through our geometric model in Supplementary Material~\ref{app_reflection_point_pos}. 
With $\thetalos \triangleq \thetacnx{0}$ denoting the \ac{AoA} of the \ac{LoS} component, the \acp{AoA} of \acp{RP} $\thetarp$ are described through~\eqref{eq_thetanl} by inserting the \ac{RP} position $\ppRP$~\eqref{eq_ppsnl}. 
While the \ac{LoS} delay $\taulos = \frac{1}{\lightspeed}  \lVert \pp - \pprsx{n} \rVert$, the \ac{RP} delay $\taurp$ is computed by inserting $\ppRP$ in~\eqref{eq_taurp}, which
%\begin{align}\label{eq_taurp}
%    \taurp = 
%    \frac{1}{\lightspeed} \left( \|\pp - \ppRP  \| + \|\ppRP - \pprsx{n}  \| \right) \,
%\end{align}
describes the propagation from the \ac{UE} over the $\nRP$-th \ac{RP} of \ac{RS} $n$ to its phase center position.
This will be referred to as \textit{\ac{RP} channel model}, where we relate the amplitudes $\alpharp$ and phases $\phirp$ to the environment geometry through the Friis equation reformulated for power-wave amplitudes,  defined in Supplementary Material~\ref{app_RP_channel}.
%Per \ac{RS} $n$, each specular surface would give rise to a single \ac{RP}. 
We consider $\Nrp$ specular surfaces, each of which would give rise to a single \ac{RP} per \ac{RS} $n$.
However, each \ac{RS} is mounted on a surface that does not produce an \ac{SMC} for that respective \ac{RS}, thus the total number of \acp{RP} including the \ac{LoS} equals $L$ for each \ac{RS}.

\subsection{Scatter Point Components}
%As for \acp{SP} (\,\ref{pgf:SP}\,)
Contrary to the unidirectional scattering (i.e., equal incident and reflecting angles) of a specular reflection at a large planar surface, the characteristic feature of a reflection at an \ac{SP} is its omnidirectional rescattering. 
In real-life scenarios, small metallic objects, corners, metal poles, or window frames often possess the characteristics of \acp{SP}~\cite{Xuhong2024MultiFeatures}.
An \ac{SP} component originates from the \ac{UE} position, impinges at the $\nSP$-th \ac{SP} at $\ppSP$ and travels to the $n$-th \ac{RS} at $\pprsx{n}$.
The \acp{AoA} of \acp{SP} $\thetasp$ are described through~\eqref{eq_thetanl} by inserting the \ac{SP} position $\ppSP$ in~\eqref{eq_ppsnl}. 
The \ac{SP} component delay is computed by inserting $\ppSP$ in \eqref{eq_taurp}.
%describes the propagation from the \ac{UE} over the $\nRP$-th \ac{RP} of \ac{RS} $n$ to its phase center position.
This will be referred to as \textit{\ac{SP} channel model}, where we relate the amplitudes $\alphasp$ and phases $\phisp$ to the environment geometry through the bistatic radar range equation reformulated for power-wave amplitudes, defined in Supplementary Material~\ref{app_SP_channel}.

\begin{figure*}[t] % Use the figure* environment to span both columns
    \centering
    \includegraphics[trim=0 30 60 0,clip,width=0.75\textwidth]{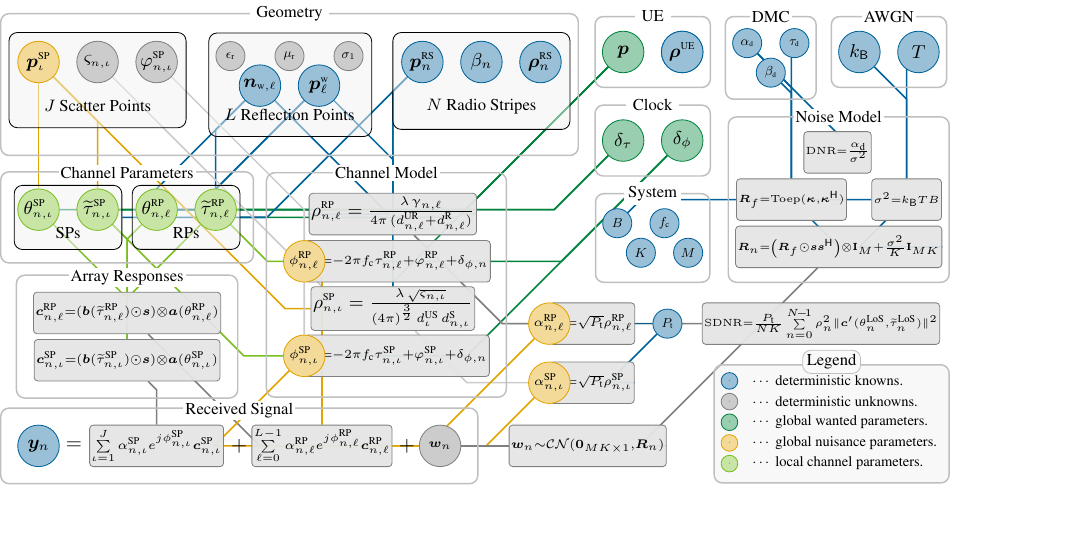}
    \vspace{-0.3cm}\caption{A non-exhaustive overview of our signal model defined in Sec.\,\ref{sec_signal_channel_model} with parameters indicated by nodes and dependencies %between parameters are 
    represented by edges. 
    We distinguish \protect\tikz \protect\node[NodeGlobalWanted, scale=0.3] {$\cdot$};
    global parameters of interest $\etabWanted$ and 
    \protect\tikz \protect\node[NodeGlobalUnwanted, scale=0.3] {$\cdot$};
    global nuisance parameters $\etabUnwanted$
    from~\eqref{eq_eta}, and
    \protect\tikz \protect\node[NodeLocal, scale=0.3] {$\cdot$};
    local per-\ac{RS} channel parameters $\etach_n$ from~\eqref{eq:etach}.
    }
    \label{fig:SignalModel}
\end{figure*}

\subsection{Dense Multipath Components}
The dense multipath term in \eqref{eq_ynk} can be modeled as a stochastic component with the distribution \cite{DMC_TAP_2014,mMIMO_CPP_TWC_2019,GreLeiWitFle:TWC2024}
\begin{align} \label{eq_dmc_stat}
    \wwdmc_n \sim \mtCN{MK}(\bm{0}, \allowbreak \rrbdmc(\etadmc) ) ~,
\end{align}
where
\begin{align}
    \wwdmc_n &\triangleq [ (\wwdmc_{\scriptscriptstyle n,0})^\trp \ \cdots \ (\wwdmc_{\scriptscriptstyle n,K-1})^\trp ]^\trp \quad \in \complexset{MK}{1}
\end{align}
denotes the \ac{DMC} observed in the spatial-frequency domain, and $ \rrbdmc \!\in\! \complexset{MK}{MK}$ is the spatial-frequency covariance matrix of the \ac{DMC}. Assuming spatially white \ac{DMC} and the Kronecker separability of the spatial and frequency domains (i.e., uncorrelated scattering between angle and delay domains), $\rrbdmc$ can be written as \cite[Eq.~(2.69)]{richter2005estimation}, \cite{DMC_TAP_2014,mMIMO_CPP_TWC_2019}
\begin{align} \label{eq_dmc_cov}
     \rrbdmc(\etadmc) = \rrb_f(\etadmc) \otimes \Imatrix_M  ~,
\end{align}
where $ \rrb_f(\etadmc) = \toep( \kappab(\etadmc), \kappab(\etadmc)^\hermit  )  \in \complexset{K}{K}$ is the frequency domain covariance matrix % with a Toeplitz structure
%\begin{align} \label{eq_rf}
 % \rrb_f(\etadmc) = \toep( \kappab(\etadmc), \kappab(\etadmc)^\hermit  ) 
%\end{align}
%In \eqref{eq_rf}, 
where $\etadmc = [ \alphad \ \betad \ \taud ]^\trp$ is the \ac{DMC} parameter vector consisting of  power $\alphad$,  normalized coherence bandwidth $\betad$, and  normalized onset time $\taud$, and $\kappab(\etadmc) \in \complexset{K}{1}$ represents the sampled version of the \ac{DMC} power spectral density \cite[Eq.~(2.61)]{richter2005estimation}
\begin{align}
    \psi_{\DMC}(f) = \frac{\alphad}{ \betad + j 2 \pi f } \ e^{-j 2 \pi f \taud} ~.
\end{align}

\subsection{Spatial-Frequency Observations at Radio Stripes}
Aggregating the received signals in \eqref{eq_ynk} over $K$ subcarriers, using the multipath model in~\eqref{eq_SMC_channel} and \ac{DMC} model in~\eqref{eq_dmc_stat}, the spatial-frequency observation matrix at the $n$-th \ac{RS} is 
\begin{align} \nonumber
    \yyb_n &\triangleq [\yy_{n,0} \ \cdots \ \yy_{n,K-1}  ] \quad \in \complexset{M}{K}
    \\ \label{eq_yy}
    %&= \sum_{\nRP=0}^{\Nrp-1} \alpharp e^{j\phirp} \aaa(\thetarp) ( \bb(\tauurp) \odot \sss )^\trp + \nonumber\\
    %&~ \sum_{\nSP=1}^{\Nsp} \alphasp e^{j\phisp} \aaa(\thetasp) ( \bb(\tauusp) \odot \sss )^\trp +
    &= \sum_{\ncomponent=0}^{\Nc-1} \alphacn e^{j\phicn} \aaa(\thetacn) ( \bb(\tauucn) \odot \sss )^\trp + 
    \wwb_n ~,
\end{align}
where 
\begin{align}
    \bb(\tau) &\triangleq   \left[ 1 \ e^{-j 2 \pi \deltaf \tau} \ \cdots \ e^{-j 2 \pi (K-1) \deltaf  \tau} \right]^\trp \in \complexset{K}{1} 
\end{align}
is the frequency domain steering vector, and
\begin{align} \label{eq_wwb}
    \wwb_n = \wwbdmc_n \odot \onematrix{M}{K} \sss^\trp + \zzb_n \quad \in \complexset{M}{K}
\end{align}
is the disturbance term consisting of  \acp{DMC} and  white noise, with $\wwbdmc_n \!=\! \veccinv{\wwdmc_n} \in \complexset{M}{K} $ and $\zzb_n \!=\! [\zz_{n,0} \ \cdots \zz_{n,K-1}] \in \complexset{M}{K}$. From Supplementary Material~\ref{app_cov_dist}, we have\footnote{The \ac{DMC} statistics may not be the same among the \acp{RS} since the distances between each \ac{RS} and the UE are different. This leads to different power, coherence bandwidth, and onset time for each \ac{RS}. See \cite{mMIMO_CPP_TWC_2019}.}
\begin{align}
    \ww_n = \vecc{ \wwb_n } \sim \mtCN{MK}(\bm{0}, \allowbreak \rrb(\etadmc, \sigma^2) ) ~,
\end{align}
where 
\begin{align} \label{eq_r_dmc}
    \rrb(\etadmc, \sigma^2) = \left( \rrb_f(\etadmc) \odot \sss \sss^\hermit \right) \otimes \Imatrix_M + \frac{\sigma^2}{K} \Imatrix_{MK} ~.
\end{align}

\subsection{SDNR Definition}\label{sec:SDNR}

For comparability of results in Sec.~\ref{sec:sim-performance-limits}, we relate the \ac{SDNR} only to the \ac{LoS} and define the average \ac{SDNR} as\footnote{$M$ does not enter in the denominator of~\eqref{eq:sdnr-definition-journal}. This effectively divides \eqref{eq:Pt} by $M$ and ensures that the positioning accuracy depends only on the \textit{angular resolution} and not the \textit{array gain}~(cf.\,Fig.\,\ref{fig:bounds-B-sweep} and Fig.\,\ref{fig:bounds-B-sweep-coh}).}
\begin{align}\label{eq:sdnr-definition-journal}
    \SDNRbar % &= \frac{0}{N}\sum_{n=0}^{N-1}\alpha_n^2 \| \bm{c}'_n \|^2 \\
     &= %\frac{\Pt}{N \lVert \bm{s}  \rVert^2 {K}}\sum_{n=0}^{N-1}\rho^2_n \  \| \bm{c}'_n \|^2 \\ 
     \frac{\Pt}{N K}\sum_{n=1}^{N} {(\rholos)}^2 \  \lVert \ccp(\thetalos, \tauulos) \rVert^2 \, ,
\end{align}
where the per-\ac{RS} angular-delay response is $\cc(\theta, \tau) \!\triangleq\!  ( \bb(\tau)\! \odot\! \sss )\! \otimes\! \aaa(\theta) \!\in\! \complexset{MK}{1}$ and $\ccp(\theta, \tau) \!\triangleq\! \rrb^{-1/2} \cc(\theta, \tau)$.
The ratio of \ac{DMC} power to thermal noise power is defined through the \ac{DNR} $\DNR \!=\! \frac{\alphad}{\sigma^2}$.
For simulations, we keep the $\SDNRbar$ in~\eqref{eq:sdnr-definition-journal} constant and set the transmit power
\begin{align}\label{eq:Pt}
      \Pt &= \frac{\SDNRbar \ N K }{\sum_{n=1}^{N}(\rholos)^2 \  \| \ccp(\thetalos, \tauulos) \|^2} 
\end{align}
to meet the desired $\SDNRbar$ for the chosen parameters.

\section{Problem Formulation%{\color{red}(PROPOSAL BENJAMIN)}
}\label{sec_prob_form}
%We formulate the joint positioning and synchronization problem for the provided signal model.
The joint positioning, synchronization and mapping problem consists in inferring a set of wanted parameters from signal observations at each \ac{RS} either using intermediate channel parameters (\textit{two-step}) or by \textit{direct} fusion of the acquired information at the \acp{RS} network level.
Given the observations $\{\yyb_n\}_{n=0}^{N-1}$ in \eqref{eq_yy} collected by all \acp{RS} from only a \emph{single snapshot} of UL transmission, the problem of interest is to estimate the \ac{UE} position $\pp$, its clock offset $\deltatau$ and its phase offsets $\deltaphin$, and the \ac{SP} positions $\ppSP$, contrary to \ac{RP} positions.\footnote{Since walls have known positions and orientations,  $\ppRP$ can be expressed through the \ac{UE} position $\pp$ (contained in $\etab$) according to our model in Supplementary Material~\ref{app_reflection_point_pos} and do not enter $\etab$ as nuisance parameters.} 
As to DMC parameters, different methods are available to estimate them, e.g.  \cite[Sec.~III-C1]{mMIMO_CPP_TWC_2019}, \cite[Sec.~6.1.8]{richter2005estimation}.  
To decouple the less investigated problem of joint localization, synchronization, and mapping supported by a network of distributed RSs from the better-understood problem of estimating DMC parameters, we assume that preliminary  estimation of $\rrbdmc(\etadmc)$ has been performed by resorting to one of those methods.
The unknown parameter vector for our estimation problem is then defined as
\begin{align}%\label{eq:etab-underbraces}
    \etab = \big[ 
        \underbrace{
            \pp^\trp \
            \deltac^\trp \ \ppspbar^\trp 
            }_{\etabWanted}
            \
            \underbrace{
            \phir^\trp \
            \alphabar^\trp 
        }_{\etabUnwanted}
    \big]^\trp 
    \quad \in \realset{\dimGlobal}{1} \, , \label{eq_eta}
\end{align}
which is divided into parameters of interest $\etabWanted$ and nuisance parameters $\etabUnwanted$, and where
\begin{subequations}\label{eq_eta_all}
\begin{align}
    \deltac     &:=[\deltatau \ \deltaphib^\trp]^\trp ~,\\
    \ppspbar    & := [{\ppSPx{1}}^\trp \ \dots \ {\ppSPx{\NspGlobal}}^\trp]^\trp    &&\in \realsetone{3 \NspGlobal} ~,\\
    \alphabcn   &:= [{\alphacnx{0}} \ \dots \ {\alphacnx{\Nc-1}}]^\trp   &&\in \realsetone{\Nc} ~,\\
    %\alphabn    &:= [{\alphabrp}^\trp  \ {\alphabsp}^\trp]^\trp              &&\in \realsetone{\Nrp+\Nsp} ~,\\
    \alphabar   &:=[\alphabnx{1}^\trp \ \dots \ \alphabnx{N}^\trp]^\trp             &&\in \realsetone{N\Nc}~,\\
    \phirx{n}     &:= [{\phicnx{1}} \ \dots \ {\phicnx{\Nc-1}}]^\trp       &&\in \realsetone{\Nc-1} ~,\\
    %\phibsp     &:= [{\phispx{0}} \ \dots \ {\phispx{\Nsp-1}}]^\trp       &&\in \realsetone{\Nsp} ~,\\
    %\phirx{n}   &:= [[{\phibrp}]_{2:\Nrp}^\trp  \ {\phibsp}^\trp]^\trp       && \in \realsetone{(\Nrp+\Nsp-1)}~, \\
    \phir       &:=[{\phirx{1}}^\trp \ \dots \ {\phirx{\Nc-1}}^\trp]^\trp               &&\in \realsetone{N(\Nc-1)}~.
\end{align}
\end{subequations}

\paragraph*{Parameters of interest}
We aim to infer the \ac{UE} position $\pp$ and synchronization parameters, $\deltac:=[\deltatau, \deltaphib^\trp]^\trp$ where we 
distinguish two different levels of phase synchronization:
\begin{itemize}[leftmargin=1.0cm]
    \item[\roundlabel{NCP}] \textit{Noncoherent} processing:\\
    $\deltac$ encompasses a time offset $\deltatau$ and a phase offset $\deltaphin$ to each \ac{RS}  $\deltaphib = [\deltaphinx{1} \cdots \deltaphinx{N}]^\trp$, thus $\deltac \in \realsetone{1+N}$.
    \item[\roundlabel{CP}] \textit{Coherent} processing:\\
    $\deltac$ encompasses a  time offset $\deltatau$ and one phase offset $\deltaphi$ to all \acp{RS}, i.e.,  $\deltac = [\deltatau \ \deltaphi]^\trp \in \realsetone{2}$.
\end{itemize}
Depending on the level of synchronization, the dimension of the global parameter vector $\etab$ is 
% \begin{align}\label{eq:dimGlobal}
%     \dimGlobal = \begin{cases}
%         4 + 3 \Nsp + 2 N( \Nrp + \Nsp) %3 + 1+N +3J + N(L+J-1) + N(L+J)   
%         & \roundlabeltxt{NCP} \\
%         5 + 3\Nsp + 2 N( \Nrp + \Nsp) - N  %3 + 1+1 +3J + N(L+J-1) + N(L+J)
%         & \roundlabeltxt{CP} \\
%     \end{cases} ~.
% \end{align}
\begin{align}\label{eq:dimGlobal}
    \dimGlobal = \begin{cases}
        \estDim+1 + 3 \Nsp + 2 N( \Nrp + \Nsp) %3 + 1+N +3J + N(L+J-1) + N(L+J)   
        & \roundlabeltxt{NCP} \\
        \estDim+2 + 3\Nsp + 2 N( \Nrp + \Nsp) - N  %3 + 1+1 +3J + N(L+J-1) + N(L+J)
        & \roundlabeltxt{CP} \\
    \end{cases} ~.
\end{align}
where $D$ indicates the number of unknown  \ac{UE} position coordinates.\footnote{Setting $\estDim=2$ requires one coordinate of the \ac{UE} antenna to be known, e.g., positioning a robot with fixed and known antenna height (sometimes termed 2.5D positioning).}
In addition, we want to map the environment in terms of dominant obstacles and use $\ppspbar \in \realsetone{3 \NspGlobal}$ to denote the vector of stacked \ac{SP} positions $\ppSP \in \realsetone{3}$.
\paragraph*{Nuisance parameters}
While $\alphabn$ contains the $\Nc$ stacked amplitudes of all components impinging at a single \ac{RS} $n$, we stack the amplitudes of the total number of $N\Nc$ components impinging on all $N$ \acp{RS} into a vector $\alphabar$.
Likewise, we stack all phases except for the \ac{LoS} $\ncomponent=0$ in a vector $\phirx{n}$ and stack the phases of the total number of \acp{RP} and \acp{SP} impinging on all \acp{RS} into a vector $\phir$. 
The phases $\philos$ of \ac{LoS} components are not entering our nuisance parameters $\etabUnwanted$, as they are implicitly contained in the remaining parameters of our global parameter vector $\etab$, according to our channel model defined in Sec.\,\ref{sec_SMC_model}.

%%%%%%%%%%%%%%%%%%%%%%%%%%%%%%%%%%%%%%%%%%%%%%%%%%%
\vspace{-0.2cm}

\section{Joint Positioning, Synchronization and Mapping in Radio Stripes Networks}\label{sec_algorithms}
We derive novel reduced-complexity algorithms to solve the joint positioning, synchronization and mapping problem  in Sec.~\ref{sec_prob_form}. We consider  phase-coherent \acp{RS}~\roundlabeltxt{CP}, namely $\deltac = [\deltatau \ \deltaphi]^\trp$, which unveils the potential of the \ac{RS} network: the spatial distribution of the \acp{RS} provides a large aperture, enabling  exploitation of spherical wavefront information under near-field conditions, bringing in turn high-resolution carrier phase information to estimate UE/SP positions.

% \end{enumerate}\
\subsection{Optimal Joint Localization, Synchronization and Mapping using the Maximum Likelihood Principle}\label{sec_coherent_mapping}
Following the Maximum Likelihood (ML) approach, we frame the positioning, synchronization, and mapping problem as a \emph{direct} joint estimation problem as
\begin{align}\label{eq_direct_pos_sc1}
         \etabhat^\text{\tiny JML} = \arg \max_{\etab} ~ p(\{\yyb_n\}_{n=1}^{N} ~ \lvert ~ \etab ) 
    \end{align}
with $p(\{\yyb_n\}_{n=1}^{N} ~ \lvert ~ \etab )$  the joint \ac{PDF} of all  observables conditioned on $\etab$, of which the vector of desired parameters is $\etabWanted = [\pp^\trp \ \deltatau \ \deltaphi \ \ppspbar^\trp]^\trp$ in the CP case.

%where $\etab_d = [\pp^\trp \ \deltatau \ \deltaphi \ (\ppSP)^\trp]^\trp$ denotes the vector containing the \emph{desired} parameters for the estimation task. 
 Assuming independent realizations of the disturbance component $\wwb_n$ in \eqref{eq_wwb} across the \acp{RS}, the log-likelihood version of the objective in \eqref{eq_direct_pos_sc1} can be written as\vspace{-0.2cm}
\begin{align} \label{eq_direct_lik}
    \log p(\{\yyb_n\}_{n=1}^{N} ~ \lvert ~ \etab ) = \sum_{n=1}^{N} \log p(\yyb_n ~ \lvert ~ \etab ) ~,
\end{align}
\vspace{-0.2cm} where \vspace{-0.15cm}
\begin{align}\nonumber
    \log \ & p(\yyb_n  ~ \lvert ~ \etab ) = - \normbig{ \rrb_n^{-1/2} \Big[ \yy_n -  \sum_{\nRP=0}^{\Nrp-1} \gamma^\RP_{n,\nRP} \, \cc(\thetarp, \tauurp) 
    \\[-0.2cm] \label{eq_direct_lik_cost}
    & \!\!\!\!\!\!\!\!\!\!\!\!\!\!\! - \sum_{\nSP=1}^{\Nsp} \gamma^\SP_{n,\nSP} \, \cc(\thetasp, \tauusp)\Big] }^2  -MK \log \pi - \log \det \rrb_n ~,
\end{align}
 with $\gamma^\RP_{n,\nRP}= \alpharp e^{j \phirp}$, $\gamma^\SP_{n,\nSP}= \alphasp e^{j \phisp}$, $L$ includes the \ac{LoS} path and the known number of walls, $J$ is assumed known (namely determined using classical information-theoretic techniques for model selection, e.g. Akaike’s method \cite{Akaike}), $\rrb_n$ is defined in \eqref{eq_r_dmc}, and $\yy_n \triangleq \vecc{\yyb_n} \in \complexset{MK}{1}$ is the vectorized form of the data matrix $\yyb_n$  \eqref{eq_yy} at $n$-th RS. %, i.e., 
% \begin{align} \label{eq_yy_n_coherent}
%     \yy_n     & \!=\! \sum_{\nRP=0}^{\Nrp-1} \gamma^\RP_{n,\nRP} \cc(\thetarp, \tauurp) + \sum_{\nSP=1}^{\Nsp} \gamma^\SP_{n,\nSP} \cc(\thetasp, \tauusp) + \ww_n ~.
%\end{align}
Neglecting constant terms in \eqref{eq_direct_lik_cost},  problem  \eqref{eq_direct_pos_sc1} becomes
\begin{align}\label{eq_direct_pos_sc1_2}
         \etabhat^\text{\tiny JML} = \arg \min_{\etab} ~ \llr^\text{\tiny JML}(\etab)  ~,
    \end{align}
where we pose $\yyp_n \triangleq \rrb_n^{-1/2} \yy_n$ and
\begin{align}
\llr^\text{\tiny JML}(\etab) \rmv\rmv&=\rmv\rmv\rmv\rmv \sum_{n=1}^{N}\rmv \normbig{  \yyp_n \rmv\rmv-\rmv\rmv\rmv\rmv \sum_{\nRP=0}^{\Nrp-1} \rmv\rmv\rmv \gamma^\RP_{n,\nRP} \, \ccp(\thetarp, \tauurp) \rmv\rmv-\rmv\rmv\rmv\rmv \sum_{\nSP=1}^{\Nsp}\rmv\rmv\rmv \gamma^\SP_{n,\nSP} \, \ccp(\thetasp, \tauusp)}^2.\label{eq_direct_pos_lik_nlos}
\end{align}

To tackle the challenging optimization problem above, we re-parameterize \eqref{eq_direct_pos_lik_nlos} in a different but equivalent  form  as
\begin{equation}\label{eq::Initial_loglikelihood_fullproblem}
     \llr^\text{\tiny JML}(\etab) = \sum_{n=1}^{N} \normbig{  \yyp_n - \bm{B}_n(\etabWanted) \bm{x}_n(\alphab_n, \phirx{n})}^2
\end{equation}
where $\bm{x}_n$ is a $(2(\Nrp+\Nsp) -1) \times 1$ \emph{real} vector stacking the real amplitude $\alphalos$ of the LOS path and both the real and imaginary parts of each complex \acp{RP} and \acp{SP} amplitudes, i.e.,
\begin{align}
\bm{x}_n & = [\alphalos, \realp{\gamma^\RP_{n,1}}, \imp{\gamma^\RP_{n,1}},\cdots, \realp{\gamma^\RP_{n,{\Nrp-1}}}, \imp{\gamma^\RP_{n,{\Nrp -1}}}, \nonumber \\ 
& \realp{\gamma^\SP_{n,1}}, \imp{\gamma^\SP_{n,1}}, \cdots, \realp{\gamma^\SP_{n,{\Nsp}}}, \imp{\gamma^\SP_{n,{\Nsp}}}]^\mathsf{T}
\end{align}
with $\bm{B}_n \in \complexset{MK}{(2(\Nrp+\Nsp) -1)}$ given by
\begin{align} \nonumber
\bm{B}_n &= [e^{j \philos}\ccp(\thetalos, \tauulos), \ccp(\thetarpx{1}, \tauurpx{1}), j \ccp(\thetarpx{1}, \tauurpx{1}), \cdots\\ \label{eq_BBn}
 &\ccp(\thetarpx{\Nrp-1}, \tauurpx{\Nrp-1}), j \ccp(\thetarpx{\Nrp-1}, \tauurpx{\Nrp-1}), \ccp(\thetaspx{1}, \tauuspx{1}), \nonumber \\
 & j \ccp(\thetaspx{1}, \tauuspx{1}), \cdots, \ccp(\thetaspx{\Nsp}, \tauuspx{\Nsp}), j \ccp(\thetaspx{\Nsp}, \tauuspx{\Nsp})
 ].
\end{align}
Then, we notice that
\begin{align}
\normbig{  \yyp_n - \bm{B}_n \bm{x}_n}^2 = &\normbig{(\yyp_{n,R} + j\yyp_{n,I}) - (\bm{B}_{n,R} + j\bm{B}_{n,I})\bm{x}_n}^2 \nonumber \\
= &\normbig{\yyp_{n,R} - \bm{B}_{n,R} \bm{x}_n + j(\yyp_{n,I} - \bm{B}_{n,I}\bm{x}_n)}^2 \nonumber \\
= & \normbig{\yyp_{n,R} - \bm{B}_{n,R} \bm{x}_n}^2 + \normbig{\yyp_{n,I} - \bm{B}_{n,I} \bm{x}_n}^2\label{eq::simplification_costfunct}
\end{align}
where $\yyp_{n,R} = \realp{\yyp_{n}}$, $\yyp_{n,I} = \imp{\yyp_{n}}$ and, similarly, $\bm{B}_{n,R} = \realp{\bm{B}_n}$, $\bm{B}_{n,I} = \imp{\bm{B}_n}$.
The expression in \eqref{eq::simplification_costfunct} is a convenient decomposition that allows us to obtain the optimal (ML) estimates of $\bm{x}_n$'s in closed-form. More specifically, by computing the derivative of $\llr(\etab)$ in \eqref{eq::Initial_loglikelihood_fullproblem} using \eqref{eq::simplification_costfunct} wrt each vector $\bm{x}_n$, $n=1,\ldots,N$, and posing it equal to zero, it follows 
\begin{equation}
(\bm{B}^\mathsf{T}_{n,R}\bm{B}_{n,R} + \bm{B}^\mathsf{T}_{n,I}\bm{B}_{n,I})\bm{x}_n = \bm{B}^\mathsf{T}_{n,R}\yyp_{n,R} + \bm{B}^\mathsf{T}_{n,I}\yyp_{n,I}
\end{equation}
since the $\bm{x}_n$'s are not linked to each other and, accordingly,  terms in the outer summation of $\llr(\etab)$ corresponding to \acp{RS} with indexes $n' \neq n$ will be constant when derived wrt $\bm{x}_n$.
Re-arranging the terms in 
\begin{equation}
\yypb_{n} = \begin{bmatrix} \yyp_{n,R} \\
\yyp_{n,I}
\end{bmatrix}, \qquad \accentset{\circ}{\bm{B}}_n = \begin{bmatrix} \bm{B}_{n,R} \\
\bm{B}_{n,I} 
\end{bmatrix} 
\end{equation}
we can finally obtain a closed-form estimate of the corresponding $\bm{x}_n$ in terms of $\BB_n$ for each individual \ac{RS} $n$ as
\begin{align} \label{eq_xhatn}
(\accentset{\circ}{\bm{B}}^\mathsf{T}_n\accentset{\circ}{\bm{B}}_n)\bm{x}_n = \accentset{\circ}{\bm{B}}_n^\trp \yypb_{n} 
\quad \longrightarrow \quad
\xxhat^\text{\tiny JML}_n = \accentset{\circ}{\bm{B}}^\dag_n \yypb_{n}
\end{align}
with $\accentset{\circ}{\bm{B}}^\dag_n = \left(\accentset{\circ}{\bm{B}}^\mathsf{T}_n \accentset{\circ}{\bm{B}}_n\right)^{-1}\accentset{\circ}{\bm{B}}^\mathsf{T}_n$.
From \eqref{eq_xhatn}, one can easily retrieve the estimates of the real \ac{LoS} amplitudes $\alphalos$'s
as well as the complex multipath amplitudes $\gamma^\RP_{n,\nRP}$'s and $\gamma^\SP_{n,\nSP}$'s as
\begin{align}
\widehat{\alpha}^\text{\tiny LoS, JML}_{n} &= [\widehat{\bm{x}}^\text{\tiny JML}_n]_1 \\  \widehat{\gamma}^\text{\tiny RP, JML}_{n,\nRP} &= [\widehat{\bm{x}}^\text{\tiny JML}_n]_{\scriptscriptstyle 2\nRP} + j[\widehat{\bm{x}}^\text{\tiny JML}_n]_{\scriptscriptstyle2\nRP+1}, \,  \ell = 1,\ldots,L-1 \\
\widehat{\gamma}^\text{\tiny SP, JML}_{n,\nSP} &= [\widehat{\bm{x}}^\text{\tiny JML}_n]_{\scriptscriptstyle 2(\Nrp\!-\!1)  + 2\nSP} + j[\widehat{\bm{x}}^\text{\tiny JML}_n]_{\scriptscriptstyle 2\Nrp + 2\nSP -1}, \, \nSP = 1,\ldots,J
\end{align}
with $[\cdot]_i$ indicating the $i$-th element of the vector argument. 

Now that we have obtained optimal estimates $\xxhat^\text{\tiny JML}_n$ of the path amplitudes in \eqref{eq_xhatn}, which correspond to the $\etabUnwanted$ part of $\etab$ in \eqref{eq_eta}, we can plug $\xxhat^\text{\tiny JML}_n$, $n=1,\ldots,N$, back into \eqref{eq::Initial_loglikelihood_fullproblem} to obtain the following compressed log-likelihood function
\begin{equation} \label{eq_direct_compressed_coherent}
     \llr^\text{\tiny JML}(\etabWanted) =
         \sum_{n=1}^{N} \normbig{  \yyp_n - \bm{B}_n(\etabWanted) \accentset{\circ}{\bm{B}}^\dag_n(\etabWanted) \yypb_n}^2 ~.
\end{equation}
Remarkably, the dimension of the estimation problem has been reduced from $D + 2+ 3\Nsp + N(2\Nrp+2\Nsp-1)$ to $D + 2 + 3\Nsp$.
By inspecting \eqref{eq_direct_compressed_coherent} and using \eqref{eq_BBn}, it is possible to highlight the dependence of $\BB_n$ (similarly of $\accentset{\circ}{\bm{B}}^\dag_n$) on  the remaining unknown parameters, namely UE position $\bm{p}$, clock offset $\delta_\tau$, phase offset $\delta_\phi$, and \ac{SP} positions $\{\ppSP\}_{\iota=1}^J$, as follows:
\begin{itemize}
    \item $\philos$ depends on $\pp$ and $\deltaphi$ through \eqref{eq_phirp} and \eqref{eq_taurp};
     \item $\tauurp$ can be written in terms of $\pp$, $\ppRP$ and $\deltatau$ through  \eqref{eq_taurp} and \eqref{eq_tau_pseudo}, and in turn RP locations $\ppRP$ can be written as a function of $\pp$ via \eqref{eq:RP-explicit} (see Supplementary Material \ref{app_reflection_point_pos});
    \item $\thetarp$ can be written in terms of $\pp$ and $\ppRP$ through \eqref{eq_thetanl}-\eqref{eq_ppsnl};
     \item $\tauusp$ depends on $\pp$, $\ppSP$ and $\deltatau$ through \eqref{eq_tau_pseudo} and \eqref{eq_taurp};
    \item $\thetasp$ depends on $\pp$ and $\ppSP$ through \eqref{eq_thetanl} and \eqref{eq_ppsnl}.
%    \item From \eqref{eq_phirp} and \eqref{eq_taurp}, $\phirp$ can be expressed in terms of $\pp$, $\ppRP$, $\deltaphi$ and $\varphirpx{\nRP}$ for $\nRP > 0$ (no need for $\ppRP$ and $\varphirpx{\nRP}$ for $\nRP = 0$). Note from \eqref{eq_eta_all} that $ \phib$ does not contain $\philos$'s.  
\end{itemize}
Interestingly, the dependencies left are only upon the
parameters of interest $\etabWanted = [\pp^\trp \ \deltatau \ \deltaphi \ \ppspbar^\trp]^\trp$. In principle, solving the estimation problem 
\begin{equation} \label{compressed_optproblem}
    \etabhat_\text{\tiny w}^\text{\tiny JML} = \arg \min_{\etab_\text{\tiny w}} ~ \llr^\text{\tiny JML}(\etab_\text{\tiny w})
\end{equation}
would require a joint optimization over the continuous support defined by $\etab_\text{\tiny w}$, which unfortunately is not feasible in closed-form, nor can be approached by  an exhaustive grid search due to the exceptionally high dimensionality of the problem. Therefore, in the following we propose a novel approach to make the estimation task feasible. Specifically, we derive novel approximate estimators that can provide good initial estimates of the parameters in $\etab_\text{\tiny w}$; then, the latter will be used to initialize an iterative, reduced-complexity optimization of the optimal compressed log-likelihood function in \eqref{eq_direct_compressed_coherent}, ultimately yielding refined estimates that we retain.

\subsection{Reduced-complexity Algorithms for Initial UE Localization, Synchronization and Mapping}\label{sec_coarse_est}
%In this section, we develop novel algorithms able to provide coarse initial estimates of the UE position, clock/phase offsets, and SPs positions at \emph{reduced complexity}. 
To tackle the formidable complexity of the problem at hand, we follow an alternative strategy, hereafter referred to as \emph{Relaxed Maximum Likelihood (RML)}. The main idea consists in decoupling the estimation of the UE-related parameters $\breve{\etab}_\text{\tiny w} = [\pp^\trp \ \deltatau \ \deltaphi]^\trp$ from the SP parameters $\{\ppSP\}_{\iota = 1}^{J}$, and relaxing some dependencies, as discussed in the following.

\subsubsection{Initial UE Localization and Synchronization}
We start by relaxing the original estimation problem in \eqref{compressed_optproblem} and propose to approximate the  signal model %in \eqref{eq_yy_n_coherent}
as consisting of \ac{LoS} and RPs contributions only, which is tantamount to neglecting, at a first stage, the presence of the $\Nsp$ paths originating from \acp{SP} \{$\ppSP\}_{\iota=1}^{\Nsp-1}$. Accordingly, \eqref{eq_direct_pos_lik_nlos} becomes 
\begin{align} \label{eq_direct_pos_lik_nlos_known_wall}
    \llr^\text{\tiny RML}(\breve{\etab}) &= \sum_{n=1}^{N} \normbig{  \yyp_n - \alphalos e^{j \philos}  \ccp(\thetalos, \tauulos)
    \nonumber \\ 
    &\qquad
    -\sum_{\nRP=1}^{L-1} \gamma^\RP_{n,\nRP} \, \ccp(\thetarp, \tauurp)
    }^2 ~
\end{align}
with $\breve{\etab} = [\underbrace{\pp^\trp \ \deltatau \ \deltaphi}_{\breve{\etab}_\text{\tiny w}} \ \underbrace{\alphab^\RP \ \phib^\RP}_{\breve{\etab}_\text{\tiny u}}]^\trp \in \realset{(D + 2+N(2\Nrp-1))}{1}$ and 
\begin{subequations}\label{eq_eta_all_reduced}
\begin{align}
    \alphab^\RP &\triangleq [{\alphabrpx{1}}^\trp \ \cdots \ {\alphabrpx{N}}^\trp]^\trp \in \realset{N\Nrp}{1} ~,
    \\
     \phib^\RP &\triangleq [[{\phibrpx{1}}]_{2:\Nrp}^\trp \ \cdots \ [{\phibrpx{N}}]_{2:\Nrp}^\trp]^\trp \in \realset{N(\Nrp-1)}{1} ~.
\end{align}
\end{subequations}

\paragraph{Estimation of Phase Offset}\label{sec_est_amp}
We now provide a  strategy to obtain a closed-form estimate of the phase offset $\deltaphi$. To this aim, we rewrite the likelihood  \eqref{eq_direct_pos_lik_nlos_known_wall} after relaxing the dependency of the \ac{LoS} phase terms $\philos$ on the unknown UE position $\pp$ and phase offset $\deltaphi$, i.e., by considering a relaxed variable $\gammalos = \alphalos e^{j \philos}$ (with no phase-coherent structure) and treating $\{\gammalos\}_{n=1}^{N}$ as  unknown complex amplitudes. This  increases by $N-1$ the number of unknowns in  $\breve{\etab}$, due to an additional unknown phase term in $\phibrp$ for each \ac{RS} $n$. We denote the new vector of unknown parameters as %$\breve{\etab}_\text{\tiny ext}$
\begin{equation}
\breve{\etab}_\text{\tiny ext} = [\pp^\trp \ \deltatau \ \deltaphib^\trp \ \bm{\gamma}^\trp_1 \ \cdots \ \bm{\gamma}^\trp_N]^\trp \in \realsetone{(D + 1+N)} \times \mathbb{C}^{N \Nrp}
\end{equation}
with
%\begin{equation}
$\bm{\gamma}_n = [\gammalos \ \gamma_{n,1} \ \cdots \ \gamma_{n,\Nrp-1}]^\mathsf{T}
$
%\end{equation}
and, accordingly, rewrite  \eqref{eq_direct_pos_lik_nlos_known_wall}  in matrix form as
\begin{equation} \label{eq_1st_llr}
     \llr^\text{\tiny RML-NCP}(\breve{\etab}_\text{\tiny ext}) = \sum_{n=1}^{N} \normbig{ \yyp_n -  \ccbp_n(\pp, \deltatau) \gammab_n }^2 ~
\end{equation}
where the additional superscript NCP highlights the non-coherent processing of the signal paths, and 
\begin{equation}\label{Cn_matrix}
 \ccbp_n(\pp, \deltatau)
     \triangleq  [ \ccp_{n, 0}(\pp, \deltatau) \ \cdots \ \ccp_{n, \Nrp-1}(\pp, \deltatau) ].
\end{equation}

The relaxed UE localization and synchronization problem can be then recast as 
\begin{equation} \label{suboptimal_UElocsync}
\widehat{\breve{\etab}}_\text{\tiny ext}^\text{\tiny RML-NCP}= \arg \min_{\breve{\etab}_\text{\tiny ext}} ~ \llr(\breve{\etab}_\text{\tiny ext}).
\end{equation}
This problem is ancillary to the ultimate estimation of $\deltaphi$: since \eqref{eq_1st_llr} is a separable LS problem, the complex amplitudes can be estimated on a per-RS basis as a function of $\pp$ and $\deltatau$ as
\begin{align} \label{eq_coarse_gain}
    \gammabhat_n(\pp, \deltatau) = \left( \ccbp_n(\pp, \deltatau) \right)^\dagger \yyp_n 
\end{align}
for $n = 1, \ldots, N$. %, with $\left( \ccbp_n(\pp, \deltatau) \right)^\dagger$ left pseudoinverse of $\ccbp_n(\pp, \deltatau)$.
From  $\gammabhat_n(\pp, \deltatau)$, we pick only the \ac{LoS} complex amplitudes and try to exploit their original structure $\widehat{\gamma}_n^\text{\tiny LoS} =\hat{\alpha}_n^\text{\tiny LoS} e^{j \hat{\phi}^\text{\tiny LoS}_n}$ to estimate the  phase offset $\deltaphi$. In fact, by inspecting \eqref{eq_phirp}, we observe that the relationship 
\begin{equation}
\hat{\phi}^\text{\tiny LoS}_n(\pp, \deltatau) = -2\pi \fc \taulos(\pp) + \deltaphi(\pp, \deltatau) \label{eq:hatphiLoS}
\end{equation}
holds for each $n=1,\ldots,N$. This suggests that, conditioned on a  tentative value of UE position (and of the clock offset for obtaining $\gammabhat_n(\pp, \deltatau)$), we can first apply a per-RS de-rotation by a complex factor as
\begin{equation}
\xi^\text{\tiny LoS}_{n}(\pp, \deltatau) \triangleq  \widehat{\gamma}_n^\text{\tiny LoS}(\pp, \deltatau) \ e^{j2\pi \fc \tau^\text{\tiny LoS}_n(\pp)}
\end{equation}
to compensate for the delay term in \eqref{eq_phirp}, leaving a dependency only on the sought $\deltaphi$ in $\xi^\text{\tiny LoS}_{n}$; then, all $\xi^\text{\tiny LoS}_{n}$'s
can be summed and an estimate of the phase offset can be readily obtained as the argument of the resulting complex number, i.e. 
\begin{equation}\label{eq_deltaphi_final}
    \hat{\delta}^\text{\tiny RML-NCP}_\phi(\pp, \deltatau) = \angle{\sum_{n=1}^{N}\xi^\text{\tiny LoS}_{n}(\pp, \deltatau)}.
\end{equation}

\paragraph{Estimation of UE Position and Clock Offset}\label{sec_est_pos_offset}
At the first stage, we neglected phase coherence among all the \acp{RS} and proposed an approach to obtain a closed-form (conditional) estimate of the phase offset $\hat{\delta}_\phi$. Now, at the second stage, we plug such an estimate back into \eqref{eq_direct_pos_lik_nlos_known_wall}, where phase coherence among \acp{RS} is fully restored and follow the same ML steps as in \eqref{eq::Initial_loglikelihood_fullproblem}-\eqref{eq_xhatn} to obtain closed-form (conditional wrt $\pp$ and $\deltatau$) estimates of $\alphab^\RP$ and $\phib^\RP$. Subsequently, the estimated vectors $\hat{\alphab}^\RP$ and $\hat{\phib}^\RP$, together with $\hat{\delta}_\phi$, can be substituted back into \eqref{eq_direct_pos_lik_nlos_known_wall}, yielding the compressed log-likelihood function
\begin{equation} \label{eq_compr_coherent}
 \llr^\text{\tiny RML}(\pp, \deltatau) =
         \sum_{n=1}^{N} \normbig{  \yyp_n - \breve{\bm{B}}_n(\pp, \deltatau) \accentset{\circ}{{\breve{\bm{B}}}}^\dag_n(\pp, \deltatau) \yypb_n}^2 ~
\end{equation}
with 
\begin{align} \nonumber
\breve{\bm{B}}_n &\triangleq [e^{j \hat{\phi}^\text{\tiny LoS}_n}\ccp(\thetalos, \tauulos) \ \ccp(\thetarpx{1}, \tauurpx{1}) \ j \ccp(\thetarpx{1}, \tauurpx{1}) \ \cdots\\ 
 &\ccp(\thetarpx{\Nrp-1}, \tauurpx{\Nrp-1}) \ j \ccp(\thetarpx{\Nrp-1}, \tauurpx{\Nrp-1})
 ],
\end{align}
where $\hat{\phi}^\text{\tiny LoS}_n$ is given in \eqref{eq:hatphiLoS} and, analogously to previous definitions in \eqref{eq_direct_compressed_coherent},
$
    \accentset{\circ}{\breve{\bm{B}}}_n = \begin{bmatrix} \breve{\bm{B}}_{n,R} \\
\breve{\bm{B}}_{n,I} 
\end{bmatrix} 
$
with $\breve{\bm{B}}_{n,R} = \realp{\breve{\bm{B}}_n}$, $\breve{\bm{B}}_{n,I} = \imp{\breve{\bm{B}}_n}$.
Interestingly, as in \eqref{eq_direct_compressed_coherent}, the dependencies left are only upon the
parameters of interest $\pp$ and $\delta_\tau$. As a result, the ultimate expression of the (approximate) UE position and clock offset estimator is
\begin{equation} \label{final_suboptimalUE}
    [\hat{\pp}^\text{\tiny RML} \ \hat{\delta}_\tau^\text{\tiny RML}]= \arg \min_{\pp, \deltatau} ~ \llr^\text{\tiny RML}(\pp, \deltatau).
\end{equation}
However, differently from \eqref{compressed_optproblem}, the approach proposed in this section further reduces the dimensionality of the estimation problem from $D + 2+3J$ to only $D + 1$. %(for a 3D scenario; $3$ for a 2D scenario). 
An extra reduction of computational cost can be finally obtained by estimating 
$\deltatau$ through the following alternative low-complexity procedure.
 
%%%%%%%%%%%%%%%%%
% clock offset estimation 
\paragraph{Low-Complexity Estimation of Clock Offset}
To reduce the dimensionality of the optimization problem in \eqref{final_suboptimalUE}, we employ a low-complexity strategy to obtain a coarse estimate of $\deltatau$ as a function of $\pp$, consisting of the following steps. First, for each RS $n$, we estimate the pseudo-delay via IFFT over subcarriers and noncoherent integration over spatial (antennas) domain. Specifically, we switch from spatial-frequency domain to delay-spatial domain by computing  $\bm{\mathcal{Y}}_n = \mathrm{IFFT}(\yyb_n^\mathsf{T})$ as the (per-column) IFFT-transformed delay-spatial observations over $N_F$ points, where $\yyb_n$ is the spatial-frequency observation at the $n$-th RS given in \eqref{eq_yy}. This is followed by a noncoherent integration across the spatial domain  (i.e., samples over the $M$ antennas) and finding the index of the maximum element %in the cost function
    \begin{equation}
    \widehat{q}_n = \arg \max_{q} \left[\sum_{m=1}^M |[\bm{\mathcal{Y}}_n]_{q,m}|^2 : 0 \leq q \leq N_F -1\right] ~,
    \end{equation}
 with $[\bm{\mathcal{Y}}_n]_{q,m}$ denoting the $(q,m)$-th entry of $\bm{\mathcal{Y}}_n$. Accordingly, a coarse estimate of the pseudo-delay $\tauu_{n}$ can be obtained by mapping the index $\hat{q}$ with the corresponding IFFT bin as
 \begin{align}
     \widehat{\tauu}_{n} = {\widehat{q}_n}/{(N_F \Delta f)} ~.
 \end{align}

Next, for a given trial position $\pp$, the clock offset can be estimated by computing the difference between the pseudo-delay and the true delay at $\pp$ based on \eqref{eq_tau_pseudo}, i.e.,
 \begin{align}
     \widehat{\delta}_{\tau,n}(\pp) = \widehat{\tauu}_{n} - \frac{1}{\lightspeed} \|\pp - \pprsx{n}  \| ~.
 \end{align}
 Finally, a coarse estimate of $\deltatau$ can be obtained simply via averaging over the \acp{RS}:
  \begin{align} \label{eq_deltatau_final}
     \widehat{\delta}_{\tau}(\pp) = \frac{1}{N} \sum_{n=1}^{N} \widehat{\delta}_{\tau,n}(\pp) ~.
 \end{align}
To determine the initialization point for \eqref{final_suboptimalUE}, one can then perform a 3D search (2D for known height of UE) via
\begin{equation} \label{final_suboptimalUE_2D}
    \hat{\pp}^\text{\tiny RML} = \arg \min_{\pp} ~ \llr^\text{\tiny RML}(\pp, \widehat{\delta}_{\tau}(\pp)),
\end{equation}
yielding $(\hat{\pp}^\text{\tiny RML}, \widehat{\delta}_{\tau}(\hat{\pp}^\text{\tiny RML}))$ as the initialization point for 4D numerical optimization in \eqref{final_suboptimalUE}.

%%%%%%%%%%%%%%%%%
\subsubsection{Initial Mapping of Scatterers Positions}
We devise an alternative strategy to obtain an initial estimate of all the \acp{SP} positions $\{\ppSP\}_{\iota=1}^J$, leveraging the initial (coarse) knowledge about UE position $\hat{\pp}^\text{\tiny RML}$ and synchronization offsets $(\hat{\delta}_\tau^\text{\tiny RML}, \hat{\delta}^\text{\tiny RML-NCP}_\phi)$ gained using the previously-proposed  algorithms. 
To this aim, we reconsider \eqref{Cn_matrix}  and observe that the matrices $\ccbp_n(\pp, \deltatau)$, which contain the angular-delay response vectors $\ccp_{n, \ell}(\pp, \deltatau) $, span a basis for the joint  $L$-dimensional \ac{LoS} and \acp{RP} space when evaluated for $\pp = \hat{\pp}^\text{\tiny RML}$ and $\deltatau = \hat{\delta}_\tau^\text{\tiny RML}$. Starting from these matrices, we propose an approach, hereafter referred to as \emph{null-space transformation (NST)}, that leverages  the subspace orthogonal to the joint \ac{LoS}/\acp{RP} space to approximately remove \ac{LoS} and \acp{RP} paths from the received signals  $\{\yyp_n\}_{n=1}^N$. Specifically, we estimate a per-RS basis for the null space (kernel) of \ac{LoS} and \acp{RP} vectors, using the coarse estimates of $\hat{\pp}^\text{\tiny RML}$ and $\hat{\delta}_\tau^\text{\tiny RML}$, as follows
\begin{equation} \label{eq:proj_matr}
    \bm{K}_n = \mathrm{null}({\ccbp}^\hermit_n(\hat{\pp}^\text{\tiny RML}, \hat{\delta}_\tau^\text{\tiny RML})) \in \complexset{MK}{MK-L}
\end{equation}
where $\mathrm{null}(\cdot)$ denotes the operator returning a basis for the null space of the matrix argument. Then, for the resulting transformed version of the observables, $\bm{\zeta}_n = \bm{K}^\hermit_n \yyp_n$, it holds  
\begin{align}\label{model_projected}
    \bm{\zeta}_n \rmv\rmv&=\rmv\rmv\rmv\rmv \sum_{\nRP=0}^{\Nrp-1} \gamma^\RP_{n,\nRP} \, \underbrace{\bm{K}^\hermit_n\ccp(\thetarp, \tauurp)}_{\approx \bm{0}} \rmv+\rmv\rmv\sum_{\nSP=1}^{\Nsp} \gamma^\SP_{n,\nSP} \, \bm{K}^\hermit_n\ccp(\thetasp, \tauusp) + \bm{v}_n \nonumber
\end{align}
%\missing{ This section is to be competed shortly.}
where $\bm{v}_n \triangleq \bm{K}^\hermit_n  \rrb_n^{-1/2} \ww_n$ and the approximation is obviously related to the fact that estimated projection matrices are used in place of the true ones. Based on $\bm{\zeta}_n$s, a reduced ML estimation problem can be formalized as
\begin{equation}
[\hat{\bm{p}}^\text{\tiny SP, NST} \ \hat{\bm{\gamma}}^\text{\tiny SP, NST}]= \arg \min_{\bm{p}^\SP, \bm{\gamma}^\SP} ~ \llr^\text{\tiny NST}(\bm{p}^\SP, \bm{\gamma}^\SP) 
\end{equation}
where $\bm{\gamma}^\SP = [{\bm{\gamma}^\SP_1}^\trp \ \cdots \ {\bm{\gamma}^\SP_N}^\trp]^\trp$, $\bm{\gamma}^\SP_n = [\gamma^\SP_{n,1} \ \cdots \ \gamma^\SP_{n,\Nsp}]^\trp$, and
\begin{equation}
\llr^\text{\tiny NST}(\bm{p}^\SP, \bm{\gamma}^\SP) = \sum_{n=1}^N \normbig{ \bm{\zeta}_n' - \sum_{\nSP = 1}^J \gamma^\SP_{n,\nSP} \, \bm{g}_n'(\thetasp, \tauusp)  }^2  \label{eq:L-NST}
\end{equation}
with $\bm{\zeta}_n' = (\bm{K}^\hermit_n \bm{K}_n)^{-1/2}\bm{\zeta}_n$ and $\bm{g}_n'(\thetasp, \tauusp) = (\bm{K}^\hermit_n \bm{K}_n)^{-1/2}\bm{K}^\hermit_n\ccp(\thetasp, \tauusp)$ resulting from the whitening of the noise, being $\mathbb{E}[\bm{v}_n\bm{v}_n^\hermit] = \bm{K}^\hermit_n \bm{K}_n$.\footnote{Notice that if $\bm{K}_n$ is selected as an orthonormal basis, $\bm{K}^\hermit_n \bm{K}_n=\bm{I}$ hence pre-whitening is not necessary (and $\bm{\zeta}_n'=\bm{\zeta}_n$).} This estimation problem is still very complex, being $(3+N)J$-dimensional; to lower its complexity, we consider a relaxation that neglects part of the mixed terms in \eqref{eq:L-NST}, i.e., 
\begin{equation}
\llr^\text{\tiny RNST}(\bm{p}^\SP, \bm{\gamma}^\SP) = \sum_{\nSP = 1}^J \sum_{n=1}^N  \normbig{ \bm{\zeta}_n' -  \gamma^\SP_{n,\nSP} \, \bm{g}_n'(\thetasp, \tauusp)  }^2  .\label{eq:L-RNST}
\end{equation}
This allows one to estimate each $ \gamma^\SP_{n,\nSP}$ separately for all $\iota =1,\ldots,J$,  i.e., $\hat{\gamma}^\text{\tiny SP, RNST}_{n,\nSP} = (\bm{g}_n'(\thetasp, \tauusp))^\dag \bm{\zeta}_n'$, and then perform a  search for the $J$ dominant dips (negative peaks) of  
$ \sum_{n=1}^N  \normbig{ \bm{\mathcal{P}}^\perp_{\bm{g}_n'} \bm{\zeta}_n'   }^2 $, where $\bm{\mathcal{P}}^\perp_{\bm{A}} = \bm{I}- \bm{A}\bm{A}^\dag$  is the projector onto the orthogonal complement to the subspace spanned by the columns of $\bm{A}$. The resulting estimates $\hat{\bm{p}}^\text{\tiny SP, RNST}$ are thus obtained through a procedure with remarkably lower complexity, amounting to a single three-dimensional search.

\section{Cram\'er-Rao Lower Bound}\label{sec_crlb}

We derive the \ac{CRB} for joint localization, synchronization and mapping with distributed \acp{RS}, i.e.\footnote{$\bm{X} \succeq \bm{0}$ denotes a positive semidefinite matrix~\cite{Kay93Estimation}.}
\begin{align}\label{eq:CRLB_def}
    \mathbb{E}_\etab \left[ (\hat{\etab} - \etab) (\hat{\etab} - \etab)^\trp \right] \succeq \fim^{-1}
\end{align}
where $\fim$ is the \ac{FIM} for the global parameter vector $\etab$. We assume that each \ac{RS} $n$ contributes independent information on $\etab$, i.e., assuming identical \ac{DMC}\footnote{Apart from the onset time $\taud$, which is coupled with the \ac{LoS} delay $\taulos$.} and noise statistics, the \ac{FIM} for the parameter vector $\etab$ is 
\begin{align}\label{eq:fim}
    \fim = \sum\limits_{n=0}^{N-1} \jacobian_n \fimetachn \jacobian_n^\trp  \quad \in \realset{\dimGlobal}{\dimGlobal} \, ,
\end{align}
which is the sum of the local channel \acp{FIM} $\fimetachn$ contributed by all $N$ \acp{RS} and propagated via the Jacobian matrices $\jacobian_n \!\in \realset{\dimGlobal}{\dimLocal} $ from local channel parameter level to global parameter level. Depending on the considered level of synchronization, the dimension $\dimGlobal$ of the global parameter vector $\etab$ changes according to~\eqref{eq:dimGlobal}. 
The dimension $\dimLocal$ of the local per-\ac{RS} channel parameter vector $\etach_n$ stays constant in both the $\roundlabeltxt{NCP}$ and the $\roundlabeltxt{CP}$ case.
%, as it captures the channel parameters, i.e., \acp{AoA}, pseudo-delays, amplitudes, and phases, of all components impinging on \ac{RS} $n$.
%The per-\ac{RS} channel parameter vector 
It is defined as 
\begin{align}\label{eq:etach}
    \etach_n = 
    \Big[
        \underbrace{{\thetabrp}^\trp {\thetabsp}^\trp}_{\thetabn^\trp} \ 
        \underbrace{\left.\tauubrp\right.^\trp \left.\tauubsp\right.^\trp}_{\tauubn^\trp} \ 
        \underbrace{{\phibrp}^\trp {\phibsp}^\trp}_{\phibn^\trp} \ 
        \underbrace{{\alphabrp}^\trp {\alphabsp}^\trp}_{\alphabn^\trp} 
    \Big]^\trp %\in \realsetone{\dimLocal}
\end{align}
and captures the stacked channel parameters for all components $\ncomponent$ impinging at \ac{RS} $n$ hence its dimension is $\dimLocal \triangleq 4 \, \Nc$. The elements of the local channel \ac{FIM} $\fimetachn \in \realset{\dimLocal}{\dimLocal}$ are defined as~\cite[Sec.~15.7]{Kay93Estimation}
\begin{align}\label{eq:fimetachn}
    \left[\fimetachn\right]_{i,j} = 2\realp{
        \frac{\partial \muyyni^\hermit}{\partial [\etach_n]_i} \rrbn^{\scriptscriptstyle-1} \frac{\partial \muyyni}{\partial [\etach_n]_j}
    }  \, ,
\end{align}
where $\muyyni \!=\! \alphacn e^{j \phicn} \cc(\thetacn, \tauucn)$ denotes the noise-free signal for each component $\ncomponent$ %\in \{1 \ \dots \ \Nc \}$, where $\Nc = \Nrp + \Nsp$ %$\Nc = 1+\Nrp + \Nsp$ is the total number of components impinging on \ac{RS} $n$, 
encompassing the \ac{LoS}, all \acp{RP}, and all \acp{SP}.
The individual entries of the local channel \ac{FIM} $\fimetachn$ from~\eqref{eq:fimetachn} can be found in Supplementary Material~\ref{app_fim}. While the dimension of the channel \ac{FIM} $\fimetachn$ stays constant in the $\roundlabeltxt{CP}$ and $\roundlabeltxt{NCP}$ cases, the dimension of the Jacobian matrices changes. The Jacobian matrices are %compute to
\begin{align}\label{eq:fim-jacobian-compact}
    \jacobian_n \!=\! \frac{\partial \etachtrp_n}{\partial \bm{\eta}} \!=\!\! 
    \begin{bmatrixs} 
        \jacobP^{\theta}_n & \jacobP^{\tauu}_n & \jacobP^{\phi}_n & \bm{0}  \\  
        \bm{0} & \bm{C}^{\tauu}_n & \bm{C}^{\phi}_n & \bm{0}  \\ 
                \jacobP^{\theta}_{\text{\tiny SP},n} & \jacobP^{\tauu}_{\text{\tiny SP},n} & \jacobP^{\phi}_{\text{\tiny SP},n} & \bm{0}  \\ 
        \bm{0} & \bm{0} & \jacobA_{\phi,n}%^{\scriptscriptstyle N(L-1+J)\times (\Nrp+\Nsp)} 
        & \bm{0} \\  
        \bm{0} & \bm{0} & \bm{0}  & \jacobA_{\alpha,n}%^{\scriptscriptstyle N(\Nrp+\Nsp)\times (\Nrp+\Nsp)}
    \end{bmatrixs} \!\in\! \realset{\dimGlobal}{\dimLocal}
    %\, ,
\end{align}
% \begin{align}\label{eq:fim-jacobian-compact}
%     \jacobian_n \!=\! \frac{\partial \etachtrp_n}{\partial \bm{\eta}} \!=\!\! 
%     \begin{bmatrixs} 
%         \jacobP^{\theta}_n & \jacobP^{\tauu}_n & \jacobP^{\phi}_n & \zeromatrix{3}{\Nc}  \\  
%         \zeromatrix{\dimClock}{\Nc} & \bm{C}^{\tauu}_n & \bm{C}^{\phi}_n & \zeromatrix{\dimClock}{\Nc}  \\ 
%                 \jacobP^{\theta}_{\text{\tiny SP},n} & \jacobP^{\tauu}_{\text{\tiny SP},n} & \jacobP^{\phi}_{\text{\tiny SP},n} & \zeromatrix{3J}{\Nc}  \\ 
%         \zeromatrix{\cdot}{\Nc} & \zeromatrix{\cdot}{\Nc} & \jacobA_{\phi,n}%^{\scriptscriptstyle N(L-1+J)\times (\Nrp+\Nsp)} 
%         & \zeromatrix{\cdot}{\Nc} \\  
%         \zeromatrix{\cdot}{\Nc} & \zeromatrix{\cdot}{\Nc} & \zeromatrix{\cdot}{\Nc}  & \jacobA_{\alpha,n}%^{\scriptscriptstyle N(\Nrp+\Nsp)\times (\Nrp+\Nsp)}
%     \end{bmatrixs} \!\in\! \realset{\dimGlobal}{\dimLocal}
%     %\, ,
% \end{align}
with individual submatrices defined in Supplementary Material~\ref{app_jacobian}.
Note that the presented \ac{CRB} represents a lower bound on the estimation error of $\hat{\etab}$ under the assumption of the correct detection and association of all components.

%\subsection{Bounds for Positioning and Synchronization}
We aim to compute the \ac{PEB} and \acp{CEB} for the parameters of interest. 
This can be done either by computing the complete \ac{CRB} matrix through inverting the large \ac{FIM} $\fim$ in~\eqref{eq:fim}, or computing the \ac{EFIM} by partitioning $\etab$ in parameters of interest $\etabWanted$ and nuisance parameters $\etabUnwanted$ as indicated in~\eqref{eq_eta}.
In the latter case, block-partitioning the \ac{FIM}
\begin{align}
    \fim = 
    \begin{bmatrix}
        \fim_{\etabWanted \etabWanted} & \fim_{\etabWanted \etabUnwanted} \\
        \fim_{\etabWanted \etabUnwanted}^\trp & \fim_{\etabUnwanted \etabUnwanted} 
    \end{bmatrix} \,,
\end{align}
the \ac{EFIM} is computed as $\efim = \fim_{\etabWanted \etabWanted} - \fim_{\etabWanted \etabUnwanted} \fim_{\etabUnwanted \etabUnwanted}^{-1} \fim_{\etabWanted \etabUnwanted}^\trp$~\cite{VanTrees2002optimumASP} from which the \ac{PEB} for $\pp$ with dimensions $\estDim$ is obtained as
\begin{align}
    \peb = \sqrt{
        \traceAuto{
            \left[ 
                \efim^{-1}
            \right]_{\scriptscriptstyle 1:\estDim,1:\estDim}
        }
    }
\end{align}
and the and \acp{CEB} are the clock offset error bound for $\deltatau$ %$\coeb$ and $\cpeb$ for the respective parameters are computed as
\begin{align}
    \coeb &= 
    \sqrt{
            \left[ 
                \efim^{-1}
            \right]_{\scriptscriptstyle D+1,D+1}
    } 
\end{align}
% and the phase offset error bound for $\deltaphi$ or $\deltaphib$ in the \roundlabeltxt{CP} or \roundlabeltxt{NCP} cases, respectively, 
% \begin{align}
%      \!\!\!\!\!\! \cpeb &= 
%     \begin{cases}
%     \begin{aligned} 
%     ~&\sqrt{
%             \traceAuto{
%                 \left[ 
%                     \efim^{-1}
%                 \right]_{\scriptscriptstyle D+2:N+D+1,D+2:N+D+1}
%             }
%         }  & \quad \roundlabeltxt{NCP} 
%         \\[3pt]
%         &\sqrt{
%             \left[ 
%                 \efim^{-1}
%             \right]_{\scriptscriptstyle D+2,D+2}
%         } & \quad \roundlabeltxt{CP} %\,,
%     \end{aligned}
%     \end{cases}.
% \end{align}
\newcommand{\Nph}{N_{\scriptscriptstyle \text{ph}}}
\newcommand{\estDimSP}{\estDim_{\scriptscriptstyle \text{SP}}}
and the phase offset error bound is defined as 
\begin{align}
     \cpeb &= \sqrt{
            \traceAuto{
                \left[ 
                    \efim^{-1}
                \right]_{\scriptscriptstyle D+2:D+1+\Nph,D+2:D+1+\Nph}
            }
        }.
\end{align}
where $\Nph$ indicates the number of phase parameters which is $\Nph=1$ or $\Nph=N$ for coherent  (\,\roundlabeltxt{CP}\,) and non-coherent processing  (\,\roundlabeltxt{NCP}\,), respectively.
For scatter point $j$ we define the corresponding \ac{SP}-\ac{PEB} as 
\begin{align}
\peb_{\scriptscriptstyle \text{SP}, j} = \sqrt{
\traceAuto{
    \left[ 
        \efim^{-1}
    \right]_{\scriptscriptstyle \mathcal{S}+(j-1)\estDimSP:\mathcal{S}+j\estDimSP,\mathcal{S}+(j-1)\estDimSP:\mathcal{S}+j\estDimSP}
}
}
\end{align}
where $\mathcal{S}=D+2+\Nph$ and $\estDimSP=3$ is the dimension of the estimated \ac{SP} position.

%%%%%%%%%%%%%%%%%%%%%%%%%%%%%%%%%%%%%%%%%%%%%%%%%%%%%%%%
%%%%%%%%%%%%%%%%%%%%%%%%%%%%%%%%%%%%%%%%%%%%%%%%%%%%%%%%
\section{Simulation Results}

%%%%%%%%%%%%%%%%%%%%%%%%%%%%%%%%%%%%%%%%%%%%%%%%%%%%%%%%
\ifthenelse{\equal{\externalizeFigures}{true}}
{
    \tikzexternaldisable    % for robustness with \ref{}...
}{}
\begin{figure*}[t] % t for top of the page
% NONCOHERENT CASE:
\setlength{\plotWidth}{0.46\linewidth}
\setlength{\plotHeight}{0.2\linewidth}
\subfigure[\roundlabeltxt{NCP}]{\scalebox{0.95}{
    %\vskip 0pt
    \input{Figures/Journal/bounds-B-sweep-RPphases}
    \label{fig:bounds-B-sweep}
    }}
\setlength{\plotWidth}{0.46\linewidth}%
\setlength{\plotHeight}{0.2\linewidth}%
\hspace{-0.3cm}\subfigure[\roundlabeltxt{CP}]{\scalebox{0.95}{
    \input{Figures/Journal/bounds-B-sweep-coh-RPphases}
    \label{fig:bounds-B-sweep-coh}
    }}
    \vspace{-0.5cm}
    \caption{
    The \roundlabeltxt{NCP} \ac{PEB} $\peb$ in \SI{}{\centi\metre} (a) and \roundlabeltxt{CP} \ac{PEB} (b) in \SI{}{\milli\metre} as a function of bandwidth $B$, varying the number of antennas $M$ at an $\SDNRbar = \SI{0}{\dB}$. We compare %\roundlabeltxt{L\,-\,-\,} 
    the \ac{LoS}-only case, with %\roundlabeltxt{LR\,-} 
    \ac{LoS} and \acp{RP}, and %\roundlabeltxt{LRS} 
    with \ac{LoS}, \acp{RP}, and \acp{SP}. 
    The latter is augmented with a scenario %{\roundlabeltxt{\scalebox{0.7}{$\scriptstyle\phirp$}}} 
    assuming known \ac{RP} phases $\phirp$.
    }\vspace{-0.5cm}
\end{figure*}
\ifthenelse{\equal{\externalizeFigures}{true}}
{
    \tikzexternalenable
}{}
%%%%%%%%%%%%%%%%%%%%%%%%%%%%%%%%%%%%%%%%%%%%%%%%%%%%%%%%

\todobox{
    \item[\tick{0}] Make the new analyses (impact of RPs vs. no RPs, improvements?)
    \item[\tick{1}] Find an analytical derivation for the limits of the bandwidth regimes!
    \item[\tick{0}] Maybe mention: \\
    "Note that by combining the spatial- and delay domains, i.e., intersecting (conical) beams with spheres, respectively, information about the \ac{UE} height can be gained even with \textit{no vertical aperture}, i.e., all antennas being at the same height."
}

In Fig. \ref{fig:scenario}, we provide an illustrative example of the considered scenario with  simulation parameters from Table~\ref{tab:sim-param}.
There are $N \!=\! 4$ \acp{RS} mounted in a room with $\Nrp\! \!=\! 4$ walls\footnote{Based on our choice of antenna polarizations, we omit modeling a floor and ceiling as the reflection coefficients $\Rp$ in~\eqref{eq:refcoeff-parallel} will be generally much lower than $\Rs$ in~\eqref{eq:refcoeff-orthogonal}~\cite[cf.\,Fig.\,1.14]{Pozar2012}.} (\ref{pgf:wall}) and $\Nsp \!=\! 2$ \acp{SP} (\,\ref{pgf:SP}\,).
The \ac{UE} (\,\ref{pgf:UE}\,) is located in the center of the room and transmits uplink pilots. 
The uplink pilots take paths via the \ac{LoS} (\,\ref{pgf:path-LoS}), paths via \acp{RP} (\,\ref{pgf:path-RPs}), and paths via \acp{SP} (\,\ref{pgf:path-SPs}) which impinge on the \acp{RS}. 
The positions of \acp{RP} (\,\ref{pgf:RP}\,) are computed geometrically according to Supplementary Material~\ref{app_reflection_point_pos}. %using geometric considerations, i.e., using the positions of the \ac{UE} and \acp{RS} as well as wall positions and orientations. %through~\eqref{eq:rp}.
The walls in the scenario depicted in Fig.\,\ref{fig:scenario} are assumed to be made of \textit{concrete} with typical relative permeability $\mu_\text{r} = 1$, relative permittivity $\epsilon_\text{r} = 6$~\cite{Olkkonen13Permittivity}, and conductivity $\mathsf{\sigma}_1 = \SI{1e-2}{\siemens\per\metre}$~\cite{Pislaru13Conductivity,Degli07Conductivity}. 
We assume linearly polarized antennas at both the \acp{RS} and \ac{UE} with polarization vectors $\polRS = \polUE = [0 \ 0 \ 1]^\trp$ aligned with the $z$-axis.

\begin{table}[t]%\vspace{-0.3cm}
			\centering
			\caption
			{List of default simulation parameters if not stated otherwise.%
                }%
\label{tab:sim-param}
			\begin{tabularx}{0.8\columnwidth}{@{}l|c|rl}%|ccc}
    		\toprule
    			\textbf{Variable} & \textbf{Symbol} & \textbf{Value} & \textbf{Unit}
                \\
   			\midrule 
                    %%%%%%%%%% Electrical Parameters %%%%%%%%%%
                    Carrier frequency & $\fc$  & \num{3.5} & \SI{}{\giga\hertz}\\
                    \ac{RS} antenna spacing & $d$ & $\lambda/2.1$  & \SI{}{\metre} \\
                    %\ac{UE} polarization v. & $\polUE$ & $[\num{0} \ \num{0} \ \num{1}]^\trp$  & - \\
                    %\ac{RS} polarization v. & $\polRS$ & $[\num{0} \ \num{0} \ \num{1}]^\trp$  & - \\
                    No. subcarriers & $K$ & $\num{20}$  & - \\
                    No. antennas & $M$ & $\num{16}$  & - \\
                    \midrule[0.1pt] 
                    %%%%%%%%%% (S)DNR %%%%%%%%%%
                    \ac{DNR} & $\DNR$ & $\num{0}$  & \SI{}{\dB} \\
                    \ac{SDNR} & $\SDNRbar$ & $\num{0}$  & \SI{}{\dB} \\
                    \midrule[0.1pt] 
                    %%%%%%%%%% Positions %%%%%%%%%%
                    \ac{UE} position & $\pp$ & $[\num{3.03} \ \num{2.87} \ \num{1}]^\trp$  & \SI{}{\metre} \\
                    \ac{RS} height & $\pprsx{n}$ & $\num{2.75}$  & \SI{}{\metre} \\
                    SP1 position & $\ppSPx{1}$ & $[\num{2} \ \num{2.2} \ \num{0.5}]^\trp$  & \SI{}{\metre} \\
                    SP2 position & $\ppSPx{2}$ & $[\num{4} \ \num{2} \ \num{1.5}]^\trp$  & \SI{}{\metre} \\
                    SP1 radius & $\radiusSPx{1}$ & \num{19.56} & \SI{}{\centi\metre} \\
                    SP2 radius & $\radiusSPx{2}$ & \num{17.57} & \SI{}{\centi\metre} \\
                    \midrule[0.1pt] 
                    %%%%%%%%%% Wall Parameters %%%%%%%%%%
                    Rel. permittivity & $\epsilon_\text{r}$ & 6  & - \\
                    Rel. permeability & $\mu_\text{r}$ & 1  & - \\
                    Conductivity & $\mathsf{\sigma}_1$ & \num{1e-2}  & \SI{}{\siemens\per\metre} \\
                    %\midrule[0.1pt] 
                    %%%%%%%%%% New parameters %%%%%%%%%%
                    %add & new & parameters  & here \\
                    %add & new & parameters  & here \\
               \bottomrule
		\end{tabularx}
\end{table}

%%%%%%%%%%%%%%%%%%%%%%%%%%
\subsection{Fundamental Performance Limits}\label{sec:sim-performance-limits}

For the scenario in Fig.\,\ref{fig:scenario} we evaluate the \ac{PEB} as a function of the bandwidth $B$, varying the %number of 
antennas $M$ and hence the aperture per \ac{RS} given a constant antenna spacing $d$.

\subsection*{\roundlabel{NCP} Bandwidth and Aperture Sweep}%
Fig.\,\ref{fig:bounds-B-sweep} depicts the \roundlabeltxt{NCP} \ac{PEB} for the cases of \ac{LoS}-only (\,\roundlabeltxt{L\,-\,-\,}\,), \ac{LoS} and \acp{RP} (\,\roundlabeltxt{LR\,-}\,) and \ac{LoS}, \acp{RP}, and \acp{SP} (\,\roundlabeltxt{LRS}\,) propagation, and for \roundlabeltxt{LRS} case with known \ac{RP} phases (\,\raisebox{0.5mm}{\roundlabeltxt{\scalebox{0.9}{$\scriptstyle\phirp$}}}\,).

For \roundlabeltxt{L\,-\,-\,} the \ac{PEB} remains almost flat for bandwidths $B<\BWhigh$.
This is because noncoherently positioning the \ac{UE} with distributed \acp{RS} is dominated by angular information $[\fim_{\theta,\theta}]_{\scriptscriptstyle\ncomponent,\ncomponent'}^{\scriptscriptstyle(n)}$ (cf.\,Table~\ref{tab:fim-terms}) until the bandwidth becomes large enough ($B\geq\BWhigh$) so delay information starts to dominate the \ac{PEB}.
We call this the \textit{high-bandwidth regime}.
Below this regime, \textit{path overlap} costs information~\cite{Shen2008PathOverlap} in the \roundlabeltxt{LR\,-} and \roundlabeltxt{LRS} cases w.r.t. the \ac{LoS}-only case through large off-diagonal elements in $\fimetachn$. 
We define the \textit{low-bandwidth regime} as the region $B\leq\BWlow$, where the \ac{LoS} overlaps with \acp{RP} in both the angular and delay domains. 
The \textit{mid-bandwidth regime} $\BWlow < B < \BWhigh$ is characterized by a bandwidth large enough to separate the \ac{LoS} from \acp{RP} in the delay domain.
%
%Leveraging the \ac{FIM} terms in Table~\ref{tab:fim-terms}, 
We next derive simple expressions for $\BWlow$ and $\BWhigh$ that capture the characteristics of the bandwidth regimes.

    Assuming a white covariance matrix $\rrbn$ the proportionality $[\fim_{\theta,\theta}]_{\scriptscriptstyle\ncomponent,\ncomponent'}^{\scriptscriptstyle(n)}\!\propto\!2\realp{\bb^\hermit(\taurpbar) \bb(\taulosbar)/K}$ holds, where we define the average \ac{LoS} delay as $\taulosbar\!:=\!\frac{1}{N}\sum_{n=1}^N\taulos$, and the average \ac{RP} delay as $\taurpbar\!:=\!\frac{1}{N(\Nrp-1)}\sum_{n=1}^N\sum_{\nRP=1}^{\Nrp-1}\taurp$. 
    For $[\fim_{\theta,\theta}]_{\scriptscriptstyle\ncomponent,\ncomponent'}^{\scriptscriptstyle(n)}$, this expression corresponds to the delay-domain \ac{RP} overlap cost relative to the delay-domain \ac{LoS} component information.
    It can be further expressed as $\frac{1}{K} \sum_{k=0}^{K-1} \cos (2\pi \deltaf k \Delta \tau)$, with $\Delta \tau \!:=\!\taurpbar\!-\!\taulosbar$, and is plotted in Fig.\,\ref{fig:overlap-temporal} (\ref{pgf:delay-overlap-exact}). 
    Using $\deltaf \triangleq \frac{B}{K}$, we approximate this result as $\frac{1}{2}\left(1+\cos(2\pi B \frac{K-1}{K} \Delta \tau)\right)$ which is indicated by the dashed curve (\ref{pgf:delay-overlap-approx}) in Fig.\,\ref{fig:overlap-temporal}.
    We choose the bandwidth threshold between the low and mid-bandwidth regimes (dotted line) to lie where this path overlap cost has dropped to $F = \SI{-3}{\dB}= \frac{1}{\sqrt{2}}$ which results in
    \begin{align}
        \BWlow = \frac{K \, \arccos \left( 2 F -1 \right)}{2 \pi \Delta \tau (K-1)} \approx \SI{17.78}{\mega\hertz} \, .
    \end{align}
%}

    The high-bandwidth regime is characterized by the delay information $[\fim_{\tauu,\tauu}]_{\scriptscriptstyle 1,1'}^{\scriptscriptstyle(n)} \!\propto\! \lVert \bbdplain(\taulos) \rVert^2$ dominating the \ac{PEB} over the angular information $[\fim_{\theta,\theta}]_{\scriptscriptstyle1,1'}^{\scriptscriptstyle(n)} \!\propto\! \lVert \aaadplain(\thetalos) \rVert^2$. 
    Note that $\bbdplain\!=\!\frac{\partial \bb}{\partial \tau}$ from \eqref{eq:bbd} and  $\aaadplain\!=\!\frac{\partial \aaa}{\partial \theta}$ from \eqref{eq:aaap}. 
    The Jacobians that map to position $\pp$ have the proportionalities $\jacobP^{\theta}_n \!\propto\! \frac{1}{c \tauulos}$ and $ \jacobP^{\tauu}_n \!\propto\! \frac{1}{c}$. 
    Equating $\frac{1}{c^2 \taulosbar^2}\lVert \aaadplain(0) \rVert^2 \! =\! \frac{1}{c^2} \lVert \bbdplain(\taulos) \rVert^2$ and abbreviating $\Summ\! \triangleq\!\sum_{m=\frac{M\!-\!1}{2}}^{\frac{M\!-\!1}{2}}m^2 \!=\! \frac{M (M^2-1)}{12}$ and $\Sumk\! \triangleq \!\sum_{k=0}^{K\!-\!1} k^2\!=\!\frac{2 K^3-3K^2+K}{6}$, we solve for $B$ contained in $\bbdplain$ to compute the high bandwidth regime threshold (for $M\!=\!12$) as
    \begin{align}
        \BWhigh = \frac{K d}{\taulosbar \lambda} \sqrt{\frac{\Summ}{\Sumk}} \approx \SI{207.9}{\mega\hertz} \, .
    \end{align}
%}
Following the discussion of our \ac{RP} channel model in Sec.\,\ref{app_RP_channel}, the case \raisebox{0.5mm}{\roundlabeltxt{\scalebox{0.9}{$\scriptstyle\phirp$}}} assumes known \ac{RP} component phases $\phirp$.
Although this is clearly unrealistic in practice, we present the respective curves in Fig.\,\ref{fig:bounds-B-sweep} to illustrate the substantial theoretical gains in positioning accuracy achievable by exploiting (perfect) knowledge of per-\ac{RS} component phases. % than nuisance (clock) phase parameters.
This effectively transforms the angle or delay-based \roundlabeltxt{NCP} positioning into carrier-phase-based positioning.
\subsection*{\roundlabel{CP} Bandwidth and Aperture Sweep}
Fig.\,\ref{fig:bounds-B-sweep-coh} depicts the \ac{PEB} $\peb$ in the \roundlabeltxt{CP} setting for the three different multipath cases.
Having knowledge of the clock phases at distributed \acp{RS} allows the phase-coherent processing of the impinging components.
While our channel model in Sec.\,\ref{sec_signal_channel_model} assumes that each \ac{RP} and each \ac{SP} introduces an unknown phase that varies for every \ac{RS} $n$, there are $N$ \ac{LoS} components impinging at the infrastructure which only need to be used to estimate a single\footnote{Note that in the \roundlabeltxt{NCP} case, $N$ \ac{LoS} component phases need to be used to estimate $N$ phase offsets contained in $\deltaphib$ which costs the information they could have contributed on the \ac{UE} position.} phase offset $\deltaphi$ to the \ac{UE} (common to all \acp{RS}). 
Hence the remaining phases of $N\!-\!1$ \ac{LoS} components contribute a significant amount of information about the \ac{UE} position due to the high curvature of the likelihood function around the true position $\pp$ but come at the price of a harder estimation problem due to its shape, a point better discussed in Sec. \ref{sec:cost-function} (cf. Fig.\,\ref{fig_coh_2d})~\cite{Wymeersch23CarrierPhasePos}.
While the \ac{PEB} is generally much lower in the \roundlabeltxt{CP} case than in the \roundlabeltxt{NCP} case (note the vertical axis scaling in Fig.\,\ref{fig:bounds-B-sweep} and Fig.\,\ref{fig:bounds-B-sweep-coh}), we again observe a performance degradation in the presence of path overlap with \acp{RP} and \acp{SP}, 
%Like in the \roundlabeltxt{NCP} case, path overlap impacts the \ac{PEB} also in the \roundlabeltxt{CP} case 
due to the coupling term depicted in Fig.\,\ref{fig:overlap-temporal}.
The \ac{PEB} remains almost flat in the \ac{LoS}-only case \roundlabeltxt{L\,-\,-\,}. 
In the \roundlabeltxt{LR,-} case, path overlap with \acp{RP} causes slight \ac{PEB} degradation, while in the \roundlabeltxt{LRS} case, overlap with \acp{SP} leads to significantly greater degradation.
In the \raisebox{0.5mm}{\roundlabeltxt{\scalebox{0.9}{$\scriptstyle\phirp$}}} case, the gain of knowing \ac{RP} phases is marginal as the \ac{PEB} is dominated by information from \ac{LoS} phases. 
%\subfigure[]{
\begin{figure}
    \centering
    \setlength{\plotWidth}{0.45\linewidth}
    \setlength{\plotHeight}{0.25\linewidth}
    \subfigure[]{\scalebox{0.95}
    {
    %\vskip 0pt
    \hspace{-4mm}
    \input{Figures/Journal/overlap-temporal}
    %\vspace{-0.2cm}\caption{abc
    %}
    \label{fig:overlap-temporal}
    }}\hspace{-2mm}%
    \setlength{\plotWidth}{0.39\linewidth}
    \setlength{\plotHeight}{0.25\linewidth}
    \subfigure[]{\scalebox{0.95}
    {
    %\vskip 0pt
    \input{Figures/Journal/overlap-spatial}
    %\vspace{-0.5cm}
    %\caption{abc
    %}
    \label{fig:overlap-spatial}
    }}
    \vspace{-3mm}
    \caption{%In the low- and mid-bandwidth regimes, angular information dominated the \ac{PEB}. 
    Path overlap in the delay domain causes large correlations in $[\fim_{\theta,\theta}]_{\scriptscriptstyle\ncomponent,\ncomponent'}^{\scriptscriptstyle(n)}$. 
    Angular overlap causes large correlations in $[\fim_{\theta,\phi}]_{\scriptscriptstyle 0,\ncomponent'}^{\scriptscriptstyle(n)}$.}\label{fig:component-overlap}\vspace{-0.5cm}
\end{figure}
\begin{figure*}[t] % t for top of the page
    \centering
    \begin{minipage}[t]{0.32\linewidth}\centering
    \vskip 0pt	% THIS IS OF UTTERMOST NECESSITY TO MAKE THE TOP "[t]" PARAMETER WORK!
    %\vspace{-0.1cm}%
        \setlength{\plotWidth}{1\linewidth}
        \input{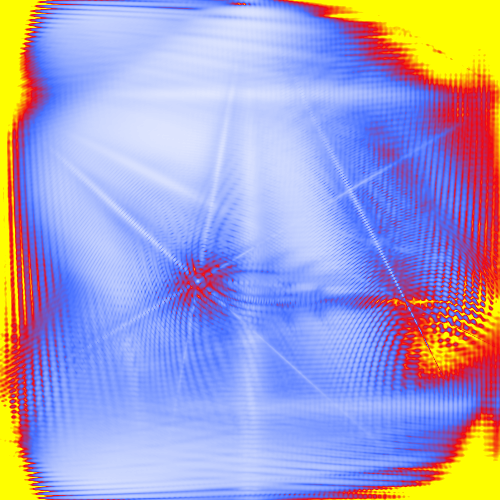}
        \vspace{-0.25cm}
        \caption{\ac{PEB} in the low-bandwidth regime: At $B\!=\!\SI{3}{\mega\hertz}$, path overlap of \ac{LoS} with \acp{SP} and \acp{RP} degrades positioning accuracy, particularly in room corners. 
        Angular information dominates the \ac{PEB}.}\label{fig:heatmap-low}
    \end{minipage}
    \hfill%\hspace{0.2cm}
    \begin{minipage}[t]{0.32\linewidth}\centering
    \vskip 0pt	% THIS IS OF UTTERMOST NECESSITY TO MAKE THE TOP "[t]" PARAMETER WORK!
    %\vspace{-0.1cm}%
        \setlength{\plotWidth}{1\linewidth}
        \input{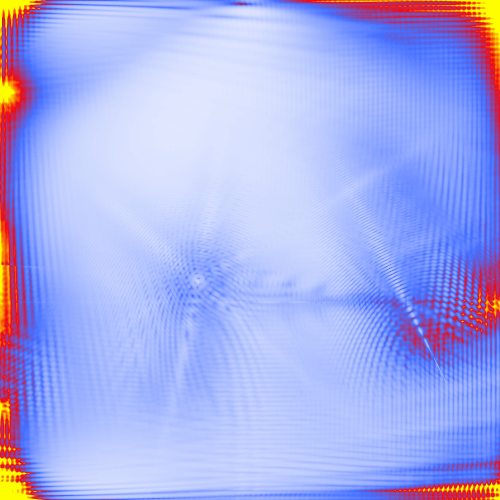}
        \vspace{-0.25cm}
        \caption{\ac{PEB} in the mid-bandwidth regime: At $B\!=\!\SI{30}{\mega\hertz}$, path overlap with the \acp{RP} is resolved but path overlap with \acp{SP} still degrades positioning accuracy. Angular information dominates the \ac{PEB}.}\label{fig:heatmap-mid}
    \end{minipage}
    \hfill%\hspace{0.2cm}
    \begin{minipage}[t]{0.32\linewidth}\centering
    \vskip 0pt	% THIS IS OF UTTERMOST NECESSITY TO MAKE THE TOP "[t]" PARAMETER WORK!
    %\vspace{-0.1cm}%
        \setlength{\plotWidth}{1\linewidth}
        \input{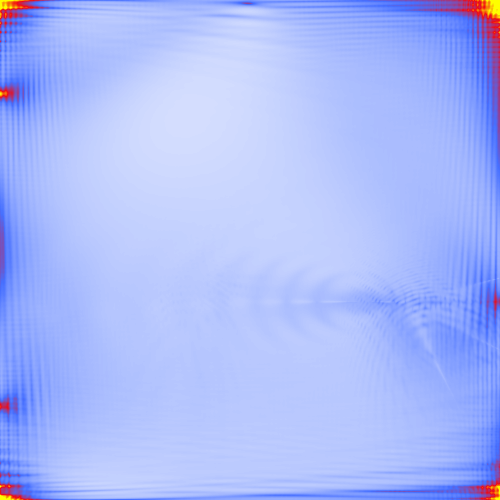}
        \vspace{-0.25cm}
        \caption{\ac{PEB} in the high-bandwidth regime: At $B\!=\!\SI{300}{\mega\hertz}$, the path overlap of the \ac{LoS} with the both the \acp{RP} and the \acp{SP} is resolved. Delay information dominates the \ac{PEB}.}\label{fig:heatmap-high}
    \end{minipage}
\vspace{-0.3cm}%
\end{figure*}

\subsection*{\roundlabel{NCP} Spatial Sweep}
The impact of angular path overlap gets more conceivable when evaluating the \ac{PEB} $\peb$ spatially in the form of heatmaps.
% ---------- low-bandwidth regime ----------
We show a heatmap for $\peb$ in the low-bandwidth regime in Fig.\,\ref{fig:heatmap-low}.
The ``checked pattern'' is caused by the information contributed by the carrier phase. 
It is observable that there are ``white stripes'' connecting the centers of \acp{RS} with \acp{SP} and \acp{RP} that are located at straight lines originating at \acp{RS} and orthogonally intersecting walls.
These straight lines indicate positions of the \ac{UE}, where the \ac{LoS} and respective \ac{SP} or \ac{RP} components overlap completely in the \textit{angular domain}.
A positioning performance degradation due to path overlap is observable just outside these straight lines, where components start to separate in the angular domain. 
%\draft{
This can be explained by larger off-diagonal elements (i.e., coupling terms) $[\fim_{\theta,\phi}]_{\scriptscriptstyle 0,\ncomponent'}^{\scriptscriptstyle(n)}$ of the \ac{LoS} \acp{AoA} %$\thetalos$ 
and \ac{NLoS} component phases %$\phirp$ and $\phisp$ 
in the channel \ac{FIM}.
%$\big[\fimetachn\big]_{i,j}$ with $i=1$ and $2\leq j \leq \Nrp+\Nsp$ in the channel \ac{FIM} 
%according to the terms derived in Appendix~\ref{app_fim}. 
Assuming a white \ac{DMC} covariance matrix, we would have %$[\fim_{\theta,\phi}]_{\scriptscriptstyle 0,\ncomponent'}^{\scriptscriptstyle(n)} \propto 2\realp{j \aaapx{1}^\hermit \aaanc}$
$[\fim_{\theta,\phi}]_{\scriptscriptstyle 0,\ncomponent'}^{\scriptscriptstyle(n)} \propto \realp{j \frac{\partial \aaa^\hermit (\thetalos)}{\partial \thetalos} \aaa(\thetacn)}$, a coupling term depicted in Fig.\,\ref{fig:overlap-spatial} (\ref{pgf:angular-overlap}) for $\thetacn = 0$ and varying $\thetalos$.
A ``white stripe'' corresponds to the null where both angles are equal. 
It is surrounded by two larger lobes.
%}
In room corners, the path overlap of the \ac{LoS} with \acp{RP} in both the angular- and delay domains causes a strong performance degradation. % at such locations. 

% ---------- mid-bandwidth regime ----------
Apart from the room corners %(see Fig.\,\ref{fig:heatmap-mid}), 
the impact of path overlap with the \acp{RP} is resolved in the mid-bandwidth regime, as they are now well separated from the \ac{LoS} in the \textit{delay domain}. 
While the \acp{RP} are located at greater distances s.t. they arrive ``later'' at the \acp{RS} in the \textit{delay domain} than the \acp{SP}, which are following the \ac{LoS} much closer in the delay domain. 
Hence, the impact of path overlap with \acp{SP} is still visible in the mid-bandwidth regime as is well-observable in Fig.\,\ref{fig:heatmap-mid}.

% ---------- high-bandwidth regime ----------
In the high-bandwidth regime, the bandwidth is large enough to ensure all components are well separated in the delay domain.
The path overlap is largely resolved and the \ac{PEB} is spatially smooth.
Delay information dominates in this bandwidth regime and improves the \ac{PEB} as can be seen from Fig.\,\ref{fig:heatmap-high}. % (note the different color bar scaling).

%%%%%%%%%%%%%%%%%%%%%%%%%%%%%%%%%%%%%%%
%%% Estimator Results
\subsection{Performance of Algorithms}\label{sec:performance}
In this part, we evaluate the performance of the proposed joint localization, synchronization and mapping algorithms from Sec.~\ref{sec_algorithms} with phase-coherent \acp{RS}~\roundlabeltxt{CP}. 
We consider a UE located at $\pp = [\num{3.0297} ~ \num{2.8735} ~ \num{1}]^\trp \, \SI{}{\metre}$ (an off-grid value to exclude any advantage for the algorithm during grid-based optimization), with  clock offset $\deltatau = \SI{16.66}{\nano\second}$ corresponding to $\SI{5}{\metre}$ in range, and phase offset $\deltaphi = 45\degree$. 
To lower the computation time, we assume known height of the UE and consider a single \ac{SP} located at $\ppSP = [\num{2.5} ~ \num{3.5} ~ \num{1.7}]^\trp \, \SI{}{\metre}$. 
The bandwidth, number of \acp{RS}, and number of antennas per \ac{RS} are set to $B = 10 \, \rm{MHz}$, $N=4$ and $M = 8$, respectively. 
Performance obtained through $100$ independent Monte Carlo trials is shown for \textit{(i)} the  RML estimator in \eqref{final_suboptimalUE}, which operates with reduced complexity as outlined in Sec.~\ref{sec_coarse_est}, and \textit{(ii)} the optimal JML in \eqref{compressed_optproblem}  initialized by the RML, RML-NCP in \eqref{suboptimal_UElocsync} and RNST in \eqref{eq:L-RNST}, as described in Sec.~\ref{sec_coherent_mapping}.\footnote{More specifically, \emph{(i)} refers to coarse estimation of UE position $\hat{\pp}^\text{\tiny RML}$ and clock offset $ \widehat{\delta}_{\tau}(\hat{\pp}^\text{\tiny RML})$ via \eqref{final_suboptimalUE_2D} and \eqref{eq_deltatau_final}, respectively, using coarse estimate of phase offset $\hat{\delta}^\text{\tiny RML-NCP}_\phi(\hat{\pp}^\text{\tiny RML}, \widehat{\delta}_{\tau}(\hat{\pp}^\text{\tiny RML}))$ via \eqref{eq_deltaphi_final}; \emph{(ii)} uses the same estimation and, in addition, coarse SP position estimates  $\hat{\bm{p}}^\text{\tiny SP, RNST}$ via \eqref{eq:L-RNST} to initialize the optimal JML that considers the true ML cost function \eqref{compressed_optproblem}.}

%%%%%%%%%%%%%%%%%%%%%%%%%%%%%%%%%%%%%%%%%%
% RMSEs and CRBs
\subsubsection{RMSE Analysis}
Fig.~\ref{fig_rmses} depicts the \ac{RMSE} performance of the considered estimators, along with the corresponding \ac{CRB}, for estimation of the \ac{UE} location in Fig.~\ref{fig_rmse_loc} and clock offset in Fig.~\ref{fig_rmse_clock}.\footnote{In computing the \ac{RMSE}, error values have been cleaned up from a few outliers (via the standard inter-quartile range method). These are due to the very spiky landscape of minima in the \ac{UE} location coordinates, as will be discussed in the next subsection.} %, phase offset and SP location. 
It is apparent that the overall accuracy for the \ac{UE} position is in the sub-meter regime and rapidly reaches centimeter-level values, even at intermediate \ac{SDNR} levels, before asymptotically decreasing to sub-millimeter accuracy towards high \ac{SDNR} levels.
More specifically, the proposed RML is able to achieve remarkable performance, with \ac{RMSE} equal to a few centimeters already at 5 dB \ac{SDNR}, despite its low complexity. This, in turn, provides satisfactory initializations for the JML estimator, which attains the \ac{CRB} already at 15 dB \ac{SDNR} for both the UE location, PEB \roundlabeltxt{CP} in Fig.~\ref{fig_rmse_loc}, and clock offset, CEB in Fig.~\ref{fig_rmse_clock}.
Towards low \ac{SDNR} levels, both the RML and JML approach the \roundlabeltxt{NCP} \ac{PEB}, with the JML estimators deviation from the \roundlabeltxt{CP} \ac{PEB} coinciding with the deviation from the \ac{CEB} in Fig.~\ref{fig_rmse_clock}. This can be attributed to the increased distortion of the estimated \ac{LoS} phases at low \ac{SDNR} levels, resulting in increasing phase offset estimation errors, thus inhibiting fully coherent performance.
The same trend can be observed in the phase offset estimation performance, omitted due to lack of space. 
The results demonstrate the effectiveness of the JML in utilizing all the available information from the coherent processing, significantly improving upon the initial estimates provided by the RML, so achieving high-quality localization and synchronization using distributed \acp{RS}. 

%%%%%%%%%%%%%%%%%%%%%%%%%%%%%%%%%%%%%%%%%%
% UE location, clock offset, phase offset and SP location RMSEs
% \begin{figure}
%         \begin{center}
%         %\vspace{-0.22in}
%         \subfigure[]{
% 			 \label{fig_rmse_loc}
% 			 \includegraphics[width=0.35\textwidth]{Figures/Estimator/UE_loc_rmse_v3.eps}  
% 		}
%         \subfigure[]{
% 			 \label{fig_rmse_clock}
% 			 \includegraphics[width=0.35\textwidth]{Figures/Estimator/clock_offset_rmse_v3.eps}
% 		}		
%   % \subfigure[]{
% 		% 	 \label{fig_rmse_phase}
% 		% 	 \includegraphics[width=0.45\textwidth]{Figures/Estimator/phase_offset_rmse_v2.eps}
% 		% }	
%   % \subfigure[]{
% 		% 	 \label{fig_rmse_sp_pos}
% 		% 	 \includegraphics[width=0.45\textwidth]{Figures/Estimator/sp_pos_rmse_v2.eps}
% 		% }	
% 		\end{center}
% 		\vspace{-0.1in}
%         \caption{RMSE on the estimation of \subref{fig_rmse_loc} UE location $\pp$ and \subref{fig_rmse_clock} clock offset $\deltatau$, %\subref{fig_rmse_phase} phase offset $\deltaphi$ and \subref{fig_rmse_sp_pos} SP location $\ppSP$, 
%         achieved by RML and JML
%  along with CRBs as a function of SDNR.}  
%         \label{fig_rmses}
%         \vspace{-0.1in}
% \end{figure}

\begin{figure}
    \begin{center}
    \subfigure[]{
        \label{fig_rmse_loc}
        \setlength{\figurewidth}{0.4\textwidth}
        \setlength{\figureheight}{0.2\textwidth}
        \def\datapath{Figures/Estimator_figures/UE_loc_rmse_v3/}
 
\pgfplotsset{every axis/.append style={
  label style={font=\footnotesize},
  legend style={font=\scriptsize},
  tick label style={font=\footnotesize},
  xticklabel={
    % test if the x value is below zero
    \ifdim \tick pt < 0pt
      % if yes, calculate the absolute value
      \pgfmathparse{abs(\tick)}%
      % and print first a minus sign in a zero-width box, followed by the absolute value
      \llap{$-{}$}\pgfmathprintnumber{\pgfmathresult}
   \else
     % if no, print as usual
      \pgfmathprintnumber{\tick}
   \fi
}}}
 
\definecolor{mycolor1}{rgb}{0.00000,0.44700,0.74100}%
\definecolor{mycolor2}{rgb}{0.85000,0.32500,0.09800}%
\definecolor{mycolor3}{rgb}{0.46667,0.67451,0.18824}%

\begin{tikzpicture}[baseline,trim axis left]%baseline,

\begin{axis}[%
width=\figurewidth,
height=0.969\figureheight,
at={(0\figurewidth,0\figureheight)},
scale only axis,
tick align=inside,		       % [B] ticks oriented inside 
axis line style = thick,	   % [B] thick axis lines 
xmin=-10,
xmax=30,
xlabel={SDNR in \SI{}{\dB}},
ymode=log,
ymin=0.0001,
ymax=0.55, %1,
yminorticks=true,
ylabel={UE localization RMSE in \SI{}{\metre}},
% axis background/.style={fill=white},
xmajorgrids,
ymajorgrids,
yminorgrids,
grid style=dotted,
legend style={at={(1,1)}, legend columns=2, anchor=north east, legend cell align=left, align=left, draw=white!15!black},
fill=white,line cap = round, fill opacity=0.8, text opacity = 1, draw opacity = 1
%extraAxisOptions,
% unit vector ratio=1 1 1
]
\addplot [color=IEEEblue, dashdotted, line width=\linewidthA, mark size=\marksizeA, mark=o, mark options={solid, IEEEblue}]
  %table[]{\datapath/UE_loc_rmse_v3-1.tsv};
    table[row sep=crcr]{%
-10	0.425354140420415\\
-5	0.253050893560027\\
0	0.136770516248591\\
5	0.0478863215753764\\
10	0.00315407125153175\\
15	0.00318772100221351\\
20	0.00307554150488881\\
25	0.00311108707169547\\
30	0.00311377447246325\\
35	0.00321727563214668\\
40	0.00309882332857686\\
45	0.00336564577374136\\
50	0.00302672103545781\\
};
\addlegendentry{RML}

\addplot [color=IEEEdarkorange, line width=\linewidthA, mark size=\marksizeA, mark=+, mark options={solid, IEEEdarkorange}]
  %table[]{\datapath/UE_loc_rmse_v3-2.tsv};
    table[row sep=crcr]{%
-10	0.512224776120022\\
-5	0.216628310716346\\
0	0.122903640828765\\
5	0.0538556585113796\\
10	0.039681130396986\\
15	0.0017205278515171\\
20	0.000431513629664685\\
25	0.000224636355389316\\
30	0.000145456411137068\\
35	7.77191991366284e-05\\
40	5.02577696322485e-05\\
45	2.38138093061632e-05\\
50	1.35636550829974e-05\\
};
\addlegendentry{JML}

\addplot [color=IEEElightgreen, dashed, mark size=\marksizeA, line width=\linewidthA]
  table[]{\datapath/UE_loc_rmse_v3-3.tsv};
\addlegendentry{PEB \roundlabeltxt{CP} }

%CP computed separately kept for comparison: the lines were identical when generating the figure (chcked by Thomas)
% \addplot [color=red, dotted, line width=\linewidthA]
%   table[]{\datapath/crb_c_nc-1.tsv};
% \addlegendentry{PEB (CP)}

\addplot [color=IEEElightgreen, dash pattern=on 0.01pt off 1.5pt, line width=\linewidthA]
  table[]{\datapath/crb_c_nc-2.tsv};
\addlegendentry{PEB \roundlabeltxt{NCP} }

\end{axis}
\end{tikzpicture}%
    }
    \vspace{-4mm}
    
    \subfigure[]{
        \label{fig_rmse_clock}
        \setlength{\figurewidth}{0.4\textwidth}
        \setlength{\figureheight}{0.2\textwidth}   
        \def\datapath{Figures/Estimator_figures/clock_offset_rmse_v3/}
 
\pgfplotsset{every axis/.append style={
  label style={font=\footnotesize},
  legend style={font=\scriptsize},
  tick label style={font=\footnotesize},
  xticklabel={
    % test if the x value is below zero
    \ifdim \tick pt < 0pt
      % if yes, calculate the absolute value
      \pgfmathparse{abs(\tick)}%
      % and print first a minus sign in a zero-width box, followed by the absolute value
      \llap{$-{}$}\pgfmathprintnumber{\pgfmathresult}
   \else
     % if no, print as usual
      \pgfmathprintnumber{\tick}
   \fi
}}}
 
\definecolor{mycolor1}{rgb}{0.00000,0.44700,0.74100}%
\definecolor{mycolor2}{rgb}{0.85000,0.32500,0.09800}%
\definecolor{mycolor3}{rgb}{0.46667,0.67451,0.18824}%

\begin{tikzpicture}[baseline,trim axis left] %baseline,

\begin{axis}[%
width=\figurewidth,
height=0.969\figureheight,
at={(0\figurewidth,0\figureheight)},
scale only axis,
tick align=inside,		       % [B] ticks oriented inside 
axis line style = thick,	   % [B] thick axis lines 
xmin=-10,
xmax=30,
xlabel={SDNR in \SI{}{\dB}},
ymode=log,
ymin=0.01,
ymax=10,
yminorticks=true,
ylabel={Clock offset RMSE in \SI{}{\metre} },
% axis background/.style={fill=white},
xmajorgrids,
ymajorgrids,
yminorgrids,
grid style=dotted,
legend style={at={(1,1)}, legend columns=2, anchor=north east, 
legend cell align=left, align=left, draw=white!15!black},
fill=white,line cap = round, fill opacity=0.8, text opacity = 1, draw opacity = 1
%extraAxisOptions,
% unit vector ratio=1 1 1
]
\addplot [color=IEEEblue, dashdotted, line width=\linewidthA, mark size=\marksizeA, mark=o, mark options={solid, IEEEblue}]
  %table[]{\datapath/clock_offset_rmse_v3-1.tsv};
  table[row sep=crcr]{%
-10	3.12587719188273\\
-5	2.58776630475108\\
0	1.7866297574085\\
5	0.946540655305516\\
10	0.638531586130696\\
15	0.452643114784582\\
20	0.41916630620081\\
25	0.425800738739483\\
30	0.419648275779256\\
35	0.4100230292475\\
40	0.413824040779333\\
45	0.413161415221361\\
50	0.413201481696823\\
};
\addlegendentry{RML}

\addplot [color=IEEEdarkorange, line width=\linewidthA, mark size=\marksizeA, mark=+, mark options={solid, IEEEdarkorange}]
  %table[]{\datapath/clock_offset_rmse_v3-2.tsv};
  table[row sep=crcr]{%
-10	3.71617060063898\\
-5	1.73414686141836\\
0	0.933152110886036\\
5	0.534378398140361\\
10	0.233234987251047\\
15	0.122988164015247\\
20	0.0743887908107787\\
25	0.0405537629708458\\
30	0.0265011329314279\\
35	0.0122568299270111\\
40	0.00736521897784823\\
45	0.00337405620051558\\
50	0.0020353313985345\\
};
\addlegendentry{JML}

\addplot [color=IEEElightgreen, dashed, line width=\linewidthA]
  %table[]{\datapath/clock_offset_rmse_v3-3.tsv};
  table[row sep=crcr]{%
-10	2.12233044689911\\
-5	1.19347411600741\\
0	0.671139815973263\\
5	0.377409653502756\\
10	0.212233044690344\\
15	0.119347411600423\\
20	0.0671139815972729\\
25	0.0377409653502166\\
30	0.0212233044690133\\
35	0.0119347411600077\\
40	0.00671139815973895\\
45	0.00377409653502229\\
50	0.00212233044690186\\
};
\addlegendentry{CEB}

\end{axis}
\end{tikzpicture}%
    }
    % \vspace{-3mm}
    
    % \subfigure[]{
    %     \label{fig_rmse_phase}
    %     \setlength{\figurewidth}{0.4\textwidth}
    %     \setlength{\figureheight}{0.175\textwidth}   
    %     \def\datapath{Figures/Estimator_figures/phase_offset_rmse_vs_snr_v2/}
    %     \input{Figures/Estimator_figures/phase_offset_rmse_vs_snr_v2/phase_offset_rmse_vs_snr_v2}
    %     % \includegraphics[width=0.35\textwidth]{Figures/Estimator/clock_offset_rmse_v3.eps}
    % }
    
    \end{center}
    \vspace{-5mm}
    \caption{RMSE on the estimation of \subref{fig_rmse_loc} \ac{UE} location $\pp$ and \subref{fig_rmse_clock} clock offset $\deltatau$, achieved by RML and JML along with \acp{CRB} as a function of \ac{SDNR}.}  
    \label{fig_rmses}
    \vspace{-6mm}
\end{figure}

%%%%%%%%%%%%%%%%%%%%%%%%%%%%%%%%%%%%%%%%%%%% 

%%%%%%%%%%%%%%%%%%%%%%%%%%%%%%%%%%%%%%%%%%
% spiky cost functions
\subsubsection{Cost Functions Analysis}\label{sec:cost-function}
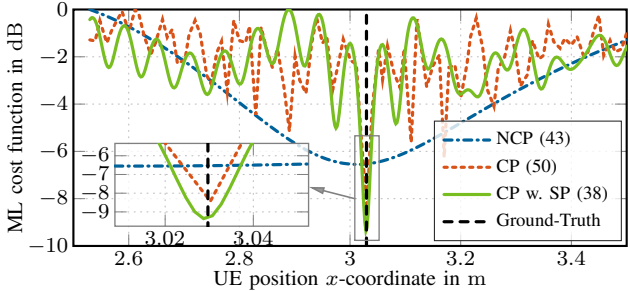
\begin{figure}
\centering
%\vspace{-0.1in}
\setlength{\figurewidth}{0.4\textwidth}%  
\setlength{\figureheight}{0.18\textwidth}   
\def\datapath{Figures/Estimator_figures/1d_costs_all_comparison/}
 
\pgfplotsset{every axis/.append style={
  label style={font=\footnotesize},
  legend style={font=\scriptsize},
  tick label style={font=\footnotesize},
  xticklabel={
    % test if the x value is below zero
    \ifdim \tick pt < 0pt
      % if yes, calculate the absolute value
      \pgfmathparse{abs(\tick)}%
      % and print first a minus sign in a zero-width box, followed by the absolute value
      \llap{$-{}$}\pgfmathprintnumber{\pgfmathresult}
   \else
     % if no, print as usual
      \pgfmathprintnumber{\tick}
   \fi
}}}
 
\definecolor{mycolor1}{rgb}{0.00000,0.44700,0.74100}%
\definecolor{mycolor2}{rgb}{0.85098,0.32549,0.09804}%
\definecolor{mycolor3}{rgb}{0.46667,0.67451,0.18824}%
\begin{tikzpicture}[baseline]

\begin{axis}[%
width=\figurewidth,
height=0.957\figureheight,
at={(0\figurewidth,0\figureheight)},
scale only axis,
tick align=inside,		       % [B] ticks oriented inside 
axis line style = thick,	   % [B] thick axis lines 
xmin=2.5,
xmax=3.5,
xlabel={UE position $x$-coordinate in \SI{}{\metre}},
ymin=-10,
ymax=0,
ylabel={ML cost function in \SI{}{\dB}},
% axis background/.style={fill=white},
xmajorgrids,
ymajorgrids,
grid style=dotted,
line cap = round,
legend style={at={(1,0)}, anchor=south east, legend cell align=left, align=left, draw=white!15!black},
fill=white,line cap = round, fill opacity=0.8, text opacity = 1, draw opacity = 1
%extraAxisOptions,
% unit vector ratio=1 1 1
]
\addplot [color=IEEEblue, dashdotted, line width=\linewidthA]
  table[]{\datapath/1d_costs_all_comparison-1.tsv};
% \addlegendentry{Non-coherent: (46)}
\addlegendentry{NCP \eqref{eq_1st_llr}}

\addplot [color=mycolor2, dotted, line width=\linewidthA]
  table[]{\datapath/1d_costs_all_comparison-2.tsv};
% \addlegendentry{Coherent: (53)}
\addlegendentry{CP \eqref{eq_compr_coherent}}

\addplot [color=IEEElightgreen, line width=\linewidthA]
  table[]{\datapath/1d_costs_all_comparison-3.tsv};
% \addlegendentry{Coherent with SPs: (40)}
\addlegendentry{CP w. SP \eqref{eq_direct_compressed_coherent}}

\addplot [color=black, dashed, line width=\linewidthA]
  table[]{\datapath/1d_costs_all_comparison-4.tsv};
\addlegendentry{Ground-Truth}

\addplot [color=black!50, line width = 0.5pt]
  table[row sep=crcr]{%
3.00860507762133	-9.76357523047663\\
3.05253338915903	-9.76357523047663\\
3.05253338915903	-5.34061551100695\\
3.00860507762133	-5.34061551100695\\
3.00860507762133	-9.76357523047663\\
};
\addplot [color=black!50,fill=black!50, line width=0.6pt, forget plot, line cap = round,  %
{-Stealth[inset=0pt, scale=1.05, angle'=25]}]
  table[row sep=crcr]{%
3.00860507762133	-8\\
2.925	-7.5\\
};
% %%%%%%%%%% %%%%%%%%%% %%%%%%%%%% %%%%%%%%%% %%%%%%%%%%

\end{axis}

\begin{axis}[%s
width=0.35\figurewidth,
height=0.335\figureheight,
at={(0.075\figurewidth,0.08\figureheight)},
scale only axis,
xmin=3.00860507762133,
xmax=3.05253338915903,
ymin=-9.76357523047663,
ymax=-5.34061551100695,
axis background/.style={fill=white},
xmajorgrids,
ymajorgrids,
xtick distance={0.02},
grid style=dotted,
fill=white,line cap = round, fill opacity=0.8, text opacity = 1, draw opacity = 1,
axis line style={color=black!50, line width = 0.5pt}
%extraAxisOptions,
% unit vector ratio=1 1 1
]
\addplot [color=IEEEblue, dashdotted, line width=\linewidthA, forget plot]
  table[]{\datapath/1d_costs_all_comparison-5.tsv};
\addplot [color=mycolor2, dotted, line width=\linewidthA, forget plot]
  table[]{\datapath/1d_costs_all_comparison-6.tsv};
\addplot [color=IEEElightgreen, line width=\linewidthA, forget plot]
  table[]{\datapath/1d_costs_all_comparison-7.tsv};
\addplot [color=black, dashed, line width=\linewidthA, forget plot]
  table[]{\datapath/1d_costs_all_comparison-8.tsv};
\end{axis}

% \begin{axis}[%
% width=1.29\figurewidth,
% height=1.29\figureheight,
% at={(-0.168\figurewidth,-0.197\figureheight)},
% scale only axis,
% xmin=0,
% xmax=1,
% ymin=0,
% ymax=1,
% axis line style={draw=none},
% ticks=none,
% axis x line*=bottom,
% axis y line*=left,
% %fill=white,line cap = round, fill opacity=0.8, text opacity = 1, draw opacity = 1
% %extraAxisOptions,
% % unit vector ratio=1 1 1
% ]
% \draw[-{Stealth}, color=black] (axis cs:0.55,0.279) -- (axis cs:0.454,0.326);
% \end{axis}

\end{tikzpicture}%
\vspace{-1mm}
\caption{ML cost functions with respect to the $x$-coordinate of UE position in \roundlabeltxt{NCP} \eqref{eq_1st_llr}, \roundlabeltxt{CP} \eqref{eq_compr_coherent} and \roundlabeltxt{CP} (with \ac{SP}) \eqref{eq_direct_compressed_coherent} at $\SDNRbar = 20$ dB.} 
\label{fig_1d_cost}
%\vspace{-0.1in}
\end{figure}

To illustrate the impact of phase coherence on the behavior of cost functions employed in the proposed estimation algorithms, we report in Fig.~\ref{fig_2d_costs} and Fig.~\ref{fig_1d_cost} examples of both noncoherent and coherent cost functions relative to the UE position. As seen from Fig.~\ref{fig_2d_costs}, the RML-NCP cost function \eqref{eq_1st_llr} \roundlabeltxt{NCP} exhibits a smooth surface whereas the RML cost function  \eqref{eq_compr_coherent} \roundlabeltxt{CP} leads to a spiky profile  with numerous local minima near the true UE position, despite being a relaxation of the JML. Indeed, spikiness results from leveraging \ac{LoS} carrier phase information in \eqref{eq_phirp} in \roundlabeltxt{CP}, which links the UE position $\pp$ in \eqref{eq_taurp} to the \ac{LoS} phase in \eqref{eq_phirp} through the term $2\pi\fc \taurp$; hence, the coherent cost function RML, and \emph{a fortiori} the JML, show wavelength-level fluctuations  (i.e., $\SI{0.086}{\metre}$). 

In contrast, when phase coherence is ignored as in the RML-NCP \eqref{eq_1st_llr}, the \roundlabeltxt{NCP} cost  depends on $\pp$ only through the subcarrier-dependent phase shifts in \eqref{eq_SMC_channel}, resulting in fluctuations on the level of inverse of bandwidth (i.e., $\SI{30}{\metre}$). 
To further highlight the distinctions observed between \roundlabeltxt{NCP} and \roundlabeltxt{CP} processing, we also show the 1D cost functions relative to $x$-coordinate of UE position in Fig.~\ref{fig_1d_cost}, with and without the presence of SP, which reveal the effect of incorporating carrier phase information into position estimation. 
In summary, %the noncoherent formulation is suitable for initial coarse estimation while 
\roundlabeltxt{CP} enables high-accuracy estimates via the use of a cost function highly sensitive to small position perturbations, which strongly motivates the proposed multi-step approach in Sec.~\ref{sec_algorithms}. The drawback is that, in a few cases, the search routine may fail in locating the correct minimum, so leading to error outliers that, in fact, have been removed in the RMSE computations discussed above. 
Nonetheless, in the following, we provide a deeper error analysis by providing the complete \ac{ECDF}.

%%%%%%%%%%%%%%%%%%%%%%%%%%%%%%%%%%%%%%%%%%
% 2D cost function wrt X and Y-coordinates
% \begin{figure}
%         \begin{center}
%         \subfigure[]{
% 			 \label{fig_noncoh_2d}
% 		\includegraphics[width=0.23\textwidth]{Figures/Estimator/cost_noncoherent_20_dB_pos.eps}
% 		}
%         \hspace{-0.3cm}% \hfill 
%         \subfigure[]{
% 			 \label{fig_coh_2d}
% 		\includegraphics[width=0.23\textwidth]{Figures/Estimator/cost_coherent_20_dB_pos.eps}
% 		}
% 		\end{center}
% 		\vspace{-0.3cm}
%         \caption{ML cost functions with respect to the $x$- and $y$-coordinates of UE position in  \subref{fig_noncoh_2d} \roundlabeltxt{NCP} \eqref{eq_1st_llr} and \subref{fig_coh_2d} \roundlabeltxt{CP} \eqref{eq_compr_coherent}  at $\SDNRbar = 20$ dB.} 
%         \label{fig_2d_costs}
%         %\vspace{-0.15in}
% \end{figure}
\begin{figure}
\begin{center}
\subfigure[\roundlabeltxt{NCP}]{
\label{fig_noncoh_2d}
\setlength{\figurewidth}{0.41\linewidth}
\setlength{\figureheight}{0.35\linewidth}   
\def\datapath{Figures/Estimator_figures/cost_noncoherent_20_dB_pos_img/}
\input{Figures/Estimator_figures/cost_noncoherent_20_dB_pos_img/cost_noncoherent_20_dB_pos_img}
}
%\hspace{0.45cm}% 
% \hfill 
\subfigure[\roundlabeltxt{CP}]{
\label{fig_coh_2d}
\setlength{\figurewidth}{0.41\linewidth}
\setlength{\figureheight}{0.35\linewidth}   
\def\datapath{Figures/Estimator_figures/cost_coherent_20_dB_pos_img/}
\input{Figures/Estimator_figures/cost_coherent_20_dB_pos_img/cost_coherent_20_dB_pos_img}
}
\end{center}
\vspace{-4mm}
\caption{ML cost functions with respect to the $x$- and $y$-coordinates of UE position in  \subref{fig_noncoh_2d} \roundlabeltxt{NCP} \eqref{eq_1st_llr} and \subref{fig_coh_2d} \roundlabeltxt{CP} \eqref{eq_compr_coherent}  at $\SDNRbar = 20$ dB. The cutting plane shown in Fig.~\ref{fig_1d_cost} is indicated with a dashed line.} 
\label{fig_2d_costs}
%\vspace{-0.15in}
\end{figure}
%%%%%%%%%%%%%%%%%%%%%%%%%%%%%%%%%%%%%%%%%%

%%%%%%%%%%%%%%%%%%%%%%%%%%%%%%%%%%%%%%%%%%
% 1D cost function wrt X-coordinate
% \begin{figure}
% 	\centering
%     %\vspace{-0.1in}
% 	\includegraphics[width=0.8\linewidth]{Figures/Estimator/1d_costs_all_comparison.eps}
% 	\vspace{-0.1in}
% 	\caption{ML cost functions with respect to the $x$-coordinate of UE position in \roundlabeltxt{NCP} \eqref{eq_1st_llr}, \roundlabeltxt{CP} \eqref{eq_compr_coherent} and \roundlabeltxt{CP} (with \ac{SP}) \eqref{eq_direct_compressed_coherent} at $\SDNRbar = 20$ dB.} 
% 	\label{fig_1d_cost}
% 	%\vspace{-0.1in}
% \end{figure}
%%%%%%%%%%%%%%%%%%%%%%%%%%%%%%%%%%%%%%%%%%

%%%%%%%%%%%%%%%%%%%%%%%%%%%%%%%%%%%%%%%%%%
% CDF analysis
\subsubsection{Empirical Cumulative Distribution Function Analysis} 
\begin{figure}[t]
\begin{center}
%\vspace{-0.22in}

\subfigure[]{\centering
\label{fig_cdfs_pos}
\setlength{\figurewidth}{0.4\textwidth}
\setlength{\figureheight}{0.14\textwidth}   
\def\datapath{Figures/Estimator_figures/cdf_ue_pos_two_SNRs_v2/}
 
\pgfplotsset{every axis/.append style={
  label style={font=\footnotesize},
  legend style={font=\scriptsize},
  tick label style={font=\footnotesize},
%   xticklabel={
%     % test if the x value is below zero
%     \ifdim \tick pt < 0pt
%       % if yes, calculate the absolute value
%       \pgfmathparse{abs(\tick)}%
%       % and print first a minus sign in a zero-width box, followed by the absolute value
%       \llap{$-{}$}\pgfmathprintnumber{\pgfmathresult}
%    \else
%      % if no, print as usual
%       \pgfmathprintnumber{\tick}
%    \fi
% }
}}
 
\definecolor{mycolor1}{rgb}{0.00000,0.44700,0.74100}%
\definecolor{mycolor2}{rgb}{0.85098,0.32549,0.09804}%
% \colorlet{mycolor1}{blue}
% \colorlet{mycolor2}{red}
\colorlet{mycolor1}{IEEEblue}
\colorlet{mycolor2}{IEEEdarkorange}%{IEEEred}

\begin{tikzpicture}[baseline,trim axis left]

\begin{axis}[%
width=\figurewidth,
height=0.879\figureheight,
at={(0\figurewidth,0\figureheight)},
scale only axis,
tick align=inside,		       % [B] ticks oriented inside 
axis line style = thick,	   % [B] thick axis lines 
xmode=log,
xmin=0.0001,
xmax=1,
xminorticks=true,
xlabel={Absolute UE location error in \SI{}{\metre}},
ymin=0,
ymax=1,
ylabel={ECDF},
% axis background/.style={fill=white},
xmajorgrids,
xminorgrids,
ymajorgrids,
grid style=dotted,
legend style={at={(0,1)},anchor=south west,legend columns=2,legend cell align=left, align=left, xshift = -2pt, line width = 0.8pt, draw=white!15!black,/tikz/column sep=0.24cm},
%,/tikz/column width=3cm},
%extraAxisOptions,
% unit vector ratio=1 1 1
line cap = round
]
\addplot [color=mycolor1, mark=o, mark repeat=20, mark phase=2, mark options={solid, mycolor1}, mark size=\marksizeA, dashdotted, line width=\linewidthA,line cap = round]
  table[]{\datapath/cdf_ue_pos_two_SNRs_v2-1.tsv};
\addlegendentry{RML (SDNR = \SI{0}{\dB})}

\addplot [color=mycolor2, dashed, mark=+, mark repeat=20, mark phase=2, mark options={solid, mycolor2}, mark size=\marksizeA, line width=\linewidthA,line cap = round]
  table[]{\datapath/cdf_ue_pos_two_SNRs_v2-2.tsv};
\addlegendentry{JML (SDNR = \SI{0}{\dB})}

\addplot [color=mycolor1, dotted, line width=\linewidthA,line cap = round]
  table[]{\datapath/cdf_ue_pos_two_SNRs_v2-3.tsv};
\addlegendentry{RML (SDNR = \SI{20}{\dB})}

\addplot [color=mycolor2, line width=\linewidthA,line cap = round]
  table[]{\datapath/cdf_ue_pos_two_SNRs_v2-4.tsv};
\addlegendentry{JML (SDNR = \SI{20}{\dB})}

\end{axis}

\end{tikzpicture}%
\vspace{0.5cm}
}\vspace{-2mm}

\subfigure[]{\centering
\label{fig_cdfs_clock}
\setlength{\figurewidth}{0.4\textwidth}
\setlength{\figureheight}{0.14\textwidth}   
\def\datapath{Figures/Estimator_figures/cdf_clock_two_SNRs_v2/}
 
\pgfplotsset{every axis/.append style={
  label style={font=\footnotesize},
  legend style={font=\scriptsize},
  tick label style={font=\footnotesize},
%   xticklabel={
%     % test if the x value is below zero
%     \ifdim \tick pt < 0pt
%       % if yes, calculate the absolute value
%       \pgfmathparse{abs(\tick)}%
%       % and print first a minus sign in a zero-width box, followed by the absolute value
%       \llap{$-{}$}\pgfmathprintnumber{\pgfmathresult}
%    \else
%      % if no, print as usual
%       \pgfmathprintnumber{\tick}
%    \fi
% }
}}
 
\definecolor{mycolor1}{rgb}{0.00000,0.44700,0.74100}%
\definecolor{mycolor2}{rgb}{0.85098,0.32549,0.09804}%
% \colorlet{mycolor1}{blue}
% \colorlet{mycolor2}{red}
\colorlet{mycolor1}{IEEEblue}
\colorlet{mycolor2}{IEEEdarkorange}%{IEEEred}

\begin{tikzpicture}[baseline,trim axis left]%

\begin{axis}[%
width=\figurewidth,
height=0.879\figureheight,
at={(0\figurewidth,0\figureheight)},
scale only axis,
tick align=inside,		       % [B] ticks oriented inside 
axis line style = thick,	   % [B] thick axis lines 
xmode=log,
xmin=0.001,
xmax=100,
xminorticks=true,
xlabel={Absolute Clock Offset error in m},
ymin=0,
ymax=1,
ylabel={ECDF},
% axis background/.style={fill=white},
xmajorgrids,
xminorgrids,
ymajorgrids,
grid style=dotted,
legend style={at={(0,1)},anchor=south west,legend columns=2,legend cell align=left, align=left, draw=white!15!black,line width = 0.8pt, xshift = -2pt, draw=white!15!black,/tikz/column sep=0.24cm},
%extraAxisOptions,
% unit vector ratio=1 1 1
line cap = round
]
\addplot [color=mycolor1, mark=o, mark repeat=20, mark phase=10, mark options={solid, mycolor1}, mark size=\marksizeA, dashdotted, line width=\linewidthA]% [color=blue, dashdotted, line width=1.5pt]
  table[]{\datapath/cdf_clock_two_SNRs_v2-1.tsv};
%\addlegendentry{RML (SDNR = \SI{0}{\dB})}

\addplot [color=mycolor2, mark=+, mark repeat=20, mark phase=10, mark options={solid, mycolor2}, mark size=\marksizeA, dashed, line width=\linewidthA]% [color=red, line width=1.5pt]
  table[]{\datapath/cdf_clock_two_SNRs_v2-2.tsv};
%\addlegendentry{JML (SDNR = \SI{0}{\dB})}

\addplot [color=mycolor1, dotted, line width=\linewidthA]
  table[]{\datapath/cdf_clock_two_SNRs_v2-3.tsv};
%\addlegendentry{RML (SDNR = \SI{20}{\dB})}

\addplot [color=mycolor2, line width=\linewidthA]
  table[]{\datapath/cdf_clock_two_SNRs_v2-4.tsv};
%\addlegendentry{JML (SDNR = \SI{20}{\dB})}

\end{axis}
\end{tikzpicture}%
}\vspace{-2mm}

\subfigure[]{\centering
\label{fig_cdfs_sp}
\setlength{\figurewidth}{0.4\textwidth}
\setlength{\figureheight}{0.14\textwidth}   
\def\datapath{Figures/Estimator_figures/cdf_sp_pos_two_SNRs/}
 
\pgfplotsset{every axis/.append style={
  label style={font=\footnotesize},
  legend style={font=\scriptsize},
  tick label style={font=\footnotesize},
  % xticklabel={
  %   % test if the x value is below zero
  %   \ifdim \tick pt < 0pt
  %     % if yes, calculate the absolute value
  %     \pgfmathparse{abs(\tick)}%
  %     % and print first a minus sign in a zero-width box, followed by the absolute value
  %     \llap{$-{}$}\pgfmathprintnumber{\pgfmathresult}
  %  \else
  %    % if no, print as usual
  %     \pgfmathprintnumber{\tick}
  %  \fi
  %   }
}}
 
\definecolor{mycolor1}{rgb}{0.00000,0.44700,0.74100}%
\definecolor{mycolor2}{rgb}{0.85000,0.32500,0.09800}%
%
% \colorlet{mycolor1}{blue}
% \colorlet{mycolor2}{red}
\colorlet{mycolor1}{IEEEblue}
\colorlet{mycolor2}{IEEEdarkorange}%{IEEEred}

\begin{tikzpicture}[baseline, trim axis left]

\begin{axis}[%
width=\figurewidth,
height=0.879\figureheight,
at={(0\figurewidth,0\figureheight)},
scale only axis,
tick align=inside,		       % [B] ticks oriented inside 
axis line style = thick,	   % [B] thick axis lines 
xmode=log,
xmin=0.01,
xmax=1000,
xminorticks=true,
xlabel={Absolute SP Location error in m},
ymin=0,
ymax=1,
ylabel={ECDF},
% axis background/.style={fill=white},
xmajorgrids,
xminorgrids,
ymajorgrids,
grid style=dotted,
legend style={at={(0,1)},anchor=south west,legend columns=2,legend cell align=left, align=left, draw=white!15!black, line width = 0.8pt, xshift = -2pt, draw=white!15!black,/tikz/column sep=0.24cm},
%extraAxisOptions,
% unit vector ratio=1 1 1
line cap = round
]
\addplot [color=mycolor1, mark=o, mark repeat=30, mark phase=2, mark options={solid, mycolor1}, mark size=\marksizeA, dashdotted, line width=\linewidthA]%[color=blue, dashdotted, line width=1.5pt]
  table[]{\datapath/cdf_sp_pos_two_SNRs-1.tsv};
%\addlegendentry{RML (SDNR = \SI{0}{\dB})}

\addplot [color=mycolor2, mark=+, mark repeat=25, mark phase=5, mark options={solid, mycolor2}, mark size=\marksizeA, dashed, line width=\linewidthA]%[color=red, line width=1.5pt]
  table[]{\datapath/cdf_sp_pos_two_SNRs-2.tsv};
%\addlegendentry{JML (SDNR = \SI{0}{\dB})}

\addplot [color=mycolor1, dotted, line width=\linewidthA]
  table[]{\datapath/cdf_sp_pos_two_SNRs-3.tsv};
%\addlegendentry{RML (SDNR = \SI{20}{\dB})}

\addplot [color=mycolor2, line width=\linewidthA]
  table[]{\datapath/cdf_sp_pos_two_SNRs-4.tsv};
%\addlegendentry{JML (SDNR = \SI{20}{\dB})}

\end{axis}
\end{tikzpicture}%
}
\end{center}
\vspace{-4mm}
\caption{ECDFs of absolute error on estimation of  \subref{fig_cdfs_pos} UE location, \subref{fig_cdfs_clock} clock offset and \subref{fig_cdfs_sp} SP location, achieved by RML and JML for two SDNR levels.}  
\label{fig_cdfs}
\vspace{-4mm}
\end{figure}
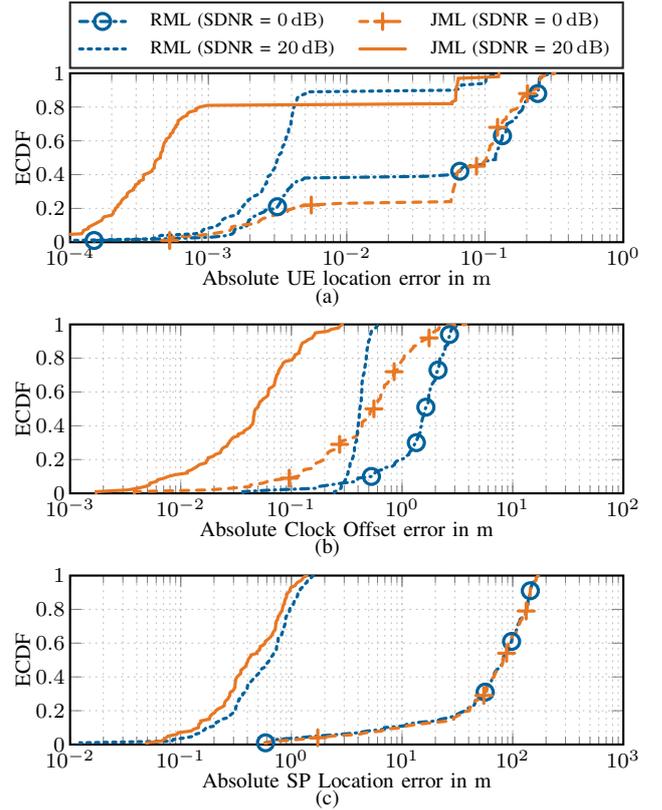

In Fig.~\ref{fig_cdfs_pos} and Fig.~\ref{fig_cdfs_clock}, we present the \ac{ECDF} curves for the estimation errors of the UE location $\pp$ and the clock offset $\deltatau$, achieved by the RML and JML estimators at two different SDNR levels. Such curves  confirm the same findings obtained by analyzing the RMSE, but provide more insights on the error distribution.
We observe in particular that RML and JML provide cm-level and mm-level accuracy most of the times, respectively, but some outliers in estimation of the UE location are also present due to the spiky nature of the cost functions, as discussed above (cf. Fig.~\ref{fig_2d_costs} and Fig.~\ref{fig_1d_cost}).
For the same reason, the JML may not outperform (or even perform slightly worse than) the RML, though this mostly happens at very low SDNR. In contrast, clock offset errors have a smoother ECDF profile, because the dependence of \eqref{eq_compr_coherent} and \eqref{eq_direct_compressed_coherent} on $\deltatau$ is  only through subcarrier-level phase progressions, as apparent from \eqref{eq_SMC_channel} and \eqref{eq_tau_pseudo}.

%%%%%%%%%%%%%%%%%%%%%%%%%%%%%%%%%%%%%%%%%%
% UE location and clock offset error CDF
% \begin{figure}
%         \begin{center}
%         %\vspace{-0.22in}
%         \subfigure[]{
% 			 \label{fig_cdfs_pos}
% 			 \includegraphics[width=0.4\textwidth]{Figures/Estimator/cdf_ue_pos_two_SNRs_v2.eps}  
% 		}
%         \subfigure[]{
% 			 \label{fig_cdfs_clock}
% 			 \includegraphics[width=0.4\textwidth]{Figures/Estimator/cdf_clock_two_SNRs_v2.eps}
% 		}		
% 		\end{center}
% 		\vspace{-0.1in}
%         \caption{ECDFs of absolute error on \subref{fig_cdfs_pos} UE location estimation and \subref{fig_cdfs_clock} clock offset estimation, achieved by RML and JML  at two different SDNR levels.}  
%         \label{fig_cdfs}
%         \vspace{-0.1in}
% \end{figure}

% \tikzset{external/remake next=true}

%%%%%%%%%%%%%%%%%%%%%%%%%%%%%%%%%%%%%%%%%%%% 

To complete the analysis, we also report the ECDF curves for the mapping, i.e., the SP location errors. Fig.~\ref{fig_cdfs_sp} confirms that, also for such variables, the proposed JML and RML provide satisfactory performance, yielding sub-meter accuracy at high SDNRs. Since the SP location estimation does not rely on carrier phase information (due to unknown scattering-induced phase shift as seen from \eqref{eq_phirp}), the JML cost function exhibits a smooth behavior with respect to SP location, resulting in outlier-free ECDF curves. This also enables JML to outperform RML in most cases, similar to the clock offset estimation. Finally, we highlight that the proposed algorithms directly extend to the estimation of multiple SPs, by searching for multiple dominant dips (negative peaks) in the RNST cost function \eqref{eq:L-RNST}. Performance results confirm the same findings as above (figures are omitted due to lack of space).

\section{Conclusion}\label{sec:conclusion}
In this work, we have studied the potential of ELAA systems implemented as distributed RS networks, particularly for joint localization, synchronization, and mapping in sub-6 GHz uplink channels. To correctly capture the complex propagation characteristics of these environments, including specular reflections from walls, scattering from objects, and diffuse multipath, we first presented a detailed generative model able to describe signals and channels over the distributed RS network. We then formulated the joint localization, synchronization and mapping problem, leveraging the near-field effects of distributed RSs in both the presence or absence of phase synchronization between the UE and the RS network. Through a comprehensive Fisher information analysis, we showed that a noncoherently operating, distributed \ac{RS} network mainly relies on \ac{AoA} information in the low and mid-bandwidth regimes, where path overlap with multipath components impairs accuracy.
In the high bandwidth regime, it relies instead on delay information.
On the other hand, a coherently operating, distributed \ac{RS} network operates through carrier-phase based positioning where large performance gains come at the cost of a more involved estimation problem. 
Relaxing the assumption of a phase-coherent infrastructure, we showed that accurate environment information can likewise allow an otherwise noncoherent \ac{RS} network to coherently process the \ac{LoS} and \acp{SMC} per \ac{RS}, which constitutes a possible future research direction. We derived the ML estimator for joint localization, synchronization, and mapping and showed that through a suitable decomposition,  optimal (in ML sense) estimates of the channel amplitudes can be obtained in closed form. This allowed us to significantly reduce the dimension of the estimation problem and end up with a log-likelihood function depending only on the parameters of interest. To overcome the still-too-high dimensionality of the problem, we then devised a three-step algorithm that suitably decouples the estimation of UE position and synchronization parameters from the SP position parameters, providing their initial estimates at a significantly reduced cost. Simulation analyses demonstrated the effectiveness of the proposed methods in capitalizing on near-field effects and a distributed network of phase-coherent RSs to achieve satisfactory localization, synchronization, and mapping performance. 
Specifically, a simple deployment with \num{4} RSs, each equipped with an \num{8}-element ULA, proved sufficient for our algorithms to provide high-quality localization, offering  submeter level accuracy even at low $\SDNRbar$ values and  mm-level accuracy already at mid  $\SDNRbar$ values.

%%%%%%%%%%%%%%%%%%%%%%%%%%%%%%%%%%%%%%%%%%%%%%%%%%%%%%%%
%%%%%%%%%%%%%%%%%%%%%%%%%%%%%%%%%%%%%%%%%%%%%%%%%%%%%%%%
\begin{appendices}

\ifthenelse{\equal{\useExternalSupplementary}{true}} % ---------- BEGIN OF EXTERNALIZED SUPPLEMENTARY MATERIAL ----------
{

\section{Rotation Matrix}\label{app_rot_mat}
The rotation matrix representing a counterclockwise azimuth rotation around the $z$-axis is given by
\begin{align}
    \mmb(\beta) = \begin{bmatrix} \cos \beta & -\sin \beta & 0 \\
    \sin \beta & \cos \beta & 0 \\
    0 & 0 & 1
    \end{bmatrix} ~.
\end{align}

%%%%%%%%%%%%%%%%%%%%%%%%%%%%%%%%%%%%%%%%
\section{Geometry-Based Channel Model}\label{app_channel_model}

\subsection{Positions of Reflection Points}\label{app_reflection_point_pos}
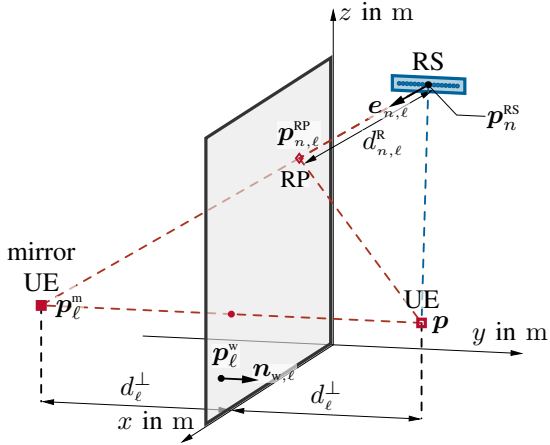
\begin{figure}[htb]
    \centering
    \vskip 0pt
    \setlength{\plotWidth}{0.9\linewidth}
\definecolor{mycolor1}{rgb}{0.72941,0.04706,0.18431}%
\definecolor{mycolor2}{rgb}{0.00000,0.38431,0.60784}%
\definecolor{mycolor3}{rgb}{0.66270,0.20390,0.12160}%

\pgfplotsset{every axis/.append style={
  label style={font=\footnotesize},
  legend style={font=\footnotesize},
  tick label style={font=\footnotesize},
  xticklabel={
    % test if the x value is below zero
    \ifdim \tick pt < 0pt
      % if yes, calculate the absolute value
      \pgfmathparse{abs(\tick)}%
      % and print first a minus sign in a zero-width box, followed by the absolute value
      \llap{$-{}$}\pgfmathprintnumber{\pgfmathresult}
   \else
     % if no, print as usual
      \pgfmathprintnumber{\tick}
   \fi
}}}

\begin{tikzpicture}

\begin{axis}[%
width=0.877\plotWidth,
height=0.75\plotWidth,
at={(0\plotWidth,0\plotWidth)},
scale only axis,
plot box ratio=1.714 1.143 1,
xmin=0,
xmax=6,
tick align=outside,
ymin=-2,
ymax=2,
zmin=0,
zmax=3.5,
line cap = round,               % [B] round line caps
view={105}{10},
%axis background/.style={fill=white},
%axis x line*=bottom,
%axis y line*=left,
%axis z line*=left,
% xmajorgrids,
% ymajorgrids,
% zmajorgrids
grid = none,        % [B]
ticks=none,         % [B]
%axis lines=center
%axis line style={draw=none} % [B]
axis lines=center,              % [B]
xlabel style={xshift = -13pt, yshift = -10pt},
xlabel={$x$ in \SI{}{\metre}},
ylabel style={xshift = 7pt},
ylabel={$y$ in \SI{}{\metre}},
zlabel={$z$ in \SI{}{\metre}},
view={105}{10}, 
%rotate around x=1.7,            % [B] ROLL AROUND x-AXIS: compensate for y-axis being not horizontal with the 2-parametric view{}{}-setting
axis line style={->, arrows = {-Stealth[inset=0pt, scale=1.05, angle'=25]}, line width = 0.35pt},  % Add arrows at the ends
tick align=inside,            % [B]
xticklabel style={xshift=-0.2cm,yshift=0.5cm},      % [B] plot above
yticklabel style={xshift=0.1cm,yshift=0.4cm},      % [B] plot above
zticklabel style={yshift=0.1cm}     % [B] plot above
%grid style=dotted,                       % [B]
]
\addplot3 [color=mycolor1, line width=1.2pt, only marks, mark size=1.4pt, mark=square, mark options={solid, mycolor1}]
 table[row sep=crcr] {%
4	2	1\\
};
 \addplot3 [color=black, dashed, line width=0.7pt]
 table[row sep=crcr] {%
4	2	1\\
4	2	0\\
};

\addplot3 [stealth-stealth,color=black, line width=0.35pt,
{Stealth[inset=0pt, scale=1.05, angle'=25]-Stealth[inset=0pt, scale=1.05, angle'=25]}
]
 table[row sep=crcr] {
4	0	0\\
4	-2	0\\
};

\addplot3 [color=mycolor3, dashed, line width=0.7pt]
 table[row sep=crcr] {%
4	0	1\\
4	-2	1\\
};

\addplot3 [color=mycolor3, dashed, line width=0.7pt]
 table[row sep=crcr] {%
1.33333333333333	-0.001	2.16666666666667\\
4	-2	1\\
};

\addplot3 [color=mycolor2, line width=1.2pt, only marks, mark size=0.3pt, mark=*, mark options={solid, mycolor2}]
 table[row sep=crcr] {%
0.001	1.30612244897959	2.75\\
0.001	1.26530612244898	2.75\\
0.001	1.22448979591837	2.75\\
0.001	1.18367346938776	2.75\\
0.001	1.14285714285714	2.75\\
0.001	1.10204081632653	2.75\\
0.001	1.06122448979592	2.75\\
0.001	1.02040816326531	2.75\\
0.001	0.979591836734694	2.75\\
0.001	0.938775510204082	2.75\\
0.001	0.897959183673469	2.75\\
0.001	0.857142857142857	2.75\\
0.001	0.816326530612245	2.75\\
0.001	0.775510204081633	2.75\\
0.001	0.73469387755102	2.75\\
0.001	0.693877551020408	2.75\\
};
 
\addplot3[area legend, line width=1.2pt, draw=white!20!black, fill=white!90!black, fill opacity=0.6, forget plot]
table[row sep=crcr] {%
x	y	z\\
0.1	0	3\\
5	0	3\\
5	0	0.03\\
0.1	0	0.03\\
}--cycle;
\addplot3 [color=mycolor1, line width=1.2pt, only marks, mark size=1.4pt, mark=square*, mark options={solid, mycolor1}]
 table[row sep=crcr] {%
4	-2	1\\
};
 \addplot3 [color=black, dashed, line width=0.7pt]
 table[row sep=crcr] {%
4	-2	1\\
4	-2	0\\
};

 \node[align=center,fill=white, opacity=0.6, text opacity = 1]
at (axis cs:4,-2,1.4) {mirror\\UE};
 \node[align=center,fill=white, opacity=0.6, text opacity = 1]
at (axis cs:4,-1.7,0.99) {$\ppm$};

\addplot3 [color=mycolor1, line width=1.2pt, only marks, mark size=0.7pt, mark=*, mark options={solid, mycolor1}]
 table[row sep=crcr] {%
4	0	1\\
};
 \addplot3 [color=black, line width=1.2pt, only marks, mark size=0.7pt, mark=*, mark options={solid, black}]
 table[row sep=crcr] {%
4.4	0	0.4\\
};
 \node[align=center,fill=white, opacity=0.6, text opacity = 1]
at (axis cs:4.4,0.05,0.64) {$\ppwl$};

 \node[align=center,fill=white, opacity=0.6, text opacity = 1]
at (axis cs:4.4,0.56,0.48) {$\nw$};
\addplot3 [color=mycolor1, line width=0.7pt, only marks, mark size=1.7pt, mark=diamond, mark options={solid, mycolor1}]
 table[row sep=crcr] {%
1.33333333333333	0	2.16666666666667\\
};
 \node[align=center,fill=white, opacity=0.6, text opacity = 1]
at (axis cs:1.333,0,2.407) {$\ppRP$};
\node[align=center,fill=white, opacity=0.6, text opacity = 1]
at (axis cs:1.5,0,1.975) {\small RP};
\addplot3 [color=mycolor1, line width=1.2pt, only marks, mark size=1.4pt, mark=square*, mark options={solid, mycolor1}]
 table[row sep=crcr] {%
4	-2	1\\
};

\addplot3[area legend, line width=1.2pt, draw=mycolor2, fill=mycolor2, fill opacity=0.3, forget plot]
table[row sep=crcr] {%
x	y	z\\
-0.00999999999999998	1.38112244897959	2.675\\
-0.01	0.618877551020408	2.675\\
-0.01	0.618877551020408	2.825\\
-0.00999999999999998	1.38112244897959	2.825\\
}--cycle;

 \node[align=center,fill=white, opacity=0.6, text opacity = 1]
at (axis cs:0,1,2.99) {RS};

\addplot3 [stealth-stealth,color=black, line width=0.35pt,
{Stealth[inset=0pt, scale=1.05, angle'=25]-Stealth[inset=0pt, scale=1.05, angle'=25]}
]
 table[row sep=crcr] {
0	1.05	2.7\\
1.33333333333333	0.05	2.11111111\\
};

\node[align=center,fill=white, opacity=0.6, text opacity = 1]
at (axis cs:0.66,0.7,2.23) {$\distR$};

\addplot3 [color=mycolor3, dashed, line width=0.7pt]
 table[row sep=crcr] {%
0	1	2.75\\
1.33333333333333	0	2.16666666666667\\
};
\addplot3 [color=mycolor2, dashed, line width=0.7pt]
 table[row sep=crcr] {%
0	1	2.75\\
4	2	1\\
};

 \addplot3 [color=mycolor3, dashed, line width=0.7pt]
 table[row sep=crcr] {%
0	1	2.75\\
1.33333333333333	0	2.16666666666667\\
};

 \addplot3 [color=mycolor3, dashed, line width=0.7pt]
 table[row sep=crcr] {%
1.33333333333333	0	2.16666666666667\\
4	2	1\\
};

 \node[align=center,fill=white, opacity=0.6, text opacity = 1]
at (axis cs:4,2,1.24) {UE};

 \node[align=center,fill=white, opacity=0.6, text opacity = 1]
at (axis cs:4,2.2,0.99) {$\pp$};

\addplot3 [stealth-stealth,color=black, line width=0.35pt,
{Stealth[inset=0pt, scale=1.05, angle'=25]-Stealth[inset=0pt, scale=1.05, angle'=25]}
]
 table[row sep=crcr] {
4	0	0\\
4	2	0\\
};

\node[align=center,fill=white, opacity=0.6, text opacity = 1]
at (axis cs:4,-1,0.2) {$d_{\scriptscriptstyle \nRP}^{\perp}$};

\node[align=center,fill=white, opacity=0.6, text opacity = 1]
at (axis cs:4,1,0.2) {$d_{\scriptscriptstyle \nRP}^{\perp}$};

\addplot3 [-stealth,color=black, line width=0.7pt,%
{-Stealth[inset=0pt, scale=1.05, angle'=25]}
%{->, arrows = {-Stealth[inset=0pt, scale=1.05, angle'=25]}
]
 table[row sep=crcr] {%
4.4	0	0.4\\
4.4	0.4	0.4\\
};

\node[align=center,fill=white, opacity=0.6, text opacity = 1]
at (axis cs:0.3853,0.6885,2.5383) {$\er$};
\addplot3 [-stealth,color=black, line width=0.7pt,%
{-Stealth[inset=0pt, scale=1.05, angle'=25]}
]
 table[row sep=crcr] {%
0.1	1	2.75\\
0.4153	0.6885	2.5683\\
};

\addplot3 [color=mycolor1, line width=1.2pt, only marks, mark size=0.7pt, mark=*, mark options={solid, black}]
 table[row sep=crcr] {%
0	1	2.75\\
};

\addplot3 [color=black, line width=0.3pt]
 table[row sep=crcr] {%
0.01	1	2.75\\
0.01	1.3	2.45\\
0.01	1.6	2.45\\
};

\node[align=center,fill=white, opacity=0.6, text opacity = 1]
at (axis cs:0.01,1.8,2.45) {$\pprs_n$};

\addplot3 [color=mycolor3, dashed, line width=0.7pt]
 table[row sep=crcr] {%
4	2	1\\
4	0	1\\
};

\end{axis}
\end{tikzpicture}%
    \vspace{-0.2cm}\caption{Geometric relations between \ac{RS} position $\pprsx{n}$, \ac{UE} position $\pp$, mirror \ac{UE} position $\ppm$, and \ac{RP} position $\ppRP$.}
     \label{fig:geometrySMC}
\end{figure}
Fig.\,\ref{fig:geometrySMC} shows the geometric relations for computing \ac{RP} positions.
Geometrically, infinite planar specular surfaces can be described through an arbitrary point $\ppwl \in \realsetone{3}$ on the surface, as well as a surface normal vector $\nw \in \realsetone{3}$. 
For a given \ac{UE} position $\pp$ and \ac{RS} position $\pprsx{n}$, the position of \ac{RP} $\nRP > 0$ computes to\footnote{Contrary to \cite{Ge20slam5G,Wymeersch18downlinkPos}, our reference position is the \ac{RS} position $\pprsx{n}$ rather than the mirror \ac{UE} position $\ppm$ resulting in different unit vectors $\er$.}\,\cite[eq.\,(31)]{Ge20slam5G},\,\cite[eq.\,(4)]{Wymeersch18downlinkPos}
\begin{align}\label{eq:RP-explicit}
    \ppRP = \pprsx{n} + 
    %\underbrace{
    \frac{
    \left( 
        \ppwl - \pprsx{n}
    \right)^\trp \nw}%
    {\left(\ppm - \pprsx{n}\right)^\trp \nw}
    %}_{= l_{n,\ell}}
    \ \left(\ppm - \pprsx{n}\right) 
\end{align}
where $\ppm$ is computed by mirroring the physical \ac{UE} across walls $\nRP>0$ resulting in mirror \ac{UE} positions
\begin{align}\label{eq:ppm}
\ppm &= \pp - 2 
    \nw 
    \underbrace{
        \left( \left(\pp - \ppwl\right)^\trp \nw \right) 
    }_{
        := d_{\scriptscriptstyle \nRP}^{\perp}
    }
\end{align}
from which the reflected signals virtually impinge at the $n$\textsuperscript{th} \ac{RS}.
The mirror source position is computed by moving from the \ac{UE} position $\pp$ two times the projected distance to the wall $d_{\scriptscriptstyle \nRP}^{\perp}$ in the negative direction of the wall normal $\nw$.
Defining a a unit vector $\er= (\ppm - \pprsx{n})/\lVert \ppm - \pprsx{n} \rVert$ that points from the $n$\textsuperscript{th} \ac{RS} at $\pprsx{n}$ to the $\nRP$\textsuperscript{th} mirror \ac{UE} at $\ppm$,~\eqref{eq:RP-explicit} can be rewritten to
\begin{align}\label{eq:RP-explicit2}
    \ppRP = \pprsx{n} + 
    \underbrace{
    \frac{
    \left( 
        \ppwl - \pprsx{n}
    \right)^\trp \nw}%
    {\er^\trp \nw}
    }_{:= \distR}
    \ \er 
\end{align}
such that its geometric interpretation is that we are moving from the \ac{RS} position $\pprsx{n}$ 
into the direction of the mirror source direction $\er$ by the length $\distR$.

\subsection{The Reflection Point Channel}\label{app_RP_channel}
The amplitudes $\alpharp$ of the \ac{LoS} $\nRP=0$ and \ac{RP} $\nRP>0$ paths can be modeled\footnote{Despite the position dependence of $\alphacn$ in our generative model, we treat them as position-independent nuisance parameters in our inference model.} by reformulating the Friis transmission equation for power-wave amplitudes %~\cite{Deutschmann22InitialAccess} 
through~\cite[eq.\,(2-119)]{balanis2005}
\begin{align}\label{eq:alpharp}
    \alpharp  = 
    \begin{cases}
    \frac{\sqrt{\Pt}\lambda}{4\pi \, \distU} ~ |{\polRS}^\trp \polUE |  & \nRP = 0 \\
    \frac{\sqrt{\Pt}\lambda}{4\pi \, (\distUR + \distR)} \, | \underbrace{(\polp \Rp + \pols \Rs)^\trp \polUE }_{:=\refcoeff} | & \nRP > 0
    \end{cases}
    \, ,
\end{align}
where $\distU = \|\pp - \pprsx{n}\|$ denotes the scalar \ac{LoS} distance between the \ac{UE} and the $n$\textsuperscript{th} \ac{RS}, 
and where $\distUR = \|\pp - \ppRP  \|$ denotes the scalar distance from the \ac{UE} position to the $\nRP$\textsuperscript{th} \ac{RP} of \ac{RS} $n$, and $\distR = \|\ppRP - \pprsx{n}\|$ denotes the scalar distance from the $\nRP$\textsuperscript{th} \ac{RP} of \ac{RS} $n$ to the phase center position of \ac{RS} $n$.
Note that we assume isotropic transmit and receive antennas, i.e., any antenna gains $G_\mathrm{t}\!=\!G_\mathrm{r}\!\triangleq\!1$. 
We further assume linearly polarized antennas with polarization unit-vectors $\polRS \in \realset{3}{1}$ at \ac{RS} $n$, and $\polUE\in \realset{3}{1}$ at the \ac{UE}.
%, and $\polRP$ the polarization at \ac{RP} $\nRP>0$ according to~\eqref{eq:polr}. 
%parallel to the $z$-axis and hence no polarization losses.
We use $\refcoeff$ to denote the reflection coefficient at a specular surface (e.g., a wall $\nRP >0$) including polarization losses, which we model as follows:

We compute the reflection coefficients for a downlink perspective which likewise holds for the uplink due to channel reciprocity.
At the distance $\rangeR = \ppRP - \pprsx{n}$ from \ac{RS} $n$, the electric field incident at the surface $\nRP$ can be described as
\begin{align}
    \Ei = \Eo \poli e^{-j \wavenumber \er^\trp \rangeR} \quad \in \complexset{3}{1}
\end{align}
with field strength $\Eo$ %, polarization unit vector $ \poli$, 
and wave vector number $\wavenumber = \frac{2\pi}{\lambda}$.
We decompose the reflected electric field%\footnote{In the following equations, we omit the parametric dependencies on $n$ and $\nRP$ for notational brevity.}
\begin{align}
    \Er = \Eo %(\underbrace{\polp \Rp + \pols \Rs}_{:=\polr \refcoeff}) e^{-j \wavenumber \er^\trp \rangeR}
    (\polp \Rp + \pols \Rs) e^{-j \wavenumber \er^\trp \rangeR}
\end{align}
into a component with polarization $\polp = ({\poli}^\trp \nw)\nw$ parallel to the surface normal $\nw$, and an orthogonal complement $\pols = \bm{\Pi}_{\nRP}^{\scriptscriptstyle\perp} \poli$ with $\bm{\Pi}_{\nRP}^{\scriptscriptstyle\perp} := \Imatrix - \nw \nw^\trp$.
Through the Fresnel equations, the parallel component experiences a reflection coefficient~\cite[(4-125)]{balanis2005},\,\cite[(1.137a)]{Pozar2012}
\begin{align}\label{eq:refcoeff-parallel}
    \Rp = \frac{Z_1 \cos \anglet - Z_0 \cos \anglei}{Z_1 \cos \anglet + Z_0 \cos \anglei} \qquad \in \complexsett ,
\end{align}
while the orthogonal one experiences a reflection coefficient~\cite[(4-128a)]{balanis2005},\,\cite[(1.143a)]{Pozar2012}
\begin{align}\label{eq:refcoeff-orthogonal}
    \Rs = \frac{Z_1 \cos \anglei - Z_0 \cos \anglet}{Z_1 \cos \anglei + Z_0 \cos \anglet} \qquad \in \complexsett ,
\end{align}
both of which depend on the angle of \textit{incidence} $\anglei$ (which equals the angle of \textit{reflection}), and the angle of \textit{refraction} $\anglet = \arcsin\left( (\epsilon_\text{r} \mu_\text{r})^{-\frac{1}{2}} \sin \anglei \right)$ given by Snell's law~\cite[(7.36)]{Jackson1999}.
The intrinsic impedance of free-space is $Z_0=\sqrt{\frac{\mu_0}{\epsilon_0}}\approx 377 \SI{}{\ohm} \in \mathbb{R}$, $\epsilon_0$ and $\mu_0$ denoting the vacuum permittivity and permeability, respectively.
The intrinsic impedance of a general material (e.g., the surface) is
\begin{align}\label{eq:impedance}
    Z_1=\sqrt{\frac{j \omega \mu_1}{\mathsf{\sigma}_1 + j \omega \epsilon_1}} \quad \in \mathbb{C} \, ,
\end{align}
given the material permeability $\mu_1 = \mu_0 \, \mu_\text{r}$, permittivity $\epsilon_1 = \epsilon_0 \, \epsilon_\text{r}$, and the conductivity $\mathsf{\sigma}_1$.
%with the permeability $\mu_1=\mu_0 \, \mu_r$ (with $\mu_r$ denoting the relative permeability), the permittivity $\epsilon_1 = \epsilon_0 \, \epsilon_r$ (with $\epsilon_r$ denoting the relative permittivity), and the conductivity $\mathsf{\sigma}_1$.
%We compute the coefficient $\refcoeff$ and the polarization unit-vector $\polr$ of the reflected wave as
%\begin{align}
%    \refcoeff &:= \lVert \polp \Rp + \pols \Rs \rVert \, e^{j\angle\left( \polp \Rp + \pols \Rs \right)}\\%\exp\left(j\angle \polp \Rp + \pols \Rs \right)\\
%    \polr &:= \left( \polp \Rp + \pols \Rs \right)/\refcoeff    \label{eq:polr}
%\end{align}

While the magnitude of the reflection coefficient $\left|\refcoeff\right|$ affects the amplitude $\alpharp$, we assume that the phase $\angle\refcoeff$ is absorbed into the unknown phase offset $\varphi_{\scriptscriptstyle n,\nRP}$ in \eqref{eq_phirp}. 
% Note that in the case of \textit{lossless} materials, i.e., $\sigma_1 = 0$, the intrinsic impedance in~\eqref{eq:impedance} becomes \textit{real-valued} and consequently the reflection coefficient in~\eqref{eq:reflection-coeff} becomes \textit{real-valued} too.
Although this phase takes very small values for low-loss materials (i.e., $\mathsf{\sigma}_1\ll1$), we still assume that (e.g. geometrical) imperfections of the specular surfaces may cause unknown phase shifts that need to be estimated as nuisance parameters.

However, suppose the wall geometry is very well known and the walls are lossless (or of known material). 
In that case, the phase shifts $\angle\refcoeff$ can be computed and turn \roundlabeltxt{NCP} into carrier-phase-based positioning by coherently processing the \ac{LoS} and \acp{SMC} per \ac{RS} and thereby achieving large performance gains as is depicted in Fig.~\ref{fig:bounds-B-sweep} in the main paper.

\subsection{The Scatter Point Channel}\label{app_SP_channel}
The amplitudes $\alphasp$ of \ac{SP} paths can be modeled by reformulating the bistatic radar range equation for power-wave amplitudes through~\cite{Deutschmann22InitialAccess}
\begin{align}\label{eq:alphasp}
    \alphasp  = 
     \frac{\sqrt{\Pt}\lambda}{(4\pi)^{ \frac{3}{2}} \, \distUS\,\distS } \, | {\polRS}^\trp \polUE | \sqrt{\rcs}
    \, ,
\end{align}
where $\distUS = \|\pp - \ppSP\|$ denotes the scalar distance from the \ac{UE} position to the $\nSP$\textsuperscript{th} \ac{SP}, and $\distS = \|\ppSP - \pprsx{n}\|$ denotes the scalar distance from the $\nSP$\textsuperscript{th} \ac{SP} to the phase center position of \ac{RS} $n$.
%% directional / 
While the unidirectional rescattering of \acp{RP} leads to an addition of distances in the denominator of~\eqref{eq:alpharp}, the omnidirectional rescattering of \acp{SP} leads to a multiplication of distances in the denominator of~\eqref{eq:alphasp} which often results in low amplitudes $\alphasp$.
%% no polarization losses
For simplicity, we assume no polarization rotation for reflections at the \acp{SP}. 
%% isotropic rescattering / rcs model
We consider perfect electrically conducting spheres with radius with $\radiusSPx{\nSP}$ as \acp{SP} %~(\,\ref{pgf:SP}\,) 
in the scenario of Fig.\,\ref{fig:scenario} of the main paper.
For such an \ac{SP}, Mie scattering describes the \ac{RCS}.
The \textit{monostatic} \ac{RCS} is $\rcs \approx {\radiusSPx{\nSP}}^2\,\pi$ given that we are operating in the \textit{optical region} characterized by $\frac{2\pi\radiusSPx{\nSP}}{\lambda}>10$~ as described in~\cite[p.\,11.5]{skolnik90RadarHandbook}. 
The \textit{bistatic} \ac{RCS} is generally dependent on incidence and reflected angles.
Although a receiving \acp{RS} could experience a much larger \textit{forward-scattering} \ac{RCS} when the \ac{SP} is located on a straight line between the \ac{UE} and the \ac{RS}, the \ac{RCS} does not vary significantly in all other directions~\cite{Strifors04RCS,Strifors98RCS}.
Hence we consider isotropic rescattering, i.e., we approximate that the \ac{RCS} $\rcs$ stays constant for all incident and reflected angles. %, which is only an approximation in our bistatic (i.e., spatially separated \ac{UE} and \acp{RS}) setup.

%%%%%%%%%%%%%%%%%%%%%%%%%%%%%%%%%%%%%%%%
\section{Covariance Matrix of the Disturbance Component in \eqref{eq_wwb}}\label{app_cov_dist}
Using the statistical characterization of the \ac{DMC} in \eqref{eq_dmc_stat} and \eqref{eq_dmc_cov}, and relying on the properties of the Kronecker and Hadamard products, the covariance matrix of $\wwb_n$ in \eqref{eq_wwb} can be computed as
\begin{align} \nonumber
    & \E \left\{ \vecc{ \wwb_n } \vecc{ \wwb_n }^\hermit  \right\} 
    \\ \nonumber
    &= \E \Big\{ \big[ \wwdmc_n \odot ( \sss \otimes \boldone_M ) + \vecc{\zzb_n} \big]
    \\ \nonumber &~~~~\times\big[ \wwdmc_n \odot ( \sss \otimes \boldone_M ) + \vecc{\zzb_n} \big]^\hermit \Big\}
    \\ \nonumber
    &=\E \Big\{ \big[ \wwdmc_n \odot ( \sss \otimes \boldone_M ) \big] \big[ \wwdmc_n \odot ( \sss \otimes \boldone_M ) \big]^\hermit \Big\}
    \\ \nonumber
    &~~+ \E \Big\{ \vecc{\zzb_n} \vecc{\zzb_n}^\hermit  \Big\}
    \\ \nonumber
    &=\E \Big\{ \wwdmc_n (\wwdmc_n)^\hermit \odot ( \sss \otimes \boldone_M )( \sss \otimes \boldone_M )^\hermit \Big\} + \sigma^2 \Imatrix_{MK}
    \\ \nonumber
    &= \left( \rrb_f(\etadmc) \otimes \Imatrix_M \right) \odot \left(  \sss \sss^\hermit \otimes \boldone_M \boldone_M^\trp \right)  + \sigma^2 \Imatrix_{MK}
    \\
    &=  \left( \rrb_f(\etadmc) \odot \sss \sss^\hermit \right) \otimes \Imatrix_M + \sigma^2 \Imatrix_{MK} ~.
\end{align}

\section{Cram\'er-Rao Lower Bound}\label{app_crlb} 

\begin{table}[tb]
			\centering
			\caption
			{Individual local channel {FIM} terms.%
                }%
\label{tab:fim-terms}
                \setlength{\tabcolsep}{4pt} % Default is 6pt
			\begin{tabularx}{1.0\columnwidth}{@{}c|c|cc}%|ccc}
    			\toprule
    			\textbf{FIM entry} & \textbf{Term} & \hspace{-1mm}\textbf{Row} $i$ & \hspace{-1mm}\textbf{Column} $j$     % "hack" :')
                \\
   			\midrule 
                %%%%%%%%%% theta %%%%%%%%%%
                %% theta - theta:
    			$[\fim_{\theta,\theta}]_{\scriptscriptstyle\ncomponent,\ncomponent'}^{\scriptscriptstyle(n)}$ & 
    			    $2\realp{
                            \gammanc^{\ast} \gammancp 
                            \dot{\cc}_{\scriptscriptstyle\theta,\ncomponent}^{\scriptscriptstyle(n)\hermit} 
                            \rrbn^{\scriptscriptstyle-1}  
                            \dot{\cc}_{\scriptscriptstyle\theta,\ncomponent'}^{\scriptscriptstyle(n)}
                        }$  
                        & ${\scriptstyle\ncomponent + 1}$
                        & ${\scriptstyle\ncomponent + 1}$\\
                %% theta - tau:
    			$[\fim_{\theta,\tauu}]_{\scriptscriptstyle\ncomponent,\ncomponent'}^{\scriptscriptstyle(n)}$ & 
    			    $2\realp{
                            \gammanc^{\ast} \gammancp 
                            \dot{\cc}_{\scriptscriptstyle\theta,\ncomponent}^{\scriptscriptstyle(n)\hermit} 
                            \rrbn^{\scriptscriptstyle-1}  
                            \dot{\cc}_{\scriptscriptstyle\tauu,\ncomponent'}^{\scriptscriptstyle(n)}
                        }$  
                        & ${\scriptstyle\ncomponent + 1}$
                        & ${\scriptstyle \Nc + \ncomponent + 1}$\\
                %% theta - phi:
    			$[\fim_{\theta,\phi}]_{\scriptscriptstyle\ncomponent,\ncomponent'}^{\scriptscriptstyle(n)}$ & 
    			    $2\realp{
                            j \gammanc^{\ast} \gammancp 
                            \dot{\cc}_{\scriptscriptstyle\theta,\ncomponent}^{\scriptscriptstyle(n)\hermit} 
                            \rrbn^{\scriptscriptstyle-1}  
                            {\cc}_{\scriptscriptstyle\ncomponent'}^{\scriptscriptstyle(n)}
                        }$  
                        & ${\scriptstyle\ncomponent + 1}$
                        & ${\scriptstyle 2\Nc + \ncomponent + 1}$\\
                %% theta - alpha:
    			$[\fim_{\theta,\alpha}]_{\scriptscriptstyle\ncomponent,\ncomponent'}^{\scriptscriptstyle(n)}$ & 
    			    $2\realp{
                            \gammanc^{\ast} e^{j\phi_{\scriptscriptstyle n,\ncomponent'}}
                            \dot{\cc}_{\scriptscriptstyle\theta,\ncomponent}^{\scriptscriptstyle(n)\hermit} 
                            \rrbn^{\scriptscriptstyle-1}  
                            {\cc}_{\scriptscriptstyle\ncomponent'}^{\scriptscriptstyle(n)}
                        }$  
                        & ${\scriptstyle\ncomponent + 1}$
                        & ${\scriptstyle 3\Nc + \ncomponent + 1}$\\
                \midrule
                %%%%%%%%%% tau %%%%%%%%%%
                %% tau - tau:
                $[\fim_{\tauu,\tauu}]_{\scriptscriptstyle\ncomponent,\ncomponent'}^{\scriptscriptstyle(n)}$ & 
    			    $2\realp{
                            \gammanc^{\ast} \gammancp 
                            \dot{\cc}_{\scriptscriptstyle\tauu,\ncomponent}^{\scriptscriptstyle(n)\hermit} 
                            \rrbn^{\scriptscriptstyle-1}  
                            \dot{\cc}_{\scriptscriptstyle\tauu,\ncomponent'}^{\scriptscriptstyle(n)}
                        }$  
                        & ${\scriptstyle \Nc+\ncomponent + 1}$
                        & ${\scriptstyle \Nc+\ncomponent + 1}$\\
                %% tau - phi:
                $[\fim_{\tauu,\phi}]_{\scriptscriptstyle\ncomponent,\ncomponent'}^{\scriptscriptstyle(n)}$ & 
    			    $2\realp{
                            j \gammanc^{\ast} \gammancp 
                            \dot{\cc}_{\scriptscriptstyle\tauu,\ncomponent}^{\scriptscriptstyle(n)\hermit} 
                            \rrbn^{\scriptscriptstyle-1}  
                            {\cc}_{\scriptscriptstyle\ncomponent'}^{\scriptscriptstyle(n)}
                        }$   
                        & ${\scriptstyle \Nc + \ncomponent + 1}$
                        & ${\scriptstyle 2\Nc + \ncomponent + 1}$\\
                %% tau - alpha:
                $[\fim_{\tauu,\alpha}]_{\scriptscriptstyle\ncomponent,\ncomponent'}^{\scriptscriptstyle(n)}$ & 
    			    $2\realp{
                            \gammanc^{\ast} e^{j\phi_{\scriptscriptstyle n,\ncomponent'}}
                            \dot{\cc}_{\scriptscriptstyle\tauu,\ncomponent}^{\scriptscriptstyle(n)\hermit} 
                            \rrbn^{\scriptscriptstyle-1}  
                            {\cc}_{\scriptscriptstyle\ncomponent'}^{\scriptscriptstyle(n)}
                        }$   
                        & ${\scriptstyle \Nc + \ncomponent + 1}$
                        & ${\scriptstyle 3\Nc + \ncomponent + 1}$\\
                \midrule
                %%%%%%%%%% phi %%%%%%%%%%
                %% phi - phi:
                $[\fim_{\phi,\phi}]_{\scriptscriptstyle\ncomponent,\ncomponent'}^{\scriptscriptstyle(n)}$ & 
    			    $2\realp{
                            \gammanc^{\ast} \gammancp 
                            {\cc}_{\scriptscriptstyle\ncomponent}^{\scriptscriptstyle(n)\hermit} 
                            \rrbn^{\scriptscriptstyle-1}  
                            {\cc}_{\scriptscriptstyle\ncomponent'}^{\scriptscriptstyle(n)}
                        }$   
                        & ${\scriptstyle 2\Nc + \ncomponent + 1}$
                        & ${\scriptstyle 2\Nc + \ncomponent + 1}$\\
                %% phi - alpha:
                $[\fim_{\phi,\alpha}]_{\scriptscriptstyle\ncomponent,\ncomponent'}^{\scriptscriptstyle(n)}$ & 
    			    $2\realp{-j
                            \gammanc^{\ast} e^{j\phi_{\scriptscriptstyle n,\ncomponent'}}
                            {\cc}_{\scriptscriptstyle\ncomponent}^{\scriptscriptstyle(n)\hermit} 
                            \rrbn^{\scriptscriptstyle-1}  
                            {\cc}_{\scriptscriptstyle\ncomponent'}^{\scriptscriptstyle(n)}
                        }$    
                        & ${\scriptstyle 2\Nc + \ncomponent + 1}$
                        & ${\scriptstyle 3\Nc + \ncomponent + 1}$\\
                \midrule
                %%%%%%%%%% alpha %%%%%%%%%%
                %% alpha - alpha:
                $[\fim_{\alpha,\alpha}]_{\scriptscriptstyle\ncomponent,\ncomponent'}^{\scriptscriptstyle(n)}$ & 
    			    $2\realp{
                            e^{-j{\phi_{\scriptscriptstyle n,\ncomponent}}} e^{j{\phi_{\scriptscriptstyle n,\ncomponent'}}}
                            {\cc}_{\scriptscriptstyle\ncomponent}^{\scriptscriptstyle(n)\hermit} 
                            \rrbn^{\scriptscriptstyle-1}  
                            {\cc}_{\scriptscriptstyle\ncomponent'}^{\scriptscriptstyle(n)}
                        }$ 
                        & ${\scriptstyle 3\Nc + \ncomponent + 1}$
                        & ${\scriptstyle3\Nc + \ncomponent + 1}$\\
               \bottomrule
		\end{tabularx}
\end{table}

\subsection{FIM Terms}\label{app_fim}
The individual terms of the local per-\ac{RS} channel \ac{FIM} from~\eqref{eq:fimetachn} are summarized in Table~\ref{tab:fim-terms}.
We use the indices $i$ and $j$ to describe the rows and columns of $\big[\fimetachn\big]_{i,j}$, respectively, depending on the component numbers $\ncomponent,\, \ncomponent'$. 
The entries in Table~\ref{tab:fim-terms} describe the upper triangular part of the channel \ac{FIM} while the lower triangular part is to be completed by exploiting the symmetry property, i.e., $\big[\fim\big]_{i,j} \triangleq \big[\fim\big]_{j,i}$ for any \ac{FIM} $\fim$.

For ease of notation, we define the following abbreviations.
For ${\cc}_{\scriptscriptstyle\ncomponent}^{\scriptscriptstyle(n)}\!:=\!\cc(\thetacn, \tauucn)\!=\!(\bb(\tauucn) \odot \sss) \otimes \aaa(\thetacn)$, 
we define the partial derivatives w.r.t. \ac{AoA} $\dot{\cc}^{\scriptscriptstyle(n)}_{\scriptscriptstyle\theta,\ncomponent} := \frac{\partial {\cc}_{\scriptscriptstyle\ncomponent}^{\scriptscriptstyle(n)}}{\partial \theta}= (\bbnc \odot \sss) \otimes \aaap$, 
and the partial derivatives w.r.t. pseudo-delay $\dot{\cc}^{\scriptscriptstyle(n)}_{\scriptscriptstyle\tauu,\ncomponent}\!:=\!\frac{\partial {\cc}_{\scriptscriptstyle\ncomponent}^{\scriptscriptstyle(n)}}{\partial \tauu}\!=\!(\bbd \odot \sss) \otimes \aaanc$, 
with $\aaap\!:=\!\frac{\partial \aaanc}{\partial \theta}$ and $\bbd\!:=\!\frac{\partial \bbnc}{\partial \tauu_n}$ abbreviating 
% \begin{align}\label{eq:aaap}
%     \aaap &=  \frac{j2\pi d \cos \thetacn}{\lambda} 
%     \left[
%         %0, e^{j \wavenumber d \sin\theta} , \hdots , (M-1)e^{j \wavenumber d (M-1) \sin\theta} 
%         -\frac{M\!-\!1}{2} \ \cdots \ \frac{M\!-\!1}{2}
%     \right]^\trp \odot \aaanc
% \end{align}
\begin{align}\label{eq:aaap}
    \aaap &=  \frac{j2\pi d \cos \thetacn}{\lambda} 
    \big[
        %0, e^{j \wavenumber d \sin\theta} , \hdots , (M-1)e^{j \wavenumber d (M-1) \sin\theta} 
        0 \ \cdots \ M\!-\!1
    \big]^\trp \odot \aaanc
\end{align}
    %  ~ \text{and} \\
and
\begin{align}\label{eq:bbd}
    \bbd  &= -j2\pi \deltaf \
    \big[
        0 \ \cdots \ K-1
    \big]^\trp \odot \bbnc \ .
\end{align}
Lumping together the real-valued amplitude and phase into a complex-valued amplitude, we further use the short notation $\gammanc := \alpha_{\scriptscriptstyle n,\ncomponent} \, e^{j \phi_{\scriptscriptstyle n,\ncomponent} }$.

%$\gammanc \ \refcoeff$

\subsection{Jacobian Terms}\label{app_jacobian}
In the following, we assume three-dimensional positioning of the \ac{UE}, i.e., $D = \dim\left(\pp\right) = 3$, while results for $D<3$ can be obtained by removing the respective row $i\in\{1 \ \cdots \ 3\}$ and column $j=i$ of the \textit{known} component $\left[\pp\right]_i$ from the global \ac{FIM} $\fim$.
We make several %notational 
abbreviations for ease of notation.
%For geometric operations, 
For the propagation of information in local channel parameters (i.e., $\etach$) to global parameters (i.e., $\etab$), we define the Householder matrix 
\begin{align}
    \house_\nRP = %\left(
    \Imatrix - 2\, \nw \, \nw^\trp 
    %\right) 
    \quad \in \realset{3}{3}
\end{align}
for all \acp{RP} (i.e., $\nRP>0$) where $\nw$ is a unit-vector orthogonal to the surface of the $\nRP$\textsuperscript{th} wall.
Note that we define $\house_0 := \Imatrix$ for the \ac{LoS} components $\nRP = 0$.

We denote the Cartesian vectorial distance from the $n$\textsuperscript{th} \ac{RS} position at $\pprsx{n}$ to the $\nRP$\textsuperscript{th} mirror \ac{UE} position from~\eqref{eq:ppm} in global coordinates as 
\begin{align}
    \rangem &= \ppm - \pprsx{n} \,.
\end{align}
Note that we define ${\boldsymbol{r}}_{\scriptscriptstyle n,0}^{\text{\tiny{m}}} := \pp - \pprsx{n}$ for the \ac{LoS} components $\nRP = 0$.
Likewise, we define the vectorial distance from the \ac{UE} position $\pp$ to the $\nSP$\textsuperscript{th} \ac{SP} position $\ppSP$ as 
\begin{align}
    \rangeUS 
    %\range_{\nSP} % &= [\rnl{x} \ \rnl{y} \ \rnl{z} ]^\trp \\
    &= \ppSP - \pp
\end{align}
and the vectorial distance from the $n$\textsuperscript{th} \ac{RS} to the $\nSP$\textsuperscript{th} \ac{SP} position as 
\begin{align}
    \rangeS %\range_{n,\nSP} % &= [\rnl{x} \ \rnl{y} \ \rnl{z} ]^\trp \\
    &= \ppSP - \pprsx{n} \,.
\end{align}
%$\range_{n,\nRP} = [\rnl{x} \ \rnl{y} \ \rnl{z} ]^\trp$ %$ = \pp - \pprsx{n}$.
%with $\rnl{x} = p_x^{(\nRP)} - x_n^\mathrm{RS}$, $\rnl{y} = p_y^{(\nRP)} - y_n^\mathrm{RS}$, and $\rnl{z} = p_z^{(\nRP)} - z_n^\mathrm{RS}$.

Taking an arbitrary vector %$\bm{r} := [\rnl{x} \ \rnl{y} \ \rnl{z} ]^\trp$ 
$\bm{r} := [r_{x} \ r_{y} \ r_{z} ]^\trp$ in Cartesian coordinates and expressing it in cylindrical coordinates $\bm{r} := [\varrho \cos \theta \ \varrho \sin \theta \ r_z ]^\trp$, one finds its derivative w.r.t. the azimuth angle $\theta$ as
\begin{align}
    \frac{\partial }{\partial \theta} \bm{r} &= [r_{y} \ -r_{x} \ 0]^\trp \triangleq \frac{\partial }{\partial \theta} \left.\mmb(\theta)\right|_{\theta = 0} \, \bm{r} \,.
\end{align}
These derivatives correspond to the \textit{sensitivities} of the \acp{AoA} of signals incident at \acp{RS} w.r.t. the vectorial distances from the \acp{RS} to the signal origins, hence they appear in $\jacobP^\theta_n$ and $\jacobP^{\theta}_{\text{\tiny SP},n}$.
We define 
\begin{align}
    \mmbp :=\frac{\partial }{\partial \theta} \left.\mmb(\theta)\right|_{\theta = 0}  \triangleq \begin{bmatrix}
        0 & -1 & 0 \\
        1 & 0 & 0 \\
        0 & 0 & 0 \\
    \end{bmatrix} \, 
\end{align}
for notational brevity.
With the above definitions, we can express the submatrices of the Jacobian in~\eqref{eq:fim-jacobian-compact} as follows:

\vspace{0.3cm}

\roundlabeltxt{$\theta \rightarrow \pp$} The derivatives of \acp{AoA} w.r.t. the \ac{UE} position are
\begin{align}
    %\boxed{
    \jacobP^\theta_n = \frac{\partial \thetabn^\trp}{\partial \pp} = 
    \left[ 
        \frac{\partial {\thetabrp}^\trp}{\partial \pp}  
        \ , \
        \frac{\partial {\thetabsp}^\trp}{\partial \pp} 
    \right] 
    \in \realset{3}{(\Nrp+\Nsp)} \, ,
\end{align}
where we distinguish between components $\nRP$ originating from the \ac{LoS} and \acp{RP}
\begin{align} 
    \frac{\partial \thetarp}{\partial \pp}  &= \house_\nRP \  \frac{\mmbp \ \rangem}{\lVert \mmbp \ \rangem \rVert^2 }
    \quad \in \realset{3}{1}
\end{align}
and components $\nSP$ originating from \acp{SP}
\begin{align} 
    \frac{\partial \thetasp}{\partial \pp} &= \bm{0} 
    \quad \in \realset{3}{1} \, .
\end{align}

%\vspace{0.5cm} \hrule \vspace{0.5cm}  % --------- ---------
\roundlabeltxt{$\tauu \rightarrow \pp$} The derivatives of pseudo-delays w.r.t. the \ac{UE} position are
\begin{align*}
    %\boxed{
    \jacobP^{\tauu}_n = \frac{\partial \tauubn^\trp}{\partial \pp} = 
    \left[ 
        \frac{\partial \left. \tauubrp \right.^{\trp}}{\partial \pp}  
        \ , \ 
        \frac{\partial \left. \tauubsp \right.^{\trp}}{\partial \pp} 
    \right] 
    \in \realset{3}{(\Nrp+\Nsp)}
\end{align*}
where we distinguish between \acp{RP} (incl. the \ac{LoS}) and \acp{SP}
\begin{align} 
    \frac{\partial \tauurp}{\partial \pp} &= \house_\nRP \ \frac{\rangem}{\lightspeed \, \lVert \rangem \rVert }
    \quad \in \realset{3}{1} \,,
    \\
    \frac{\partial \tauusp}{\partial \pp} &= \ \frac{-\rangeUS}{\lightspeed \, \lVert \rangeUS \rVert }
    \quad \in \realset{3}{1} \,.
\end{align}

%\vspace{0.5cm} \hrule \vspace{0.5cm}  % --------- --------- 

\roundlabeltxt{$\phi \rightarrow \pp$} The derivatives of phases w.r.t. the \ac{UE} position are
\begin{align*}
    %\boxed{
    \jacobP^\phi_{n} = \frac{\partial {\phibn}^\trp}{\partial \pp} = 
    \left[ 
        \frac{\partial \left. \phibrp \right.^{\trp}}{\partial \pp}  
        \ , \ 
        \frac{\partial \left. \phibsp \right.^{\trp}}{\partial \pp} 
    \right] 
    \in \realset{3}{(\Nrp+\Nsp)}
\end{align*}
where we distinguish between \acp{RP} (incl. the \ac{LoS}) and \acp{SP}
\begin{align}
    \frac{\partial {\phirp} }{\partial \pp}  &= \house_\nRP \ \frac{-2\pi}{\lambda}\frac{\rangem}{\lVert \rangem \rVert }
    \quad \in \realset{3}{1}\,,
    \\
    \frac{\partial  {\phisp} }{\partial \pp} &=  \frac{2\pi}{\lambda} \frac{ \rangeUS}{\lVert \rangeUS \rVert }
    \quad \in \realset{3}{1} \,.
\end{align}

\roundlabeltxt{$\theta \rightarrow \ppspbar$} The derivatives of \acp{AoA} w.r.t. the \ac{SP} positions are
\begin{align*}
    %\boxed{
    \jacobP^{\theta}_{\text{\tiny SP},n} = \frac{\partial \thetabn^\trp}{\partial \ppspbar} = 
    \left[ 
        \frac{\partial {\thetabrp}^\trp}{\partial \ppspbar}  
        \ , \
        \frac{\partial {\thetabsp}^\trp}{\partial \ppspbar} 
    \right] 
    \in \realset{3\Nsp}{(\Nrp+\Nsp)}
\end{align*}
where we distinguish between \acp{RP} (incl. the \ac{LoS}) and \acp{SP}
\begin{align} 
    \frac{\partial \thetarp}{\partial \ppSPx{\nSP'}}  &= \bm{0} \quad \in \realset{3}{1} \quad \forall \, \nRP,\nSP' \,,
    \\
    \frac{\partial \thetasp}{\partial \ppSPx{\nSP'}} 
    &=  
    \begin{cases} 
    \begin{aligned} % aligned for \\[Xpt]
        & \frac{ \mmbp \ \rangeS}{\lVert \mmbp \ \rangeS \rVert^2 } 
    \quad  &\in \realset{3}{1} & \quad %\forall \, 
    \nSP=\nSP' \,,
        \\[3pt]
        & \qquad \bm{0} \quad &\in \realset{3}{1} & \quad \text{else} \,.
    \end{aligned}   % aligned for \\[Xpt]
    \end{cases}
\end{align} 

%\vspace{0.5cm} \hrule \vspace{0.5cm}  % --------- --------- 

\roundlabeltxt{$\tauu \rightarrow \ppspbar$} 
The derivatives of pseudo-delays w.r.t. the \ac{SP} positions are
\begin{align*}
    %\boxed{
    \jacobP^{\tauu}_{\text{\tiny SP},n} = \frac{\partial \tauubn^\trp}{\partial \ppspbar} = 
    \left[ 
        \frac{\partial \left. \tauubrp \right.^{\trp}}{\partial \ppspbar}  
        \ , \ 
        \frac{\partial \left. \tauubsp \right.^{\trp}}{\partial \ppspbar} 
    \right] 
    \in \realset{3\Nsp}{(\Nrp+\Nsp)}
\end{align*}
where we distinguish between \acp{RP} (incl. the \ac{LoS}) and \acp{SP}
\begin{align} 
    \frac{\partial \tauurp }{\partial \ppSPx{\nSP'}} &= \bm{0} \quad \in \realset{3}{1} \quad \forall \, \nRP,\nSP' \,,
    \\
    \frac{\partial \tauusp }{\partial \ppSPx{\nSP'}} &= 
    \begin{cases} 
    \begin{aligned} % aligned for \\[Xpt]
        \frac{\rangeS}{\lightspeed \, \lVert \rangeS \rVert } &+ \frac{\rangeUS}{\lightspeed \, \lVert \rangeUS \rVert } & \in \realset{3}{1} &\quad %\forall \, 
        \nSP=\nSP' \,,
        \\[3pt]
        &\bm{0} & \in \realset{3}{1} & \quad \text{else} \,.
    \end{aligned}   % aligned for \\[Xpt]
    \end{cases}
\end{align}

%\vspace{0.5cm} \hrule \vspace{0.5cm}  % --------- --------- 

\roundlabeltxt{$\phi \rightarrow \ppspbar$} 
The derivatives of phases w.r.t. the \ac{SP} positions are
\begin{align*}
    %\boxed{
    \jacobP^{\phi}_{\text{\tiny SP},n} = \frac{\partial \phibn^\trp}{\partial \ppspbar} = 
    \left[ 
        \frac{\partial \left. \phibrp \right.^{\trp}}{\partial \ppspbar}  
        \ , \ 
        \frac{\partial \left. \phibsp \right.^{\trp}}{\partial \ppspbar} 
    \right] 
    \in \realset{3\Nsp}{(\Nrp+\Nsp)}
\end{align*}
where we distinguish between \acp{RP} (incl. the \ac{LoS}) and \acp{SP}
\begin{align} 
    \frac{\partial \phirp }{\partial \ppSPx{\nSP'}} &= \bm{0}
    \quad \in \realset{3}{1}  \quad \forall \, \nRP,\nSP' \,,
    \\
    \frac{\partial \phisp }{\partial \ppSPx{\nSP'}} &=
    \begin{cases} 
    \begin{aligned} % aligned for \\[Xpt]
        \frac{-2\pi}{\lambda} \frac{\rangeS}{\lVert \rangeS \rVert } &+ \frac{-2\pi}{\lambda} \frac{\rangeUS}{\lVert \rangeUS \rVert }
        %\quad \in \realset{3}{1} 
        & \quad \forall \, \nSP=\nSP' \,,
        \\[3pt]
        ~  &\bm{0} \quad \in \realset{3}{1} & \quad \text{else} \,.
    \end{aligned}   % aligned for \\[Xpt]
    \end{cases}
\end{align}

%\vspace{0.5cm} \hrule \vspace{0.5cm}  % --------- --------- 

\roundlabeltxt{$\tauu \rightarrow \deltac$} The derivatives of pseudo-delays w.r.t. the clock parameters are
\begin{align*}
    %\boxed{
    \jacobC^{\tauu}_{n} = \frac{\partial \tauubn^\trp}{\partial \deltac} = 
    \left[ 
        \frac{\partial \left. \tauubrp \right.^{\trp}}{\partial \deltac}  
        \ , \ 
        \frac{\partial \left. \tauubsp \right.^{\trp}}{\partial \deltac} 
    \right] \in \binset{\dimClock}{(\Nrp+\Nsp)}\,, \\
%\end{align*}
%where we distinguish between \acp{RP} (incl. the \ac{LoS}) and \acp{SP}
%\begin{align*}
    = \begin{cases}
    \begin{aligned} % aligned for \\[Xpt]
    \Bigg[%\left[ 
            \bigg[%\left[ 
                \frac{\partial \tauubrp }{\partial \deltatau}  
                \ , \
                \frac{\partial  \tauubrp }{\partial \deltaphib^\trp}  
            \bigg]^\trp%\right]^\trp
            \ , \
            \bigg[%\left[ 
                \frac{\partial  \tauubsp }{\partial \deltatau} 
                \ , \
                \frac{\partial  \tauubsp }{\partial \deltaphib^\trp} 
            \bigg]^\trp%\right]^\trp
    \Bigg]%\right]%^\trp 
    & \quad \roundlabeltxt{NCP} 
    \\[3pt]
    \Bigg[%\left[ 
            \bigg[%\left[ 
                \frac{\partial \tauubrp}{\partial \deltatau}  
                \ , \
                \frac{\partial  \tauubrp}{\partial \deltaphi}  
            \bigg]^\trp%\right]^\trp
            \ , \
            \bigg[%\left[ 
                \frac{\partial \tauubsp}{\partial \deltatau} 
                \ , \
                \frac{\partial \tauubsp }{\partial \deltaphi }
            \bigg]^\trp%\right]^\trp
    \Bigg]%\right]%^\trp 
    & \quad \roundlabeltxt{CP}
    \end{aligned}   % aligned for \\[Xpt]
    \end{cases} \,, 
\end{align*}
where we distinguish between the \roundlabeltxt{NCP} case
with $\dimClock\!:=\!\dim\left( \deltac \right)\!= N\!+\!1$, and the \roundlabeltxt{CP} case with $\dimClock\!=\!2$.
The individual scalar derivatives are
\begin{align} 
    \frac{\partial \tauurp }{\partial \deltatau}   &= 1 \, ,\qquad 
    &\frac{\partial \tauurp }{\partial {\deltaphi}_{\scriptscriptstyle n'}} &= 0 \quad \forall \, n,n' ~,
    \\
    \frac{\partial \tauusp }{\partial \deltatau}   &= 1 \, ,
    \qquad 
    &\frac{\partial \tauusp }{\partial {\deltaphi}_{\scriptscriptstyle n'}}   &= 0 \quad \forall \, n,n' ~,
\end{align}
where $n'$ is omitted in the \roundlabeltxt{CP} case, and the Jacobian submatrix becomes
\begin{align}\label{eq:jacobctauu}
    %\Rightarrow 
    \jacobC^{\tauu}_{n} = 
    \begin{cases}
    \begin{aligned} % aligned for \\[Xpt]
        &\begin{bmatrix}
           \onematrix{1}{\Nrp}  & \onematrix{1}{\Nsp} \\
            \zeromatrix{N}{\Nrp} & \zeromatrix{N}{\Nsp}
        \end{bmatrix} \quad~
    & \roundlabeltxt{NCP} 
    \\[3pt]
        &\begin{bmatrix}
           \onematrix{1}{\Nrp}  & \onematrix{1}{\Nsp} \\
            \zeromatrix{1}{\Nrp} & \zeromatrix{1}{\Nsp}
        \end{bmatrix} \quad~
    & \roundlabeltxt{CP}
    \end{aligned}   % aligned for \\[Xpt]
    \end{cases} \,.
\end{align}
%where $\bm{0}_{(\mathcal{D}_{\text{\tiny r}} \times \mathcal{D}_{\text{\tiny c}})}$ denotes a $(\mathcal{D}_{\text{\tiny r}} \times \mathcal{D}_{\text{\tiny c}})$-matrix of zeros, i.e., with \textit{all} entries $[\bm{0}]_{i,\iota}\triangleq 0 \, \forall i,\iota$. 
%Likewise, $\bm{1}_{(\mathcal{D}_{\text{\tiny r}} \times \mathcal{D}_{\text{\tiny c}})}$ denotes a $(\mathcal{D}_{\text{\tiny r}} \times \mathcal{D}_{\text{\tiny c}})$-matrix of all ones.

%\vspace{0.5cm} \hrule \vspace{0.5cm}  % --------- --------- 
\roundlabeltxt{$\phi \rightarrow \deltac$} The derivatives of phases w.r.t. the clock parameters are
\begin{align*}
    %\boxed{
    \jacobC^{\phi}_{n} = \frac{\partial \phibn^\trp}{\partial \deltac} = 
    \left[ 
        \frac{\partial \left. \phibrp \right.^{\trp}}{\partial \deltac}  
        \ , \ 
        \frac{\partial \left. \phibsp \right.^{\trp}}{\partial \deltac} 
    \right] \in \binset{\dimClock}{(\Nrp+\Nsp)}
    \\
    = \begin{cases}
    \begin{aligned} % aligned for \\[Xpt]
    \Bigg[%\left[ 
            \bigg[%\left[ 
                \frac{\partial \phibrp }{\partial \delta_{\tau}}  
                \ , \
                \frac{\partial  \phibrp }{\partial \bm{\delta}_\phi^\trp}  
            \bigg]^\trp%\right]^\trp
            \ , \
            \bigg[%\left[ 
                \frac{\partial  \phibsp }{\partial \delta_\tau} 
                \ , \
                \frac{\partial  \phibsp }{\partial \bm{\delta}_{\phi}^\trp} 
            \bigg]^\trp%\right]^\trp
    \Bigg]%\right]%^\trp 
    & \quad \roundlabeltxt{NCP} 
    \\[3pt]
    \Bigg[%\left[ 
            \bigg[%\left[ 
                \frac{\partial \phibrp}{\partial \delta_{\tau}}  
                \ , \
                \frac{\partial  \phibrp}{\partial \delta_\phi}  
            \bigg]^\trp%\right]^\trp
            \ , \
            \bigg[%\left[ 
                \frac{\partial \phibsp}{\partial \delta_\tau} 
                \ , \
                \frac{\partial \phibsp }{\partial \delta_{\phi} }
            \bigg]^\trp%\right]^\trp
    \Bigg]%\right]%^\trp 
    & \quad \roundlabeltxt{CP}
    \end{aligned}   % aligned for \\[Xpt]
    \end{cases} \,.
\end{align*}
%where we distinguish between the \roundlabeltxt{NCP} case
%where $\dimClock = N+1$, and the \roundlabeltxt{CP} case where $\dimClock = 2$.
The individual scalar derivatives are
\begin{align*}
    \frac{\partial \phirp }{\partial {\deltaphi}_{\scriptscriptstyle n'}}   &= 
    \begin{cases}
    \begin{aligned} % aligned for \\[Xpt]
        1 & \quad %\forall 
        n=n' %\,,
        \\[3pt]
        0 & \quad \text{else} %\,,
    \end{aligned}   % aligned for \\[Xpt]
    \end{cases}  , \quad
    &\frac{\partial \phisp }{\partial {\deltaphi}_{\scriptscriptstyle n'}} = 
    \begin{cases} 
    \begin{aligned} % aligned for \\[Xpt]
        1 & \quad %\forall 
        n=n' %\,,
        \\[3pt]
        0 & \quad \text{else} %\,,
    \end{aligned}   % aligned for \\[Xpt]
    \end{cases} 
\end{align*}
\begin{align}
    \frac{\partial \phirp }{\partial \deltatau}   &= 0 \quad \forall n \quad , \quad
    \frac{\partial \phisp }{\partial \deltatau}   = 0 \quad \forall n \qquad 
\end{align}
where $n'$ is omitted in the \roundlabeltxt{CP} case. 
The Jacobian submatrix becomes
\begin{align} \label{eq:jacobcphi}
    \jacobC^{\phi}_{n} = 
    \begin{cases}
    \begin{aligned} % aligned for \\[Xpt]
        \begin{bmatrix}
            \zeromatrix{1}{\Nc} %\bm{0}_{(1\times (\Nrp+\Nsp))}  
            \\
            \bm{A}_{\delta_{\phi,n}} 
        \end{bmatrix} & \quad \roundlabeltxt{NCP} %\,,
        \\[3pt]
        \begin{bmatrix}
            \zeromatrix{1}{\Nc} %\bm{0}_{(1\times (\Nrp+\Nsp))}  
            \\
            \onematrix{1}{\Nc}
        \end{bmatrix}  & \quad \roundlabeltxt{CP} %\,, & \quad \text{else} %\,,
    \end{aligned}   % aligned for \\[Xpt]
    \end{cases}  , 
\end{align}
where $\jacobA_{\delta_{\phi,n}} \in \binset{(\dimClock-1)}{\Nc}$ is an association matrix with all elements being zero except the $n$\textsuperscript{th} row being a row of all ones.
Equivalently, $\bm{A}_{\delta_\phi,n} = \ebn \otimes \onematrix{1}{\Nc} %\bm{1}_{(1 \times \Nrp+\Nsp)}
$, with $\ebn \in \binsetone{N}$ being a unit vector where the $n$\textsuperscript{th} entry is one.

%\vspace{0.5cm} \hrule \vspace{0.5cm}  % --------- --------- 
\roundlabeltxt{$\phi \rightarrow \phir$} 
The component phases $\phibn\!\in\!\realsetone{\Nc}$ of the $n$\textsuperscript{th} \ac{RS} from~\eqref{eq:etach} map to the combined vector of phases $\phir$ from~\eqref{eq_eta_all} with $\dim\left(\phir\right) = %\sum_n\Nc-1$ 
N(\Nc-1)$
via
\begin{align*}
    %\boxed{
    \frac{\partial \phibn^\trp}{\partial \phir} = 
    \jacobA_{\phi,n}
    \quad \in \binset{N(\Nc-1)}{\Nc}
\end{align*}
where $\jacobA_{\phi,n}$ is an association matrix that associates the entries $\phicn$ (with the exception of the \ac{LoS} component $\ncomponent = 0$) in the channel parameter vector $\etach_n$ with the respective phases in $\phir$ of the global parameter vector $\etab$.
%Thus, $\jacobA_{\phi,n}$ has $(L_n-1+J)$ one-entries and zero-entries.
This association matrix can be computed as
\begin{align}\label{eq:JacobA_phi}
    \jacobA_{\phi,n} = \ebn \otimes 
    \left[
        \zeromatrix{(\Nc-1)}{1} \ , \
        \Imatrix
    \right]\,,
\end{align}
with the zero vector omitting to map the \ac{LoS}-component phase to the global parameters. % and $\Imatrix$ being an $(\Nc-1)$-identity matrix.

%\vspace{0.5cm} \hrule \vspace{0.5cm}  % --------- --------- 
\roundlabeltxt{$\alpha \rightarrow \alphabar$} 
The component amplitudes $\alphabn \in \realsetone{\Nc}$ of the $n$\textsuperscript{th} \ac{RS} map to the combined vector of amplitudes $\alphabar$ with $\dim\left(\alphabar\right) = %\sum_n\Nc$ 
N\,\Nc$
via
\begin{align*}
    %\boxed{
    \frac{\partial \alphabn^\trp}{\partial \alphabar} = 
    \jacobA_{\alpha,n}
    \quad \in \binset{N \Nc}{\Nc}
\end{align*}
where $\jacobA_{\alpha,n}$ is an association matrix that associates the entries $\alphacn$ in the channel parameter vector $\etach_n$ with the respective amplitudes in $\alphabar$ of the global parameter vector $\etab$.
%Thus, $\jacobA_{\alpha,n}$ has $(L_n+J)$ one-entries and zero-entries else.
This association matrix can be computed as
\begin{align}\label{eq:JacobA_alpha}
    \jacobA_{\alpha,n} = \ebn \otimes 
        \Imatrix \,.
\end{align}

}{}

\end{appendices}

%%%%%%%%%%%%%%%%%%%%%%%%%%%%%%%%%%%%%%%%%%%%%%%%%%%%%%%%
%%%%%%%%%%%%%%%%%%%%%%%%%%%%%%%%%%%%%%%%%%%%%%%%%%%%%%%%
\bibliographystyle{IEEEtran}
\balance
\bibliography{IEEEabrv,main}

\end{document}